\documentclass[10pt,journal]{IEEEtran} %
\usepackage[noadjust]{cite}
\usepackage[bookmarks=false]{hyperref}
\hypersetup{
    colorlinks,
    linkcolor={blue!50!black},
    citecolor={blue!50!black},
    urlcolor={blue!80!black}
}
\usepackage[font=footnotesize]{subcaption}
\usepackage[font=footnotesize]{caption}
\usepackage{caption}
\usepackage[utf8]{inputenc}
\usepackage{graphicx,url}

\usepackage{booktabs}
\graphicspath{ {./images/} }
\usepackage{multirow}
\usepackage{hhline}
\usepackage{xspace}
\usepackage{graphicx}
\usepackage{amssymb,amsmath,mathtools}
\usepackage{amsthm}
\newtheorem{theorem}{Theorem}   %
\newtheorem{remark}{Remark}
\newtheorem{lemma}{Lemma}
\newtheorem{corollary}{Corollary}
\usepackage{bm}                 %
\usepackage{pifont}
\newtheorem{definition}{Definition}
\newtheorem{assumption}{Assumption}
\theoremstyle{plain}
\newtheorem{example}{Example}
\newtheorem{scenario}{Scenario}
\newtheorem{problem}{Problem}
\usepackage{color,soul}
\usepackage[dvipsnames]{xcolor}
\usepackage{bbm}
\usepackage{stmaryrd}
\usepackage{textcomp}
\usepackage{lipsum,lineno}
\usepackage{float}
\makeatletter
\newif\if@restonecol
\makeatother
\usepackage{amsmath}
\usepackage{kbordermatrix}
\usepackage{mathrsfs}
\usepackage[shortlabels]{enumitem}
\newcommand\numeq[1]%
{\stackrel{\scriptscriptstyle(\mkern-1.5mu#1\mkern-1.5mu)}{=}}
\newcommand\numeqq[1]%
{\stackrel{\scriptscriptstyle(\mkern-1.5mu#1\mkern-1.5mu)}{\triangleq}}
\newcommand\numleq[1]%
{\stackrel{\scriptscriptstyle(\mkern-1.5mu#1\mkern-1.5mu)}{\leq}}
\newcommand\numgeq[1]%
{\stackrel{\scriptscriptstyle(\mkern-1.5mu#1\mkern-1.5mu)}{\geq}}
\newcommand\numimp[1]%
{\stackrel{\scriptscriptstyle(\mkern-1.5mu#1\mkern-1.5mu)}{\implies}}
\newcommand\norm[1]{\lVert#1\rVert}
\newcounter{MYtempeqncnt}
\usepackage[titlenumbered,ruled,linesnumbered]{algorithm2e}
\SetAlCapNameFnt{\footnotesize}
\SetAlCapFnt{\footnotesize}
\usepackage[shortlabels]{enumitem}

\let\oldnl\nl% Store \nl in \oldnl
\newcommand{\nonl}{\renewcommand{\nl}{\let\nl\oldnl}}
\usepackage{algpseudocode}

\usepackage{fancyhdr}
\usepackage{tabularx}
\usepackage{dsfont}
\usepackage{listings}
\usepackage{tcolorbox}
\lstdefinestyle{mystyle}{
    backgroundcolor=\color{backcolour},
    commentstyle=\color{codegreen},
    keywordstyle=\color{magenta},
    numberstyle=\tiny\color{codegray},
    stringstyle=\color{codepurple},
    basicstyle=\ttfamily\footnotesize,
    breakatwhitespace=false,
    breaklines=true,
    captionpos=b,
    keepspaces=true,
    showspaces=false,
    showstringspaces=false,
    showtabs=false,
    tabsize=2,
    xleftmargin=50pt,
    xrightmargin=50pt
  }
\usepackage{tikz,pgfplots}
\usepackage{pgfplots}
\usepackage{pgfplotstable}
\definecolor{gray2}{HTML}{ededed}
\definecolor{gray3}{HTML}{F5F5F5}
\definecolor{RoyalAzure}{rgb}{0.0, 0.22, 0.66}
\usetikzlibrary{shapes.geometric,backgrounds,patterns, trees}
\usetikzlibrary{3d,decorations.text,shapes.arrows,positioning,fit,backgrounds}
\usetikzlibrary{positioning, decorations.pathmorphing, shapes}
\usetikzlibrary{decorations.pathreplacing}
\usetikzlibrary{shapes.geometric,backgrounds,patterns, trees}
\usetikzlibrary{spy}
\usetikzlibrary{arrows.meta,
                bending,
                intersections,
                quotes,
                shapes.geometric}
              \usetikzlibrary{automata, positioning}
              \usepgfplotslibrary{fillbetween}
\usetikzlibrary{shapes,arrows}
\usetikzlibrary{arrows.meta}
\usetikzlibrary{positioning}
\tikzset{set/.style={draw,circle,inner sep=0pt,align=center}}
\usetikzlibrary{automata, positioning}
  \usetikzlibrary{shapes,shadows}
  \tikzstyle{abstractbox} = [draw=black, fill=white, rectangle,
  inner sep=10pt, style=rounded corners, drop shadow={fill=black,
  opacity=1}]
\tikzstyle{abstracttitle} =[fill=white]
\usetikzlibrary{calc,positioning,shapes.geometric}
\usetikzlibrary{arrows.meta,arrows}
\DeclareMathOperator*{\argmax}{arg\,max}
\DeclareMathOperator*{\argmin}{arg\,min}

\DeclareMathOperator*{\maximize}{maximize}
\DeclareMathOperator*{\minimize}{minimize}

\usetikzlibrary{matrix}
\tikzstyle{cblue}=[circle, draw, thin,fill=cyan!20, scale=0.8]
\tikzstyle{qgre}=[rectangle, draw, thin,fill=green!20, scale=0.8]
\tikzstyle{rpath}=[ultra thick, red, opacity=0.4]
\tikzstyle{legend_isps}=[rectangle, rounded corners, thin,
                       fill=gray!20, text=blue, draw]

\tikzstyle{legend_overlay}=[rectangle, rounded corners, thin,
                           top color= white,bottom color=green!25,
                           minimum width=2.5cm, minimum height=0.8cm,
                           pinegreen]
\tikzstyle{legend_phytop}=[rectangle, rounded corners, thin,
                          top color= white,bottom color=cyan!25,
                          minimum width=2.5cm, minimum height=0.8cm,
                          royalblue]
\tikzstyle{legend_general}=[rectangle, rounded corners, thin,
                          top color= white,bottom color=lavander!25,
                          minimum width=2.5cm, minimum height=0.8cm,
                          violet]
                          \usetikzlibrary{matrix}

\colorlet{myRed}{red!20}
\tikzset{
  rows/.style 2 args={/utils/temp/.style={row ##1/.append style={nodes={#2}}},
    /utils/temp/.list={#1}},
  columns/.style 2 args={/utils/temp/.style={column ##1/.append style={nodes={#2}}},
    /utils/temp/.list={#1}}}
\usetikzlibrary{backgrounds,calc,shadings,shapes.arrows,shapes.symbols,shadows}
\definecolor{switch}{HTML}{006996}

\makeatletter
\pgfkeys{/pgf/.cd,
  parallelepiped offset x/.initial=2mm,
  parallelepiped offset y/.initial=2mm
}
\pgfdeclareshape{parallelepiped}
{
  \inheritsavedanchors[from=rectangle] % this is nearly a rectangle
  \inheritanchorborder[from=rectangle]
  \inheritanchor[from=rectangle]{north}
  \inheritanchor[from=rectangle]{north west}
  \inheritanchor[from=rectangle]{north east}
  \inheritanchor[from=rectangle]{center}
  \inheritanchor[from=rectangle]{west}
  \inheritanchor[from=rectangle]{east}
  \inheritanchor[from=rectangle]{mid}
  \inheritanchor[from=rectangle]{mid west}
  \inheritanchor[from=rectangle]{mid east}
  \inheritanchor[from=rectangle]{base}
  \inheritanchor[from=rectangle]{base west}
  \inheritanchor[from=rectangle]{base east}
  \inheritanchor[from=rectangle]{south}
  \inheritanchor[from=rectangle]{south west}
  \inheritanchor[from=rectangle]{south east}
  \backgroundpath{
    \southwest \pgf@xa=\pgf@x \pgf@ya=\pgf@y
    \northeast \pgf@xb=\pgf@x \pgf@yb=\pgf@y
    \pgfmathsetlength\pgfutil@tempdima{\pgfkeysvalueof{/pgf/parallelepiped
      offset x}}
    \pgfmathsetlength\pgfutil@tempdimb{\pgfkeysvalueof{/pgf/parallelepiped
      offset y}}
    \def\ppd@offset{\pgfpoint{\pgfutil@tempdima}{\pgfutil@tempdimb}}
    \pgfpathmoveto{\pgfqpoint{\pgf@xa}{\pgf@ya}}
    \pgfpathlineto{\pgfqpoint{\pgf@xb}{\pgf@ya}}
    \pgfpathlineto{\pgfqpoint{\pgf@xb}{\pgf@yb}}
    \pgfpathlineto{\pgfqpoint{\pgf@xa}{\pgf@yb}}
    \pgfpathclose
    \pgfpathmoveto{\pgfqpoint{\pgf@xb}{\pgf@ya}}
    \pgfpathlineto{\pgfpointadd{\pgfpoint{\pgf@xb}{\pgf@ya}}{\ppd@offset}}
    \pgfpathlineto{\pgfpointadd{\pgfpoint{\pgf@xb}{\pgf@yb}}{\ppd@offset}}
    \pgfpathlineto{\pgfpointadd{\pgfpoint{\pgf@xa}{\pgf@yb}}{\ppd@offset}}
    \pgfpathlineto{\pgfqpoint{\pgf@xa}{\pgf@yb}}
    \pgfpathmoveto{\pgfqpoint{\pgf@xb}{\pgf@yb}}
    \pgfpathlineto{\pgfpointadd{\pgfpoint{\pgf@xb}{\pgf@yb}}{\ppd@offset}}
  }
}

\makeatletter
\tikzset{anchor/.append code=\let\tikz@auto@anchor\relax,
  add font/.code=%
    \expandafter\def\expandafter\tikz@textfont\expandafter{\tikz@textfont#1},
  left delimiter/.style 2 args={append after command={\tikz@delimiter{south east}
    {south west}{every delimiter,every left delimiter,#2}{south}{north}{#1}{.}{\pgf@y}}}}
\tikzstyle{sms} = [rectangle callout, draw,very thick, rounded corners, minimum height=20pt]
\makeatletter
\tikzset{anchor/.append code=\let\tikz@auto@anchor\relax,
  add font/.code=%
    \expandafter\def\expandafter\tikz@textfont\expandafter{\tikz@textfont#1},
  left delimiter/.style 2 args={append after command={\tikz@delimiter{south east}
    {south west}{every delimiter,every left delimiter,#2}{south}{north}{#1}{.}{\pgf@y}}}}
\tikzstyle{sms} = [rectangle callout, draw,very thick, rounded corners, minimum height=20pt]
\usetikzlibrary{positioning,calc}
\tikzstyle{block} = [rectangle, draw,
text width=10.5em, text centered, rounded corners, minimum height=4em]
\tikzstyle{line} = [draw, -latex]
\tikzset{
  mybackground9/.style={execute at end picture={
        \begin{scope}[on background layer]
          \draw[black,fill=black!5,rounded corners=6ex] (current bounding box.south west)
                    rectangle (current bounding box.north east);
          \node[draw,fill=white,ellipse,anchor=west,inner sep=1pt,minimum width=4ex] at (current bounding box.north
                   west){#1};
        \end{scope}
    }},
}

\tikzset{
  mybackground13/.style={execute at end picture={
        \begin{scope}[on background layer]
          \draw[black, fill=gray2, rounded corners=4ex] (current bounding box.south west)
                    rectangle (current bounding box.north east);
          \node[draw,fill=white,ellipse,anchor=west,inner sep=1pt,minimum width=4ex] at (current bounding box.north
                   west){#1};
        \end{scope}
    }},
}
\tikzset{
  mybackground14/.style={execute at end picture={
        \begin{scope}[on background layer]
          \draw[black, rounded corners=2ex] (current bounding box.south west)
                    rectangle (current bounding box.north east);
          \node[draw,fill=white,ellipse,anchor=west,inner sep=1pt,minimum width=4ex] at (current bounding box.north
                   west){#1};
        \end{scope}
    }},
}

\tikzset{
  mybackground6/.style={execute at end picture={
        \begin{scope}[on background layer]
          \draw[black,rounded corners=1ex, line width=0.15mm] (current bounding box.south west)
                    rectangle (current bounding box.north east);
          \node[draw,fill=white,ellipse,anchor=west,inner sep=1pt,minimum width=4ex] at (current bounding box.north
                   west){#1};
        \end{scope}
    }},
}

\tikzset{
  mybackground11/.style={execute at end picture={
        \begin{scope}[on background layer]
          \draw[black, fill=Black!80!Sepia!9, rounded corners=6ex] (current bounding box.south west)
                    rectangle (current bounding box.north east);
          \node[draw,fill=white,ellipse,anchor=west,inner sep=1pt,minimum width=4ex] at (current bounding box.north
                   west){#1};
        \end{scope}
    }},
}

\tikzset{
  mybackground15/.style={execute at end picture={
        \begin{scope}[on background layer]
          \draw[black, fill=Black!80!Sepia!9, rounded corners=3ex] (current bounding box.south west)
                    rectangle (current bounding box.north east);
          \node[draw,fill=white,ellipse,anchor=west,inner sep=1pt,minimum width=4ex] at (current bounding box.north
                   west){#1};
        \end{scope}
    }},
}

\tikzset{
  mybackground12/.style={execute at end picture={
        \begin{scope}[on background layer]
          \draw[black, fill=Black!40!Emerald!30, rounded corners=3ex, line width=0.3mm] (current bounding box.south west)
                    rectangle (current bounding box.north east);
        \end{scope}
    }},
}
\tikzset{
  mybackground18/.style={execute at end picture={
      \begin{scope}[on background layer]
        \draw[black, fill=gray3, rounded corners=3.5ex] (current bounding box.south west)
        rectangle (current bounding box.north east);
        \node[draw,fill=white,ellipse,anchor=west,inner sep=1pt,minimum width=4ex] at (current bounding box.north
        west){#1};
      \end{scope}
    }}
}
\tikzset{
  mybackground58/.style={execute at end picture={
        \begin{scope}[on background layer]
          \draw[black, fill=blue!40!black!5, rounded corners=1ex] (current bounding box.south west)
                    rectangle (current bounding box.north east);
          \node[draw,fill=white,ellipse,anchor=west,inner sep=1pt,minimum width=4ex, rounded corners=1ex] at (current bounding box.north
                   west){#1};
        \end{scope}
    }},
}
\tikzset{l3 switch/.style={
    parallelepiped,fill=switch, draw=white,
    minimum width=0.75cm,
    minimum height=0.75cm,
    parallelepiped offset x=1.75mm,
    parallelepiped offset y=1.25mm,
    path picture={
      \node[fill=white,
        circle,
        minimum size=6pt,
        inner sep=0pt,
        append after command={
          \pgfextra{
            \foreach \angle in {0,45,...,360}
            \draw[-latex,fill=white] (\tikzlastnode.\angle)--++(\angle:2.25mm);
          }
        }
      ]
       at ([xshift=-0.75mm,yshift=-0.5mm]path picture bounding box.center){};
    }
  },
  ports/.style={
    line width=0.3pt,
    top color=gray!20,
    bottom color=gray!80
  },
  rack switch/.style={
    parallelepiped,fill=white, draw,
    minimum width=1.25cm,
    minimum height=0.25cm,
    parallelepiped offset x=2mm,
    parallelepiped offset y=1.25mm,
    xscale=-1,
    path picture={
      \draw[top color=gray!5,bottom color=gray!40]
      (path picture bounding box.south west) rectangle
      (path picture bounding box.north east);
      \coordinate (A-west) at ([xshift=-0.2cm]path picture bounding box.west);
      \coordinate (A-center) at ($(path picture bounding box.center)!0!(path
        picture bounding box.south)$);
      \foreach \x in {0.275,0.525,0.775}{
        \draw[ports]([yshift=-0.05cm]$(A-west)!\x!(A-center)$)
          rectangle +(0.1,0.05);
        \draw[ports]([yshift=-0.125cm]$(A-west)!\x!(A-center)$)
          rectangle +(0.1,0.05);
       }
      \coordinate (A-east) at (path picture bounding box.east);
      \foreach \x in {0.085,0.21,0.335,0.455,0.635,0.755,0.875,1}{
        \draw[ports]([yshift=-0.1125cm]$(A-east)!\x!(A-center)$)
          rectangle +(0.05,0.1);
      }
    }
  },
  server/.style={
    parallelepiped,
    fill=white, draw,
    minimum width=0.35cm,
    minimum height=0.75cm,
    parallelepiped offset x=3mm,
    parallelepiped offset y=2mm,
    xscale=-1,
    path picture={
      \draw[top color=gray!5,bottom color=gray!40]
      (path picture bounding box.south west) rectangle
      (path picture bounding box.north east);
      \coordinate (A-center) at ($(path picture bounding box.center)!0!(path
        picture bounding box.south)$);
      \coordinate (A-west) at ([xshift=-0.575cm]path picture bounding box.west);
      \draw[ports]([yshift=0.1cm]$(A-west)!0!(A-center)$)
        rectangle +(0.2,0.065);
      \draw[ports]([yshift=0.01cm]$(A-west)!0.085!(A-center)$)
        rectangle +(0.15,0.05);
      \fill[black]([yshift=-0.35cm]$(A-west)!-0.1!(A-center)$)
        rectangle +(0.235,0.0175);
      \fill[black]([yshift=-0.385cm]$(A-west)!-0.1!(A-center)$)
        rectangle +(0.235,0.0175);
      \fill[black]([yshift=-0.42cm]$(A-west)!-0.1!(A-center)$)
        rectangle +(0.235,0.0175);
    }
  },
}

\usetikzlibrary{calc, shadings, shadows, shapes.arrows}
\tikzset{cross/.style={cross out, draw=black, minimum size=2*(#1-\pgflinewidth), inner sep=0pt, outer sep=0pt},
cross/.default={1pt}}
\tikzset{%
  interface/.style={draw, rectangle, rounded corners, font=\LARGE\sffamily},
  ethernet/.style={interface, fill=yellow!50},% ethernet interface
  serial/.style={interface, fill=green!70},% serial interface
  speed/.style={sloped, anchor=south, font=\large\sffamily},% line speed at edge
  route/.style={draw, shape=single arrow, single arrow head extend=4mm,
    minimum height=1.7cm, minimum width=3mm, white, fill=switch!20,
    drop shadow={opacity=.8, fill=switch}, font=\tiny}% inroute/outroute arrows
}
\newcommand*{\shift}{1.3cm}% For placing the arrows later
\newcommand*{\router}[1]{
\begin{tikzpicture}
  \coordinate (ll) at (-3,0.5);
  \coordinate (lr) at (3,0.5);
  \coordinate (ul) at (-3,2);
  \coordinate (ur) at (3,2);
  \shade [shading angle=90, left color=switch, right color=white] (ll)
    arc (-180:-60:3cm and .75cm) -- +(0,1.5) arc (-60:-180:3cm and .75cm)
    -- cycle;
  \shade [shading angle=270, right color=switch, left color=white!50] (lr)
    arc (0:-60:3cm and .75cm) -- +(0,1.5) arc (-60:0:3cm and .75cm) -- cycle;
  \draw [thick] (ll) arc (-180:0:3cm and .75cm)
    -- (ur) arc (0:-180:3cm and .75cm) -- cycle;
  \draw [thick, shade, upper left=switch, lower left=switch,
    upper right=switch, lower right=white] (ul)
    arc (-180:180:3cm and .75cm);
  \node at (0,0.5){\color{blue!60!black}\Huge #1};% The name of the router
  \begin{scope}[yshift=2cm, yscale=0.28, transform shape]
    \node[route, rotate=45, xshift=\shift] {\strut};
    \node[route, rotate=-45, xshift=-\shift] {\strut};
    \node[route, rotate=-135, xshift=\shift] {\strut};
    \node[route, rotate=135, xshift=-\shift] {\strut};
  \end{scope}
\end{tikzpicture}}

\makeatletter
\pgfdeclareradialshading[tikz@ball]{cloud}{\pgfpoint{-0.275cm}{0.4cm}}{%
  color(0cm)=(tikz@ball!75!white);
  color(0.1cm)=(tikz@ball!85!white);
  color(0.2cm)=(tikz@ball!95!white);
  color(0.7cm)=(tikz@ball!89!black);
  color(1cm)=(tikz@ball!75!black)
}
\tikzoption{cloud color}{\pgfutil@colorlet{tikz@ball}{#1}%
  \def\tikz@shading{cloud}\tikz@addmode{\tikz@mode@shadetrue}}
\makeatother

\tikzset{my cloud/.style={
     cloud, draw, aspect=2,
     cloud color={gray!5!white}
  }
}

\usetikzlibrary{backgrounds}
\usetikzlibrary{patterns}
\pgfmathdeclarefunction{gauss}{3}{%
  \pgfmathparse{1/(#3*sqrt(2*pi))*exp(-((#1-#2)^2)/(2*#3^2))}%
}
\pgfmathdeclarefunction{cdf}{3}{%
  \pgfmathparse{1/(1+exp(-0.07056*((#1-#2)/#3)^3 - 1.5976*(#1-#2)/#3))}%
}
\pgfmathdeclarefunction{fq}{3}{%
  \pgfmathparse{1/(sqrt(2*pi*#1))*exp(-(sqrt(#1)-#2/#3)^2/2)}%
}
\pgfmathdeclarefunction{fq0}{1}{%
  \pgfmathparse{1/(sqrt(2*pi*#1))*exp(-#1/2)}%
}
\renewcommand\thetable{\arabic{table}}
\usepackage{etoolbox}

\newcommand{\Crossb}{\mathbin{\tikz [x=2ex,y=2ex,line width=.14ex] \draw (0,0) -- (1,1) (0,1) -- (1,0);}}%

\newcommand{\mybox}[4]{
    \begin{figure}[H]
        \centering
    \begin{tikzpicture}
      \node[anchor=text,text width=\columnwidth-1.2cm, draw, rounded corners, line width=1pt, fill=#3, inner sep=3.5mm] (big) {\\#4};
        \node[draw, rounded corners, line width=.5pt, fill=#2, anchor=west, xshift=5mm] (small) at (big.north west) {#1};
    \end{tikzpicture}
    \end{figure}
  }

\newcommand{\figref}[1]{\hyperref[#1]{Fig.~\ref*{#1}}}
\newcommand{\Figref}[1]{\hyperref[#1]{Figure~\ref*{#1}}}
\newcommand{\tableref}[1]{\hyperref[#1]{Table~\ref*{#1}}}
\newcommand{\appendixref}[1]{\hyperref[#1]{Appendix~\ref*{#1}}}
\newcommand{\theoremref}[1]{\hyperref[#1]{Thm.~\ref*{#1}}}
\newcommand{\Theoremref}[1]{\hyperref[#1]{Theorem~\ref*{#1}}}
\newcommand{\lemmaref}[1]{\hyperref[#1]{Lemma~\ref*{#1}}}
\newcommand{\propref}[1]{\hyperref[#1]{Prop.~\ref*{#1}}}
\newcommand{\Propref}[1]{\hyperref[#1]{Proposition~\ref*{#1}}}
\newcommand{\corref}[1]{\hyperref[#1]{Cor.~\ref*{#1}}}
\newcommand{\Corref}[1]{\hyperref[#1]{Corollary~\ref*{#1}}}
\newcommand{\scenarioref}[1]{\hyperref[#1]{Scenario~\ref*{#1}}}
\newcommand{\Scenarioref}[1]{\hyperref[#1]{\textsc{scenario}~\ref*{#1}}}
\newcommand{\probref}[1]{\hyperref[#1]{Prob.~\ref*{#1}}}
\newcommand{\Probref}[1]{\hyperref[#1]{Problem~\ref*{#1}}}
\newcommand{\gameref}[1]{\hyperref[#1]{Game~\ref*{#1}}}
\newcommand{\chapterref}[1]{\hyperref[#1]{Chapter~\ref*{#1}}}
\newcommand{\sectionref}[1]{\hyperref[#1]{\S\ref*{#1}}}
\newcommand{\Algref}[1]{\hyperref[#1]{Algorithm ~\ref*{#1}}}
\newcommand{\myalgref}[1]{\hyperref[#1]{Alg.~\ref*{#1}}}
\newcommand{\Myalgref}[1]{\hyperref[#1]{Algorithm~\ref*{#1}}}
\newcommand{\defref}[1]{\hyperref[#1]{Def.~\ref*{#1}}}
\newcommand{\Defref}[1]{\hyperref[#1]{Definition~\ref*{#1}}}
\newcommand{\assumptionref}[1]{\hyperref[#1]{Assumption~\ref*{#1}}}
\newcommand{\remarkref}[1]{\hyperref[#1]{Remark~\ref*{#1}}}
\newcommand{\exampleref}[1]{\hyperref[#1]{Ex.~\ref*{#1}}}

\usepackage{fontawesome}
\usepackage{lettrine}

\newcommand{\D}{\mathrm{D}}
\newcommand{\A}{\mathrm{A}}

\newcommand{\acro}[1]{\textsc{#1}\xspace}
\newcommand{\acros}[1]{\textsc{#1}s\xspace}

\newcommand{\posg}{\acro{posg}}
\newcommand{\posgs}{\acros{posg}}

\newcommand{\pomdp}{\acro{pomdp}}
\newcommand{\apt}{\acro{apt}}

\newcommand{\idps}{\acro{idps}}

\newcommand{\sdn}{\acro{sdn}}

\newcommand{\ids}{\acro{ids}}
\newcommand{\cpu}{\acro{cpu}}

\newcommand{\openflow}{\acro{openflow}}
\newcommand{\ssh}{\acro{ssh}}
\newcommand{\irc}{\acro{irc}}
\newcommand{\smtp}{\acro{smtp}}
\newcommand{\snmp}{\acro{snmp}}
\newcommand{\ntp}{\acro{ntp}}
\newcommand{\spark}{\acro{spark}}
\newcommand{\ryu}{\acro{ryu}}
\newcommand{\debian}{\acro{debian}}

\newcommand{\samba}{\acro{samba}}
\newcommand{\jessie}{\acro{jessie}}
\newcommand{\wheezy}{\acro{wheezy}}
\newcommand{\tomcat}{\acro{tomcat}}

\newcommand{\dns}{\acro{dns}}
\newcommand{\snort}{\acro{snort}}
\newcommand{\http}{\acro{http}}
\newcommand{\hsvi}{\acro{hsvi}}
\newcommand{\cem}{\acro{cem}}
\newcommand{\nfsp}{\acro{nfsp}}

\newcommand{\netem}{\acro{netem}}
\newcommand{\apache}{\acro{apache}}
\newcommand{\ts}{\acro{ts}}
\newcommand{\mysql}{\acro{mysql}}
\newcommand{\tcpp}{\acro{tcp}}
\newcommand{\xmas}{\acro{xmas}}

\newcommand{\udp}{\acro{udp}}
\newcommand{\syn}{\acro{syn}}
\newcommand{\mongo}{\acro{mongodb}}
\newcommand{\postgres}{\acro{postgres}}
\newcommand{\telnet}{\acro{telnet}}

\newcommand{\cgroups}{\acro{cgroups}}
\newcommand{\cassandra}{\acro{cassandra}}
\newcommand{\ubuntu}{\acro{ubuntu}}
\newcommand{\vulscan}{\acro{vulscan}}
\newcommand{\hdfs}{\acro{hdfs}}
\newcommand{\ftp}{\acro{ftp}}

\newcommand{\ovs}{\acro{ovs}}
\newcommand{\dmz}{\acro{dmz}}
\newcommand{\rnd}{\acro{r\&d}}
\newcommand{\admin}{\acro{admin}}
\newcommand{\zone}{\acro{zone}}

\newcommand{\cve}{\acro{cve}}
\newcommand{\cwe}{\acro{cwe}}

\newcommand{\ppo}{\acro{ppo}}
\newcommand{\mds}{\acro{mds}}

\makeatletter
\patchcmd{\@makecaption}
  {\scshape}
  {}
  {}
  {}
\makeatother

\allowdisplaybreaks
\begin{document}
\bstctlcite{MyBSTcontrol}

\title{Automated Security Response through \\Online Learning with Adaptive Conjectures}

\author{\IEEEauthorblockN{Kim Hammar \IEEEauthorrefmark{2}, Tao Li\IEEEauthorrefmark{3}, Rolf Stadler\IEEEauthorrefmark{2}, and Quanyan Zhu\IEEEauthorrefmark{3}}\\
 \IEEEauthorblockA{\IEEEauthorrefmark{2}
   Division of Network and Systems Engineering, KTH Royal Institute of Technology, Sweden\\
 \IEEEauthorblockA{\IEEEauthorrefmark{3}
   Department of Electrical and Computer Engineering, New York University, USA\\
 }
 Email: \{kimham, stadler\}@kth.se}, \{tl2636, qz494\}@nyu.edu\\
\today
}
\maketitle
\begin{abstract}
We study automated security response for an \textsc{it} infrastructure and formulate the interaction between an attacker and a defender as a partially observed, non-stationary game. We relax the standard assumption that the game model is correctly specified and consider that each player has a probabilistic conjecture about the model, which may be misspecified in the sense that the true model has probability $0$. This formulation allows us to capture uncertainty and misconception about the infrastructure and the intents of the players. To learn effective game strategies online, we design \textbf{C}onjectural \textbf{O}nline \textbf{L}earning (\textsc{col}), a novel method where a player iteratively adapts its conjecture using Bayesian learning and updates its strategy through rollout. We prove that the conjectures converge to best fits, and we provide a bound on the performance improvement that rollout enables with a conjectured model. To characterize the steady state of the game, we propose a variant of the Berk-Nash equilibrium. We present \textsc{col} through an advanced persistent threat use case. Testbed evaluations show that \textsc{col} produces effective security strategies that adapt to a changing environment. We also find that \textsc{col} enables faster convergence than current reinforcement learning techniques.
\end{abstract}
\section{Introduction}
\begin{figure}
  \centering
  \scalebox{1.2}{
    \input{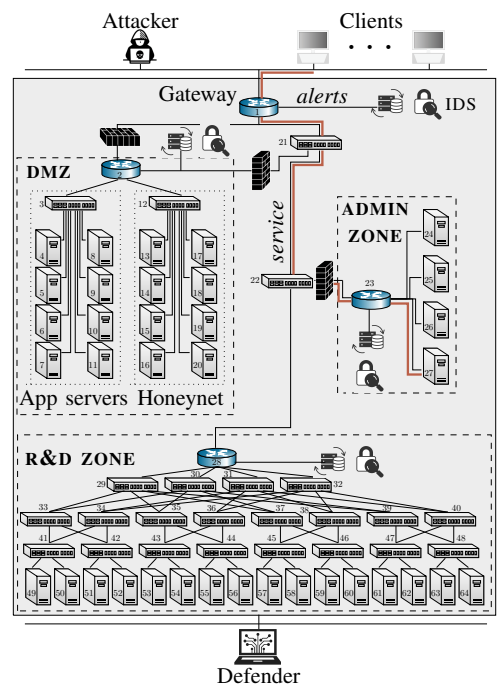}
  }
  \caption{The target infrastructure and the actors involved in the Advanced Persistent Threat (\apt) use case.}
  \label{fig:use_case}
\end{figure}

\lettrine[lines=2]{\textbf{A}}{n} organization's security strategy has traditionally been defined and updated by domain experts. Though this approach can provide basic security for an organization’s \textsc{it} infrastructure, a growing concern is that infrastructure update cycles become shorter and attacks increase in sophistication. To address this challenge, game-theoretic methods for automating security strategies have been proposed, whereby the interaction between an attacker and a defender is modeled as a game \cite{game_t_sec_survey}; see \figref{fig:use_case}. While such methods can produce optimal security strategies, they rely on unrealistic assumptions about the infrastructure. In particular, most of the current methods are limited to stationary and correctly specified games, which assume a static infrastructure that can be accurately modeled without misspecification \cite{flipit,dynamic_game_linan_zhu,tao_info,kamhoua2021game,honeypot_game,DBLP:journals/compsec/HorakBTKK19,hammar_stadler_tnsm_23,kim_gamesec23, ZHAO2020106878, 8691466}. These assumptions are unrealistic for several reasons.

First, \textsc{it} infrastructures are \textit{dynamic}: components fail, software packages are updated, new vulnerabilities are discovered, etc. As a consequence, the game between the attacker and the defender is \textit{non-stationary}. Second, attackers and defenders often have incorrect prior knowledge about the infrastructure and the opponent, which means that players generally have \textit{misspecified} models. Third, the defender has limited knowledge about an attacker's presence and actions, which means that the game has \textit{partial observability}.

\vspace{2mm}

\noindent \textbf{Motivating example: the \textsc{notpetya} attack.} \textsc{notpetya} is a malware that was used by the \textsc{sandworm} Advanced Persistent Threat (\textsc{apt}) in a worldwide attack in 2017 \cite{notpetya_us}. Security researchers initially conjectured that \textsc{notpetya} was a version of the \textsc{petya} ransomware (hence the name) \cite{notpetya_mitre_sandworm_2}. As a result, many organizations focused on traditional ransomware response strategies. However, it later became evident that the malware was not financially motivated but designed for destruction. This misspecification delayed effective responses.

\vspace{2mm}

\begin{figure}
  \centering
  \scalebox{0.78}{
    \input{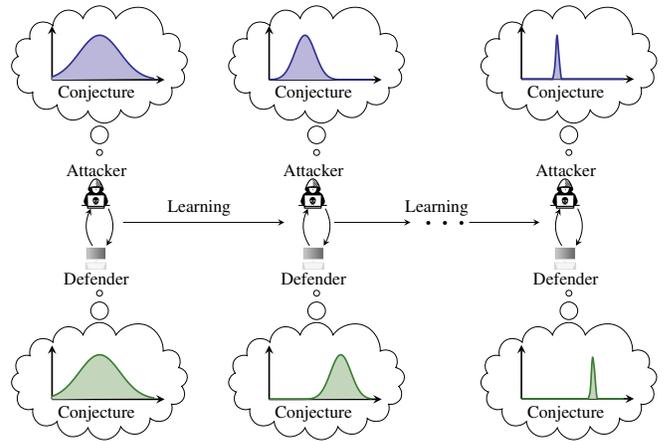}
    }
    \caption[\textbf{C}onjectural \textbf{O}nline \textbf{L}earning (\textsc{col}).]{\textbf{C}onjectural \textbf{O}nline \textbf{L}earning (\hyperref[alg:online_rollout]{\textsc{col}}): each player adapts a (possibly misspecified) conjecture about the game model through Bayesian learning.}
    \label{fig:col_highlevel}
\end{figure}

In this paper, we address the above challenges and present \textbf{C}onjectural \textbf{O}nline \textbf{L}earning (\hyperref[alg:online_rollout]{\textsc{col}}), a game-theoretic method for \textit{online learning} of security strategies that applies to dynamic \textsc{it} environments where attackers and defenders have misconceptions about the environment and the opponent's strategy. Using this method, we formulate the interaction between an attacker and a defender as a non-stationary, partially observed game. We relax the standard assumption that the game model is correctly specified and consider the case where each player has a probabilistic \textit{conjecture} about the model, i.e., a probability distribution over possible models, which may be \textit{misspecified} in the sense that the true model has probability $0$. Both players iteratively adapt their conjecture using \textit{Bayesian learning} and update their strategies using \textit{rollout}, which is a form of approximate dynamic programming \cite{bertsekas2021rollout}; see \figref{fig:col_highlevel}. We prove that the conjectures converge to best fits, and we provide a bound on the performance improvement that rollout enables with a conjectured model. To characterize the steady state of the game, we define a variant of the \hyperref[def:berk_nash]{\textit{Berk-Nash equilibrium}} \cite[Def. 1]{berk_nash}, which represents a fixed point where players act optimally given their conjectures.

While the study of learning with misspecified models has attracted long-standing interest in economics \cite{berk_nash}, engineering \cite{kagel_mechanism_design}, and psychology \cite{rabin_psychology}, it remains unexplored in the security context. Related research in the security literature include (\textit{i}) game-theoretic approaches based on bounded rationality \cite{simons_bounded_rationality, samuelson_bounded_ratinality, Rosenthal1989,8691466,9144263,bounded_rational_stackelberg_1,8362263, ijcai2018p775,9559403,8750848,behavioral_gt_1,behavioral_gt_2}; (\textit{ii}) game-theoretic approaches based on imperfect and incomplete information \cite{posg_cyber_deception_network_epidemic,hammar_stadler_cnsm_20,hammar_stadler_tnsm_23,9328143, nework_security_alpcan}; and (\textit{iii}) model-free learning techniques \cite{learning_in_games_fudenberg,young_strategic_2004,nash_q_learning,markov_game_q_littman,ge_li_zhu_infocomm_workshop,tao_info,hammar_stadler_cnsm_20,hammar_stadler_tnsm,hammar_stadler_tnsm_23,kim_gamesec23, r1_ref1}. (A review of related work can be found in \sectionref{sec:related_work}.) To our knowledge, we provide the first study of learning with misspecified models in a security context. The benefit of this approach is threefold. First, it provides a new methodology to capture uncertainty and misspecification in security games. Second, as we show in this paper, it applies to dynamic, non-stationary, and partially observed games. Third, the model conjectures produced by our method are guaranteed to converge under reasonable conditions, and the worst-case performance of the learned strategies is bounded.

We present our method (\hyperref[alg:online_rollout]{\textsc{col}}) through a use case that involves an \apt{} on an \textsc{it} infrastructure; see \figref{fig:use_case}. We emulate this infrastructure with a \textit{digital twin}, on which we run \apt actions and defender responses. (A video demonstration of the digital twin is available at \cite{csle_docs}.) During such runs, we collect measurements and logs, from which we estimate infrastructure statistics. This data is then used to instantiate simulations of the use case, based on which we evaluate the performance of \hyperref[alg:online_rollout]{\textsc{col}}. We find that \hyperref[alg:online_rollout]{\textsc{col}} produces effective security strategies that adapt to a changing environment. The simulations also show that \hyperref[alg:online_rollout]{\textsc{col}} enables faster convergence than current reinforcement learning techniques. In addition to the simulation studies, we evaluate \hyperref[alg:online_rollout]{\textsc{col}} on the digital twin and compare it against the \snort Intrusion Detection and Prevention System (\idps) \cite{snort}. The results attest that \hyperref[alg:online_rollout]{\textsc{col}} adapts to changes in the distribution of network traffic and outperforms \snort in several key metrics, e.g., percentage of blocked attack attempts and throughput of client requests.

\vspace{2mm}

\noindent\textbf{Contributions.}
\begin{enumerate}
\item We introduce a novel game-theoretic formulation for the problem of automated security response where each player (i.e., attacker or defender) has a probabilistic conjecture about the game model. This formulation allows us to capture model misspecification and uncertainty.
\item We present \hyperref[alg:online_rollout]{\textsc{col}}, a new method for online learning of game strategies where a player iteratively adapts its conjecture using Bayesian learning and updates its strategy through rollout. This method allows us to automatically adapt security strategies to changes in the environment.
\item We prove that, when using \hyperref[alg:online_rollout]{\textsc{col}}, the conjectures of both players converge, and we characterize the steady state as a variant of the \hyperref[def:berk_nash]{Berk-Nash equilibrium} \cite[Def. 1]{berk_nash}. We also provide a bound on the performance improvement that rollout enables with a conjectured model.
\item We evaluate \hyperref[alg:online_rollout]{\textsc{col}} using simulation and emulation studies based on a digital twin running $64$ virtualized servers and $10$ different types of \apt{}s \cite{csle_docs}. This evaluation provides insights into how \hyperref[alg:online_rollout]{\textsc{col}} performs under different conditions and shows that it converges faster than current reinforcement learning techniques. It also shows that \hyperref[alg:online_rollout]{\textsc{col}} outperforms the \snort \textsc{idps} in several key metrics \cite{snort}.
\end{enumerate}

\vspace{2mm}

\textbf{Software and data availability.} Source code of our platform for creating digital twins and a dataset of $6400$ \textsc{apt} traces are available in the repository at \cite{csle_docs}. This repository also includes container images that implement the \textsc{apt}s and the response actions used in our experimental evaluation.
\section{Use Case: Advanced Persistent Threat (\apt{})}\label{sec:use_case}
We consider the problem of defending an organization's \textsc{it} infrastructure against an \apt{} caused by an \textit{attacker} \cite{r1_ref4}. The operator of the infrastructure, which we call the \textit{defender}, takes measures to protect it against the attacker while providing services to a client population; see \figref{fig:use_case}. The infrastructure includes a set of servers and an Intrusion Detection System (\ids) that logs events in real-time. Clients access the services through a public gateway, which is also open to the attacker.

The attacker aims to intrude on the infrastructure over an extended period. It begins with reconnaissance to identify vulnerabilities, after which it attempts to compromise servers through exploits. Once inside the infrastructure, the attacker employs lateral movement techniques, escalates privileges, and uses advanced evasion tactics to avoid detection.

The defender monitors the infrastructure by observing \ids alerts and other statistics. It can recover potentially compromised servers (e.g., by upgrading their software), which temporarily disrupts service for clients. When deciding to take this response action, the defender balances two conflicting objectives: (\textit{i}) maintain services to its clients; and (\textit{ii}) recover compromised servers.
\section{Game Model of the apt Use Case}\label{sec:system_model}
We formulate the above use case as a zero-sum stochastic game with one-sided partial observability
\begin{align}
\Gamma \triangleq \langle \mathcal{N}, \mathcal{S}, (\mathcal{A}_{\mathrm{k}})_{\mathrm{k} \in \mathcal{N}}, f, c, \gamma,  \mathbf{b}_1, \mathcal{O}, z \rangle, \label{eq:game_def}
\end{align}
where $\mathcal{N}$ is the set of players, $\mathcal{S}$ is the set of states, $(\mathcal{A}_{\mathrm{k}})_{\mathrm{k} \in \mathcal{N}}$ are the sets of actions, and $\mathcal{O}$ is the set of observations; $f$ is the transition function, $c$ is the cost function, and $z$ is the observation function; $\gamma$ is the discount factor; and $\mathbf{b}_1$ is the initial state distribution \cite[Def. 3.1]{horak_solving_one_sided_posgs}. $\Gamma$ is a discrete-time game with two players: the ($\mathrm{D}$)efender and the ($\mathrm{A}$)ttacker.

The attacker and the defender have different observability in the game. The defender observes \ids alerts but has no certainty about an attacker's presence. (While we focus on the \ids alert metric in this paper, our model can be used with alternative sources of metrics, e.g., information flow tracking \cite{217638,r1_ref4}.) The attacker, on the other hand, has complete observability. It has access to all the information the defender has access to and the defender’s past actions. We can motivate this assumption in several ways. First, the assumption holds for insider attacks \cite{hammar_stadler_tnsm_23}. Second, it reflects that it is generally not known what information is available to the attacker \cite{horak_solving_one_sided_posgs}. Third, it reduces the computational complexity of solving the game \cite{NIPS2007_3435c378,horak_thesis}.

In the following subsections, we define the components of the game, its evolution, and the players' objectives.
\begin{figure*}[!t]
\normalsize
\setcounter{MYtempeqncnt}{\value{equation}}
\setcounter{equation}{5}
\begin{equation}
\label{eq_belief_operator}
\mathds{B}(\mathbf{b}_{t-1}, a^{(\mathrm{D})}_{t-1}, o_t, \pi_{\mathrm{A}})(s_t) \triangleq \frac{z(o_{t} \mid s_{t})\sum_{s_{t-1} \in \mathcal{S}}\sum_{a^{(\mathrm{A})}_{t-1} \in \mathcal{A}_{\mathrm{A}}}\pi_{\mathrm{A}}(a^{(\mathrm{A})}_{t-1} \mid \mathbf{b}_{t-1},s_{t-1})\mathbf{b}_{t-1}(s_{t-1})f(s_{t} \mid s_{t-1}, a^{(\mathrm{D})}_{t-1},a^{(\mathrm{A})}_{t-1})}{\sum_{a^{(\mathrm{A})}_{t-1} \in \mathcal{A}_{\mathrm{A}}}\sum_{s^{\prime},s\in \mathcal{S}}z(o_{t}\mid s^{\prime})\pi_{\mathrm{A}}(a^{(\mathrm{A})}_{t-1} \mid s, \mathbf{b}_{t-1})f(s^{\prime} \mid s, a^{(\mathrm{D})}_{t-1},a^{(\mathrm{A})}_{t-1})\mathbf{b}_{t-1}(s)}.
\end{equation}
\setcounter{equation}{1}
\hrulefill
\vspace*{4pt}
\end{figure*}

\vspace{2mm}

\noindent\textbf{Notation.} Boldface lower case letters (e.g., $\mathbf{x}=(x_1, x_2,\hdots)$) denote column vectors. $\mathbf{e}_i$ is the $i$-th standard basis vector. Upper case calligraphy letters (e.g., $\mathcal{V}$) represent sets. $\mathbb{P}$ is a probability measure. (The construction of the underlying probability space is standard and shall be omitted for brevity.) The set of probability distributions over $\mathcal{V}$ is written as $\Delta(\mathcal{V})$. $\mathbbm{1}$ is the indicator function. $\delta_i(\cdot)$ is the Dirac delta function. A random variable is written in upper case (e.g., $X$), a random vector in boldface (e.g., $\mathbf{X}$). The expectation of $f$ with respect to $X$ is written as $\mathbb{E}_X[f]$. $x \sim f$ means that $x$ is sampled from $f$. If the expression $f$ includes many random variables that depend on $\pi$, we write $\mathbb{E}_{\pi}[f]$. We use $\mathbb{P}[x|y]$ as a shorthand for $\mathbb{P}[X=x|Y=y]$ and $-k$ as a shorthand for $\mathcal{N}\setminus \{\mathrm{k}\}$. Further notation is listed in \tableref{tab:notation}.
\begin{table}
  \centering
  \scalebox{0.9}{
  \begin{tabular}{ll} \toprule
    {\textit{Notation(s)}} & {\textit{Description}} \\ \midrule
    $\Gamma, c, N$ & The game (\ref{eq:game_def}), cost function (\ref{eq:cost_fun}), and \# servers (\sectionref{sec:game_dynamics}) \\
    $\mathrm{D}, \mathrm{A}$ & The defender player and the attacker player (\ref{eq:game_def})\\
    $\mathcal{N},\mathcal{S},\mathcal{O}$ & Sets of players, states, and observations (\ref{eq:game_def})\\
    $\mathcal{A}_{\mathrm{D}},\mathcal{A}_{\mathrm{A}}$ & Sets of defender and attacker actions (\ref{eq:game_def})\\
    $t, \gamma$ & Time step and discount factor (\ref{eq:objectives})\\
    $\pi_{\mathrm{D}},\pi_{\mathrm{A}}$ & Defender and attacker strategies (\sectionref{sec:strategies})\\
    $\Pi = \Pi_{\mathrm{D}} \times \Pi_{\mathrm{A}}$ & Defender and attacker strategy spaces (\sectionref{sec:strategies})\\
    $\tilde{\pi}_{\mathrm{D}},\tilde{\pi}_{\mathrm{A}}$ & \hyperref[eq:best_responses]{Best response} strategies\\
    $\bm{\pi}^{\star}=(\pi^{\star}_{\mathrm{D}},\pi^{\star}_{\mathrm{A}})$ & Nash equilibrium strategies (\ref{eq:minmax_objective})\\
    $\mathscr{B}_{\mathrm{D}}, \mathscr{B}_{\mathrm{A}}$ & \hyperref[eq:best_responses]{Best response} correspondences\\
    $J_{\mathrm{D}}, J_{\mathrm{A}}$ & Defender and attacker objectives (\ref{eq:objectives})\\
    $f, z$ & Transition function (\ref{eq:transitions}) and observation function (\ref{eq:obs_fun}) \\
    $s_t, o_t$ & State (\ref{eq:transitions}) and observation (\ref{eq:obs_fun}) at time $t$\\
    $\mathbf{a}_t=(a_t^{(\mathrm{D})}, a_t^{(\mathrm{A})})$ & Actions at time $t$ (\sectionref{sec:actions})\\
    $S_t,O_t$ & Random variables with realizations $s_t$ (\ref{eq:transitions}) and $o_t$ (\ref{eq:obs_fun})\\
    $\mathbf{A}_{t}$ & Random vector with realization $\mathbf{a}_t$ (\sectionref{sec:actions})\\
    $\mathbf{b}_t, \mathbf{B}_t$ & Defender belief ($\mathbf{b}_t$ realizes the random vector $\mathbf{B}_t$) (\ref{eq:belief_upd})\\
    $\mathcal{B}, \mathds{B}$ & Belief space and belief operator of the defender (\ref{eq_belief_operator})\\
    $\mathbf{h}^{(\mathrm{k})}_{t}, \mathbf{h}_{t}$ & History of player $\mathrm{k}$ and joint history (\sectionref{sec:observability})\\
    $\mathbf{H}^{(\mathrm{k})}_{t}, \mathbf{H}_{t}$ & Random vectors with realizations $\mathbf{h}^{(\mathrm{k})}_{t}$ and $\mathbf{h}_{t}$ (\sectionref{sec:observability})\\
    $\mathcal{H}_t = \mathcal{H}^{(\mathrm{D})}_{t} \times \mathcal{H}^{(\mathrm{A})}_{t}$ & History spaces (\sectionref{sec:observability})\\
    $\mathsf{S}, \mathsf{C}$ & Stop and continue actions (\sectionref{sec:actions})\\
    $\mathscr{R}$ & Rollout operator for online learning (\ref{eq:rollout_operator})\\
    $\overline{\pi}_{-\mathrm{k}, t}$ & Player $\mathrm{k}$'s conjecture of player $-\mathrm{k}$'s strategy (\myalgref{alg:online_rollout})\\
    $\ell_{\D}, \ell_{\A}$ & Lookahead horizons (\myalgref{alg:online_rollout})\\
    $\overline{\ell}_{-\mathrm{k}, t}$ & Player $\mathrm{k}$'s conjecture of player $-\mathrm{k}$'s lookahead (\myalgref{alg:online_rollout})\\
    $\bm{\theta}_{t}$ & Parameter vector of $\Gamma$ at time $t$ (\ref{eq:bayesian_estimator_game_1})\\
    $\overline{\bm{\theta}}^{(\mathrm{k})}_{t}$ & Player $\mathrm{k}$'s conjecture of $\bm{\theta}_{t}$ at time $t$ (\ref{eq:bayesian_estimator_game_1})\\
    $\mathcal{L}, \Theta_{\mathrm{k}}$ & Player $\mathrm{k}$'s sets of possible conjectures of $\ell_\A$ and $\bm{\theta}$ (\ref{eq:bayesian_learning})\\
    $\mathcal{L}^{\star}, \Theta_{\mathrm{k}}^{\star}$ & Sets of consistent conjectures (\ref{eq:consistent_conjecture_sets})\\
    $\mathbf{i}_t^{(\mathrm{k})},\mathbf{I}_t^{(\mathrm{k})}$ & Information feedback of player $\mathrm{k}$ at time $t$ (\ref{eq:information_feedback})\\
    $\mu_t,\rho^{(\mathrm{k})}_t$ & Posteriors $\mathbb{P}[\ell_{\A} \mid \mathbf{h}_t^{(\D)}]$ (\ref{eq:bayesian-update}) and $\mathbb{P}[\bm{\theta}^{(\mathrm{k})}_t \mid \mathbf{h}_t^{(\mathrm{k})}]$ (\ref{eq:bayesian_estimator_game_1}) \\
    $\nu, K(\overline{\alpha},\nu)$ & Occupancy measure and discrepancy of conjecture $\overline{\alpha}$ (\ref{eq:discrepancy}) \\
    $\pi_{1,\mathrm{k}}, \pi_{t,\mathrm{k}}$ & Base and rollout strategy of player $\mathrm{k}$ at time $t$ (\ref{eq:rollout_operator}) \\
    $\bm{\pi}_{\mathbf{h}_t}$ & Strategy profile induced by \myalgref{alg:online_rollout} at time $t$ (\sectionref{sec:convergen_analysis})\\
    $\mathbb{P}^{\mathscr{R}}$ & Distribution over $\underset{t\geq 1}{\Crossb}(\mathcal{H}_t^{(\D)} \times \mathcal{H}_t^{(\A)})$ (\theoremref{thm:conjecture_convergence}) \\
   $K_{\mathcal{L}}^\star,K_{\Theta_{\mathrm{k}}}^\star$ & Minimal discrepancy values for $\mathcal{L}$ and $\Theta_{\mathrm{k}}$ (\ref{eq:consistent_conjecture_sets}) \\
    \bottomrule\\
  \end{tabular}}
  \caption{Notation.}\label{tab:notation}
\end{table}
\subsection{Actions}\label{sec:actions}
Both players can invoke two actions: ($\mathsf{S}$)top and ($\mathsf{C}$)ontinue. The action spaces are thus $\mathcal{A}_{\mathrm{D}}\triangleq \mathcal{A}_{\mathrm{A}}\triangleq \{\mathsf{S},\mathsf{C}\}$. $\mathsf{S}$ triggers a change in the game state while $\mathsf{C}$ is a passive action that does not change the state. Specifically, $a^{(\mathrm{A})}_t=\mathsf{S}$ is the attacker's compromise action, and $a^{(\mathrm{D})}_t=\mathsf{S}$ is the defender's recovery action (as defined in the use case \sectionref{sec:use_case}).
\subsection{Dynamics}\label{sec:game_dynamics}
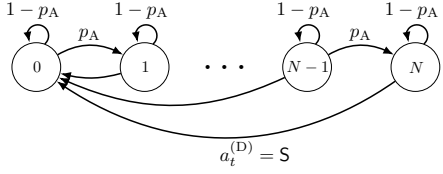
\begin{figure}
  \centering
  \scalebox{0.9}{
    \begin{tikzpicture}[fill=white, >=stealth,
    node distance=3cm,
    database/.style={
      cylinder,
      cylinder uses custom fill,
      shape border rotate=90,
      aspect=0.25,
      draw}]

\node[scale=0.8] (kth_cr) at (0,2.15)
{
  \begin{tikzpicture}

\node[scale=1] (level1) at (-1.7,-5.6)
{
  \begin{tikzpicture}
\node[draw,circle, minimum width=15mm, scale=0.6](s0) at (0,0) {};
\node[draw,circle, minimum width=15mm, scale=0.6](s1) at (2,0) {};
\node[draw,circle, minimum width=15mm, scale=0.6](s3) at (5,0) {};
\node[draw,circle, minimum width=15mm, scale=0.6](s4) at (7,0) {};

\node[inner sep=0pt,align=center, scale=0.85] (time) at (0.07,0)
{
$0$
};

\node[inner sep=0pt,align=center, scale=0.85] (time) at (2.07,0)
{
$1$
};
\node[inner sep=0pt,align=center, scale=0.85] (time) at (5.07,0)
{
$N-1$
};
\node[inner sep=0pt,align=center, scale=0.85] (time) at (7.07,0)
{
$N$
};

\node[inner sep=0pt,align=center, scale=2] (time) at (3.6,0)
{
$\hdots$
};

\draw[thick,-{Latex[length=2mm]}] (s0.70) arc (-50:220:2mm);
\draw[thick,-{Latex[length=2mm]}] (s1.70) arc (-50:220:2mm);
\draw[thick,-{Latex[length=2mm]}] (s3.70) arc (-50:220:2mm);
\draw[thick,-{Latex[length=2mm]}] (s4.70) arc (-50:220:2mm);

\draw[thick,-{Latex[length=2mm]}, bend left=30] (s0) to (s1);
\draw[thick,-{Latex[length=2mm]}, bend left=15] (s1) to (s0);
\draw[thick,-{Latex[length=2mm]}, bend left=30] (s3) to (s4);
\draw[thick,-{Latex[length=2mm]}, bend left=25] (s3) to (s0);
\draw[thick,-{Latex[length=2mm]}, bend left=35] (s4) to (s0);

\node[inner sep=0pt,align=center, scale=1] (time) at (4.1,-1.6)
{
$a^{(\mathrm{D})}_t=\mathsf{S}$
};

\node[inner sep=0pt,align=center, scale=1] (time) at (1.05,0.6)
{
$p_{\mathrm{A}}$
};
\node[inner sep=0pt,align=center, scale=1] (time) at (6.05,0.6)
{
$p_{\mathrm{A}}$
};
\node[inner sep=0pt,align=center, scale=1] (time) at (0,1.05)
{
$1-p_{\mathrm{A}}$
};
\node[inner sep=0pt,align=center, scale=1] (time) at (2,1.05)
{
$1-p_{\mathrm{A}}$
};
\node[inner sep=0pt,align=center, scale=1] (time) at (5,1.05)
{
$1-p_{\mathrm{A}}$
};
\node[inner sep=0pt,align=center, scale=1] (time) at (7,1.05)
{
$1-p_{\mathrm{A}}$
};
    \end{tikzpicture}
  };
    \end{tikzpicture}
  };

\end{tikzpicture}
    }
    \caption{State transition diagram of the game: disks represent states; arrows represent state transitions; labels indicate probabilities and conditions for state transition; the initial state is $s_1=0$.}
    \label{fig:state_transitions}
  \end{figure}

The state $s_t \in \mathcal{S}\triangleq \{0,1,\hdots,N\}$ represents the number of compromised servers at time $t$, where $s_1=0$. The transition $s_t \rightarrow s_{t+1}$ occurs with probability $f\big(s_{t+1} \mid s_t, a^{(\mathrm{D})}_t, a^{(\mathrm{A})}_t\big)$:
\begin{subequations}\label{eq:transitions}
\begin{align}
&f(S_{t+1}=0 \mid s_{t}, \mathsf{S}, a^{(\mathrm{A})}_t) \triangleq 1\label{transition_a}\\
&f(S_{t+1}=s_t \mid s_{t}, \mathsf{C}, \mathsf{C})\triangleq f(S_{t+1}=N \mid N, \mathsf{C}, \mathsf{S}) \triangleq 1\label{transition_b} \\
&f(S_{t+1}=s_t \mid s_{t}, \mathsf{C}, \mathsf{S}) \triangleq 1-p_{\mathrm{A}} \quad\quad\quad\quad s_t < N\label{transition_c}\\
&f(S_{t+1}=s_t+1 \mid s_{t}, \mathsf{C}, \mathsf{S}) \triangleq p_{\mathrm{A}} \quad\quad\quad\quad s_t < N, \label{transition_d}
\end{align}
\end{subequations}
where $p_{\mathrm{A}}$ is the probability of a successful attack. All other transitions have probability $0$; see \figref{fig:state_transitions}.

(\ref{transition_a}) defines the transition $s_t \rightarrow 0$, which occurs when the defender takes action $\mathsf{S}$. (\ref{transition_b})--(\ref{transition_c}) define the recurrent transition $s_{t+1}=s_t$, which occurs when both players take action $\mathsf{C}$ or when the attacker is unsuccessful in compromising a server, which happens with probability $1-p_{\mathrm{A}}$. Lastly, (\ref{transition_d}) defines the transition $s_t \rightarrow s_{t}+1$, which occurs with probability $p_{\mathrm{A}}$ when the attacker takes action $\mathsf{S}$ and the defender takes action $\mathsf{C}$.
%(This assumption reflects the fact that the attacker is an \apt \cite{alshamrani2019survey}.)
\subsection{Observability}\label{sec:observability}
The attacker has complete observability. It knows the game state, the defender's actions, and the defender's observations. In contrast, the defender has a finite set of observations $o_t \in \mathcal{O}$. Consequently, the \textit{information feedbacks} for the attacker and the defender at time $t$ are
\begin{align}
\mathbf{i}_t^{(\mathrm{A})} \triangleq (o_t, s_t, a^{(\mathrm{D})}_{t-1}) \quad\text{and}\quad \mathbf{i}_t^{(\mathrm{D})} \triangleq (o_t),\label{eq:information_feedback}
\end{align}
respectively, where $o_t$ is drawn from a random variable $O_t$ whose distribution depends on the clients and the state, i.e.,
\begin{align}
o_t \sim z(\cdot \mid s_t).\label{eq:obs_fun}
\end{align}
Each player $\mathrm{k}$ has \textit{perfect recall} \cite[Def. 7]{kuhn1953}, which means that it remember the play history $\mathbf{h}^{(\mathrm{k})}_t\triangleq (\mathbf{b}_1, a^{(\mathrm{k})}_{l}, \mathbf{i}_l^{(\mathrm{k})})_{l=1,2,\hdots} \in \mathcal{H}^{(\mathrm{k})}_{t}$. Based on this history, the defender computes the \textit{belief state} $\mathbf{b}_{t} \in \mathcal{B}$, which is defined as
\begin{align}
\mathbf{b}_{t}(s_{t}) &\triangleq \mathbb{P}[S_t=s_t\mid \mathbf{h}^{(\mathrm{D})}_t],\label{eq:belief_upd}
\end{align}
where $\mathbf{b}_t$ is computed recursively through the operator $\mathds{B}$ (\ref{eq_belief_operator}), which is defined at the top of the page.
%where $\pi_\A$ is the attacker strategy and $\mathds{B}$ is defined in (\ref{eq_belief_operator}).
% = \mathds{B}(\mathbf{b}_{t-1}, a^{(\mathrm{D})}_{t-1}, o_t, \pi_{\mathrm{A}})
\setcounter{equation}{6}
\subsection{Strategies and Objectives}\label{sec:strategies}
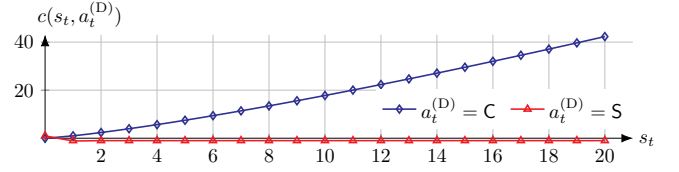
\begin{figure}
  \centering
  \scalebox{0.75}{
    \begin{tikzpicture}[
    % define a style for the dots
    dot/.style={
        draw=black,
        fill=blue!90,
        circle,
        minimum size=3pt,
        inner sep=0pt,
        solid,
    },
    ]

\node[scale=1] (kth_cr) at (0,2.15)
{
  \begin{tikzpicture}[declare function={sigma(\x)=1/(1+exp(-\x));
      sigmap(\x)=sigma(\x)*(1-sigma(\x));}]
\pgfplotstableread{
0 0.0
1 1.0
2 2.378414230005442
3 3.9482220388574776
4 5.656854249492381
5 7.476743906106103
6 9.390507480439723
7 11.3860359318845
8 13.454342644059432
9 15.588457268119896
10 17.78279410038923
11 20.032763155216593
12 22.33451661845039
13 24.684775987507244
14 27.08070988374737
15 29.519845068981457
16 32.0
17 34.51923414277182
18 37.07580859032812
19 39.66815496866703
20 42.294850537622565
}\datatablee

\pgfplotstableread{
0 1
1 -1
2 -1
3 -1
4 -1
5 -1
6 -1
7 -1
8 -1
9 -1
10 -1
11 -1
12 -1
13 -1
14 -1
15 -1
16 -1
17 -1
18 -1
19 -1
20 -1
}\datatableee
\begin{axis}[
%        ymode=log,
        xmin=0,
        xmax=21,
        ymin=-2,
        ymax=43,
%        log basis y={10},
%        width = 6cm,
        width =12cm,
        height = 3.5cm,
        axis lines=center,
        xmajorgrids=true,
        ymajorgrids=true,
        major grid style = {lightgray},
        minor grid style = {lightgray!25},
        scaled y ticks=false,
        yticklabel style={
        /pgf/number format/fixed,
        /pgf/number format/precision=5
        },
%       ticks=none,
%        xlabel={\small $b(1)$},
%        ylabel={\small $\tilde{\pi}_{i,l,\tilde{\theta}^{(i)}}(S|b(1))$},
        xlabel style={below right},
        ylabel style={above left},
        axis line style={-{Latex[length=2mm]}},
        smooth,
        legend style={at={(1,0.5)}, draw=none},
        legend columns=2,
        legend style={
                    % the /tikz/ prefix is necessary here...
                    % otherwise, it might end-up with `/pgfplots/column 2`
                    % which is not what we want. compare pgfmanual.pdf
            /tikz/column 2/.style={
                column sep=5pt,
              }
              }
              ]
\addplot[Blue,mark=diamond, mark repeat=1, name path=l1, thick, domain=0:20] table [x index=0, y index=1] {\datatablee};
\addplot[Red,mark=triangle, mark repeat=1, name path=l1, thick, domain=0:20] table [x index=0, y index=1] {\datatableee};
\legend{$a^{(\mathrm{D})}_t=\mathsf{C}$, $a^{(\mathrm{D})}_t=\mathsf{S}$}
\end{axis}
\end{tikzpicture}
};

\node[inner sep=0pt,align=center, scale=1, rotate=0, opacity=1] (obs) at (5.85,1.42)
{
  $s_t$
};
\node[inner sep=0pt,align=center, scale=1, rotate=0, opacity=1] (obs) at (-4.2,3.65)
{
  $c(s_t, a^{(\mathrm{D})}_t)$
};

  \end{tikzpicture}
  }
  \caption{An example cost function $c(s_t, a^{(\mathrm{D})}_t)$ (\ref{eq:cost_fun}) of $\Gamma$ (\ref{eq:game_def}); hyperparameters are listed in \appendixref{appendix:hyperparameters}}.
  \label{fig:cost_function}
\end{figure}

Since $\mathbf{b}_t$ is a sufficient statistic for the Markovian state $s_t$ (\ref{eq:transitions}) \cite[Def. 4.2, Lem. 5.1, Thm. 7.1]{stochastic_systems_kumar}, we can define the players' \textit{behavior Markov strategies} as $\pi_{\mathrm{D}} \in \Pi_{\mathrm{D}} \triangleq \mathcal{B} \rightarrow \Delta(\mathcal{A}_{\mathrm{D}})$ and $\pi_{\mathrm{A}} \in \Pi_{\mathrm{A}} \triangleq \mathcal{B} \times \mathcal{S} \rightarrow \Delta(\mathcal{A}_{\mathrm{A}})$ \cite[Def. 5]{kuhn1953}. Their performances are quantified using the cost function
\begin{align}
c(s_t, a^{(\mathrm{D})}_t) &\triangleq \overbrace{s^{p}_t\mathbbm{1}_{a^{(\mathrm{D})}_t\neq\mathsf{S}}}^{\text{intrusion cost}} + \overbrace{\mathbbm{1}_{a^{(\mathrm{D})}_t=\mathsf{S}}(q- r\mathbbm{1}_{s_t>0})}^{\text{response action cost}},\label{eq:cost_fun}
\end{align}
where $p \geq 1$, $q > 0$, and $r > 0$ are scalar constants satisfying $1>q-r$; see \figref{fig:cost_function}. The first term in (\ref{eq:cost_fun}) encodes the intrusion cost $s^p_t$, which increases with the number of compromised servers $s_t$. The second term encodes the stop action cost, which is $q-r$ if an intrusion occurs and $q$ otherwise.
% If no intrusion occurs or the defender takes the stop action, the first term in (\ref{eq:cost_fun}) evaluates to $0$.

The goal of the defender is to \textit{minimize} the expected cumulative cost, and the goal of the attacker is to \textit{maximize} the same quantity. Therefore, the objective functions are
\begin{subequations}\label{eq:objectives}
\begin{align}
J^{(\pi_{\mathrm{D}}, \pi_{\mathrm{A}})}_{\mathrm{D}}(\mathbf{b}_1) &\triangleq \mathbb{E}_{(\pi_{\mathrm{D}}, \pi_{\mathrm{A}})}\left[\sum_{t=1}^{\infty}\gamma^{t-1}c(S_t, A^{(\mathrm{D})}_t) \mid \mathbf{b}_1\right] \label{eq:objective_1}\\
J^{(\pi_{\mathrm{D}}, \pi_{\mathrm{A}})}_{\mathrm{A}}(\mathbf{b}_1) &\triangleq -J^{(\pi_{\mathrm{D}}, \pi_{\mathrm{A}})}_{\mathrm{D}}(\mathbf{b}_1), \label{eq:objective_2}
\end{align}
\end{subequations}
where $\gamma \in [0,1)$ is a discount factor and $\mathbb{E}_{(\pi_{\mathrm{D}}, \pi_{\mathrm{A}})}$ is the expectation over the random vectors $(\mathbf{H}^{(\mathrm{D})}_t,\mathbf{H}^{(\mathrm{A})}_t)_{t \in \{1,2,\hdots\}}$ when the game is played according to $(\pi_{\mathrm{D}},\pi_{\mathrm{A}})$.

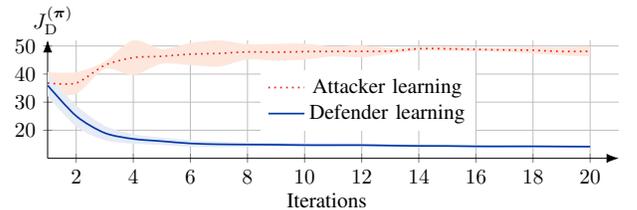
\begin{figure}[H]
  \centering
  \scalebox{0.8}{
    \begin{tikzpicture}
\pgfplotstableread{
1 36.80666666666667 32.93170378644876 40.68162954688459
2 36.80666666666667 32.93170378644876 40.68162954688459
3 43.14333333333334 41.81383607380404 44.472830592862636
4 45.86333333333334 39.677813803436955 52.04885286322972
5 46.326666666666675 43.687238687443894 48.966094645889456
6 47.123333333333335 43.19083008889159 51.05583657777508
7 47.34666666666667 42.66379802756221 52.029535305771134
8 47.919999999999995 45.20682574638705 50.63317425361294
9 47.77333333333333 44.7984466131679 50.74822005349877
10 48.01333333333334 45.167902047414074 50.85876461925261
11 48.06 46.295912380736375 49.82408761926363
12 48.10333333333333 45.997034544645395 50.20963212202127
13 48.156666666666666 47.15980806806714 49.15352526526619
14 49.04333333333333 48.40789239570375 49.67877427096291
15 49.00666666666666 48.296475488173535 49.71685784515979
16 48.776666666666664 48.12303469233639 49.43029864099694
17 48.61333333333334 47.957815878237405 49.26885078842927
18 48.5 47.35271558079065 49.64728441920935
19 48.053333333333335 46.6329509763471 49.47371569031957
20 48.153333333333336 46.34343257887548 49.96323408779119
}\attackertable

\pgfplotstableread{
1 36 32.20558729653528 39.79441270346472
2 25.07190476190476 21.27749205844004 28.866317465369484
3 19.024285714285714 16.456812325640552 21.591759102930876
4 16.89952380952381 15.224415020015162 18.57463259903246
5 16.089523809523808 14.778444615746412 17.400603003301203
6 15.293809523809522 14.125718350920527 16.461900696698518
7 14.99095238095238 14.014874843076168 15.967029918828594
8 14.901904761904762 14.066560615337126 15.737248908472397
9 14.837142857142856 14.149172145523156 15.525113568762556
10 14.705714285714285 14.045894806836642 15.365533764591927
11 14.69047619047619 14.095353104349995 15.285599276602385
12 14.695238095238096 14.139297353143256 15.251178837332937
13 14.537142857142856 14.04246524238014 15.03182047190557
14 14.411904761904763 13.910455505594596 14.91335401821493
15 14.390476190476189 13.93080750633784 14.850144874614537
16 14.264761904761905 13.804396995606522 14.725126813917287
17 14.246190476190474 13.81751543853959 14.674865513841358
18 14.26 13.82742146279512 14.692578537204879
19 14.19761904761905 13.793812726567879 14.60142536867022
20 14.156190476190474 13.80117527826112 14.511205674119829
}\defendertable

\pgfplotsset{/dummy/workaround/.style={/pgfplots/axis on top}}

\node[scale=1] (kth_cr) at (0,0)
{
\begin{tikzpicture}
  \begin{axis}
[
        xmin=1,
        xmax=21,
        ymin=10,
        ymax=52,
        width =1.25\columnwidth,
        height = 0.4\columnwidth,
        axis lines=center,
        xmajorgrids=true,
        ymajorgrids=true,
        major grid style = {lightgray},
        minor grid style = {lightgray!25},
        scaled y ticks=false,
        yticklabel style={
        /pgf/number format/fixed,
        /pgf/number format/precision=5
        },
        xlabel style={below right},
        ylabel style={above left},
        axis line style={-{Latex[length=2mm]}},
        smooth,
        legend style={at={(0.75,0.75)}},
        legend columns=1,
        legend style={
          draw=none,
            /tikz/column 2/.style={
                column sep=5pt,
              }
              }
              ]
              \addplot[Red, dotted, name path=l1, thick, domain=1:250] table [x index=0, y index=1] {\attackertable};
              \addplot[RoyalAzure,name path=l1, thick, domain=1:250] table [x index=0, y index=1] {\defendertable};
              \addplot[draw=none,Red,mark repeat=2, name path=A, thick, domain=1:250] table [x index=0, y index=2] {\attackertable};
              \addplot[draw=none,Red,mark repeat=2, name path=B, thick, domain=1:250] table [x index=0, y index=3] {\attackertable};
              \addplot[Red!10!white]  fill between [of=A and B];

              \addplot[draw=none,Red,mark repeat=2, name path=A, thick, domain=1:250] table [x index=0, y index=2] {\defendertable};
              \addplot[draw=none,Red,mark repeat=2, name path=B, thick, domain=1:250] table [x index=0, y index=3] {\defendertable};
              \addplot[RoyalAzure!10!white]  fill between [of=A and B];
    \legend{Attacker learning, Defender learning}
\end{axis}
\node[inner sep=0pt,align=center, scale=1, rotate=0, opacity=1] (obs) at (0.15,2.3)
{
$J^{(\bm{\pi})}_{\mathrm{D}}$
};
\node[inner sep=0pt,align=center, scale=1, rotate=0, opacity=1] (obs) at (4.7,-0.7)
{
Iterations
};
%\node[inner sep=0pt,align=center, scale=1, rotate=0, opacity=1] (obs) at (13.8,0.1)
%{
%  $t$
%};
\end{tikzpicture}
};
\end{tikzpicture}
  }
  \caption{\hyperref[eq:best_responses]{Best response} learning when $N=1$ using \cem \cite{cem_rubinstein} and the threshold parameterization expressed in \theoremref{thm:threshold_defender}; the curves show the mean and the $95\%$ confidence interval from evaluations with $20$ random seeds; hyperparameters are listed in \appendixref{appendix:hyperparameters}.}
  \label{fig:best_response_learning}
\end{figure}

A defender strategy $\tilde{\pi}_{\mathrm{D}} \in \Pi_{\mathrm{D}}$ is a \textit{best response} against $\pi_{\mathrm{A}}\in \Pi_{\mathrm{A}}$ if it \textit{minimizes} $J^{(\pi_{\mathrm{D}},\pi_{\mathrm{A}})}_{\mathrm{D}}$ (\ref{eq:objective_1}). Similarly, an attacker strategy $\tilde{\pi}_{\mathrm{A}}$ is a best response against $\pi_{\mathrm{D}}$ if it \textit{maximizes} $J^{(\pi_{\mathrm{D}},\pi_{\mathrm{A}})}_{\mathrm{D}}$ (\ref{eq:objective_2}). Hence, the best response correspondences are
\begin{subequations}\label{eq:best_responses}
\begin{align}
\mathscr{B}_{\mathrm{D}}(\pi_{\mathrm{A}}) &\triangleq \argmin_{\pi_{\mathrm{D}} \in \Pi_{\mathrm{D}}}J^{(\pi_{\mathrm{D}}, \pi_{\mathrm{A}})}_{\mathrm{D}}(\mathbf{b}_1)\label{eq:br_defender}\\
\mathscr{B}_{\mathrm{A}}(\pi_{\mathrm{D}}) &\triangleq \argmax_{\pi_{\mathrm{A}} \in \Pi_{\mathrm{A}}}J^{(\pi_{\mathrm{D}}, \pi_{\mathrm{A}})}_{\mathrm{D}}(\mathbf{b}_1).\label{eq:br_attacker}
\end{align}
\end{subequations}
When the infrastructure contains a single server ($N=1$), there exist \hyperref[eq:best_responses]{best responses} with \textit{threshold structure}, as stated below.
% and $z$ (\ref{eq:obs_fun}) is totally positive of order 2 (i.e., \textsc{tp}-2 \cite[Def. 10.2.1]{krishnamurthy_2016})
\begin{theorem}\label{thm:threshold_defender}
If $N=1$, then
\begin{enumerate}[(A)]
\item For any $\pi_{\mathrm{A}} \in \Pi_{\mathrm{A}}$, there exists a value $\alpha^{\star} \in [0,1]$ and a \hyperref[eq:best_responses]{best response} $\tilde{\pi}_{\mathrm{D}} \in \mathscr{B}_{\mathrm{D}}(\pi_{\mathrm{A}})$ (\ref{eq:br_defender}) that satisfies
\begin{align}
\tilde{\pi}_{\mathrm{D}}(\mathbf{b}) &= \mathsf{S} \iff \mathbf{b}(1) \geq \alpha^{\star}. \tag{10a}\label{eq:defender_threshold_strategy}
\end{align}
\item Assuming $\pi_{\mathrm{A}}(0, \mathbf{e}_{1})=\mathsf{S} \forall \pi_{\mathrm{A}} \in \Pi_{\mathrm{A}}$. Then, for any $\pi_{\mathrm{D}} \in \Pi_{\mathrm{D}}$ that satisfies (\ref{eq:defender_threshold_strategy}), there exists a value $\beta^{\star}\in [0,1]$ and a \hyperref[eq:best_responses]{best response} $\tilde{\pi}_{\mathrm{A}} \in \mathscr{B}_{\mathrm{A}}(\pi_{\mathrm{D}})$ (\ref{eq:br_attacker}) that satisfies
\begin{align}
\tilde{\pi}_{\mathrm{A}}(s, \mathbf{b}) &= \mathsf{S} \iff s=0, \text{ }\mathbf{b}(1) \leq \beta^{\star}.\tag{10b} \label{eq:attacker_threshold_strategy}
\end{align}
\end{enumerate}
\end{theorem}
\setcounter{equation}{10}
The above theorem implies that when $N=1$, the \hyperref[eq:best_responses]{best responses} can be parameterized by thresholds, which allows formulating (\ref{eq:br_defender})--(\ref{eq:br_attacker}) as parametric optimization problems. \Figref{fig:best_response_learning} shows the convergence curves when performing these optimizations with the \textbf{C}ross-\textbf{E}ntropy \textbf{M}ethod (\cem) \cite{cem_rubinstein}. We provide the proof in \appendixref{app:thm_threshold_defender}.
\subsection{Equilibria}
When the attacker and the defender play \hyperref[eq:best_responses]{best responses}, their strategy pair is a \textit{Nash equilibrium} \cite[Eq. 1]{nash51} and can be written as
\begin{align}
\bm{\pi}^{\star} \triangleq (\pi_{\mathrm{D}}^{\star}, \pi_{\mathrm{A}}^{\star}) \in \mathscr{B}_{\mathrm{D}}(\pi_{\mathrm{A}}^{\star}) \times \mathscr{B}_{\mathrm{A}}(\pi_{\mathrm{D}}^{\star}).\label{eq:minmax_objective}
\end{align}
This equilibrium solves the following minimax problem \cite{vonNeumann_1928:TGG}.
\begin{subequations}\label{eq:formal_problem}
  \begin{align}
    \minimize_{\pi_{\mathrm{D}} \in \Pi_{\mathrm{D}}}\maximize_{\pi_{\mathrm{A}} \in \Pi_{\mathrm{A}}}&\text{ } \text{ }J^{(\pi_{\mathrm{D}}, \pi_{\mathrm{A}})}_{\mathrm{D}}(\mathbf{b}_1)\\
    \mathrm{subject}\text{ }\mathrm{to} &\text{ }\text{ }s_{t+1} \sim f\big(\cdot \mid s_t, \mathbf{a}_t\big) && \forall t \label{eq:dynamics_constraint}\\
                                                                                                      &\text{ }\text{ }o_{t} \sim z\big(\cdot \mid s_t) && \forall t \label{eq:obs_constraint}\\
    &\text{ }\text{ }a^{(\mathrm{A})}_{t} \sim \pi_{\mathrm{A}}\big(\cdot \mid s_t,\mathbf{b}_t\big) && \forall t\label{eq:attacker_strategy_constraint}\\
                                                                                                      &\text{ }\text{ }a^{(\mathrm{D})}_{t} \sim \pi_{\mathrm{D}}\big(\cdot \mid \mathbf{b}_t\big) && \forall t\label{eq:defender_strategy_constraint}\\
                                                            &\text{ }\text{ }s_1 \sim \mathbf{b}_1, \label{eq:init_dist_a}
\end{align}
\end{subequations}
where (\ref{eq:dynamics_constraint}) is the dynamics constraint; (\ref{eq:obs_constraint}) describes the observations; (\ref{eq:attacker_strategy_constraint})--(\ref{eq:defender_strategy_constraint}) capture the actions; and (\ref{eq:init_dist_a}) defines the initial state distribution. (Remark: As a solution to (\ref{eq:formal_problem}) exists \cite[Thm. 2.3]{horak_thesis}, we write $\min\max$ instead of $\inf\sup$.)

\begin{figure}
  \centering
  \scalebox{0.75}{
    \begin{tikzpicture}[
    dot/.style={
        draw=black,
        fill=blue!90,
        circle,
        minimum size=3pt,
        inner sep=0pt,
        solid,
    },
    ]

\node[scale=1] (kth_cr) at (0,2.15)
{
  \begin{tikzpicture}
    \begin{axis}[
      xmin=0,
      grid=major,
      grid style={dashed},
      xmax=1,
      ymin=-0.6,
      ymax=8,
        axis lines=center,
%       ticks=none,
        xlabel style={below right},
        ylabel style={above left},
        axis line style={-{Latex[length=2mm]}},
        smooth,
        legend style={at={(0.62,0.43)}},
        legend columns=2,
        legend style={
            /tikz/column 2/.style={
                column sep=5pt,
              },
              draw=none
            },
            width=12cm,
            height=3.5cm
            ]
            \addplot[Red!50!white,mark repeat=10,mark size=1.3pt,samples=100,smooth, name path=l1, domain=0:1]   (x,{(1-x)*4.73 + x*2.73});
\addplot[Black, thick, mark repeat=10,mark size=1.3pt,samples=100,smooth, name path=l1, domain=0.33:1]   (x,{(1-x)*4.73 + x*2.73});

\addplot[Red!50!white,mark repeat=10,mark size=1.3pt,samples=100,smooth, name path=l1, domain=0:1]   (x,{(1-x)*3.77768 + x*4.83717});
\addplot[Red!50!white,mark repeat=10,mark size=1.3pt,samples=100,smooth, name path=l1, domain=0:1]   (x,{(1-x)*3.80 + x*4.63});
\addplot[Red!50!white,mark repeat=10,mark size=1.3pt,samples=100,smooth, name path=l1, domain=0:1]   (x,{(1-x)*3.93 + x*4.26});
\addplot[Red!50!white,mark repeat=10,mark size=1.3pt,samples=100,smooth, name path=l1, domain=0:1]   (x,{(1-x)*3.8 + x*4.64});
\addplot[Red!50!white,mark repeat=10,mark size=1.3pt,samples=100,smooth, name path=l1, domain=0:1]   (x,{(1-x)*3.93 + x*4.26});
\addplot[Red!50!white,mark repeat=10,mark size=1.3pt,samples=100,smooth, name path=l1, domain=0:1]   (x,{(1-x)*3.91 + x*4.28});
\addplot[Red!50!white,mark repeat=10,mark size=1.3pt,samples=100,smooth, name path=l1, domain=0:1]   (x,{(1-x)*3.86 + x*4.41});
\addplot[Red!50!white,mark repeat=10,mark size=1.3pt,samples=100,smooth, name path=l1, domain=0:1]   (x,{(1-x)*3.7 + x*4.80});
\addplot[Black,thick, mark repeat=10,mark size=1.3pt,samples=100,smooth, name path=l1, domain=0:0.33]   (x,{(1-x)*3.7 + x*4.80});
\addplot[Black, thick, mark repeat=10,mark size=1.3pt,samples=100,smooth, name path=l1, domain=0.33:1]   (x,{(1-x)*4.73 + x*2.73});
\legend{alpha vectors, $J^{\star}_{\mathrm{D}}(\mathbf{b}(1))$ (\ref{eq:formal_problem})}
\end{axis}
\node[inner sep=0pt,align=center, scale=1, rotate=0, opacity=1] (obs) at (11,0.05)
{
  $\mathbf{b}(1)$
};
\end{tikzpicture}
};
  \end{tikzpicture}
  }
  \caption{The equilibrium value (\ref{eq:formal_problem}) of $\Gamma$ when $N=1$ (computed with the \hsvi algorithm \cite[Alg. 1]{horak_bosansky_hsvi}); $J_{\D}^{\star}(\mathbf{b}(1))=\min_{i}[1-\mathbf{b}(1), \mathbf{b}(1)]^T\bm{\alpha}^{(i)}$, where $\bm{\alpha}^{(i)}$ is an \textit{alpha vector} \cite[Def. 1]{smallwood_1}; see \appendixref{appendix:hyperparameters} for the hyperparameters.}
  \label{fig:value_fun}
\end{figure}
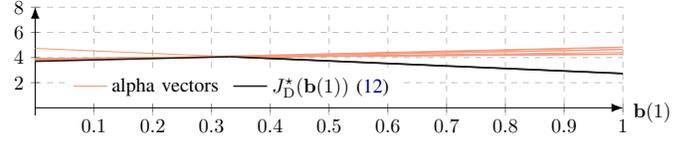

While any strategy pair $\bm{\pi}^{\star}$ that satisfies (\ref{eq:minmax_objective}) is a Nash equilibrium, (\ref{eq:formal_problem}) implies that $\bm{\pi}^{\star}$ together with $\mathds{B}$ (\ref{eq_belief_operator}) can form a stronger equilibrium, namely a \textit{perfect Bayesian equilibrium}.
\begin{definition}[Perfect Bayesian equilibrium \protect{\cite{fudenberg,bayesian_perfect_equilibria}}]\label{def:psbe}
$\bm{\pi}^{\star}$ (\ref{eq:minmax_objective}) and $\mathds{B}$ (\ref{eq_belief_operator}) is a perfect Bayesian equilibrium iff
    \begin{enumerate}[leftmargin=*]
    \item $\bm{\pi}^{\star}$ is a Nash equilibrium in $\Gamma|_{\mathbf{h}^{(\mathrm{D})}_t}$ $\forall \mathbf{h}^{(\D)}_t\in \mathcal{H}_t^{(\mathrm{D})}$,\\ where $\Gamma|_{\mathbf{h}^{(\mathrm{D})}_t}$ is the subgame starting from $\mathds{B}(\mathbf{h}^{(\mathrm{D})}_t,\pi^{\star}_\A)$.
    \item For any $\mathbf{h}^{(\D)}_t \in \mathcal{H}_t^{(\mathrm{D})}$ with $\mathbb{P}[\mathbf{h}_t^{(\D)} \mid \bm{\pi}^{\star},\mathbf{b}_1]>0$, then
\begin{align*}
\mathds{B}(\mathbf{h}^{(\D)}_{t}, \pi^{\star}_{\A})=\mathds{B}(\mathds{B}(\mathbf{h}^{(\D)}_{t-1}, \pi^{\star}_{\A}), \pi^{\star}_{\mathrm{D}}(\mathds{B}(\mathbf{h}^{(\D)}_{t-1}, \pi^{\star}_{\A})), o_t, \pi^{\star}_{\mathrm{A}}).
\end{align*}
    \end{enumerate}
\end{definition}
\begin{theorem}\label{thm:equilibrium_existence}
Given the instantiation of $\Gamma$ described in \sectionref{sec:system_model}, the following holds.
\begin{enumerate}[(A),leftmargin=*]
\item $|\mathscr{B}_{\mathrm{D}}(\pi_{\mathrm{A}})|>0$ and $|\mathscr{B}_{\mathrm{A}}(\pi_{\mathrm{D}})| > 0$ $\forall (\pi_{\mathrm{A}},\pi_{\mathrm{D}})$.
%\item $\Gamma$ has a mixed Nash equilibrium.
\item $\Gamma$ has a \hyperref[def:psbe]{perfect Bayesian equilibrium}.
  \end{enumerate}
\end{theorem}
\begin{proof}
Since $\Gamma$ is finite and $\gamma \in [0,1)$, (A) follows from \cite[Thms. 7.6.1-7.6.2]{krishnamurthy_2016}\cite[Thm. 6, p. 160]{Banach1922}. We omit the proof as it is standard in Markov decision theory. Readers familiar with the literature are encouraged to consult \cite{krishnamurthy_2016} for the details. We prove (B) by construction. For any \textit{reachable} subgame, a Nash equilibrium $\bm{\pi}^{\star}$ exists \cite[\S 3]{posg_equilibria_existence_finite_horizon}\cite[Thm. 2.3]{horak_thesis}. For any \textit{unreachable} subgame $\Gamma|_{\mathbf{h}^{(\mathrm{D})}_j}$, we can construct another Nash equilibrium $\bm{\pi}^{\star,j}$. This follows because the proofs in \cite[\S 3]{posg_equilibria_existence_finite_horizon}\cite[Thm. 2.3]{horak_thesis} do not depend on $\mathbf{b}_1$. By combining $\bm{\pi}^{\star}$ with the equilibria of all unreachable subgames, we obtain a \hyperref[def:psbe]{perfect Bayesian equilibrium}.
\end{proof}
\Figref{fig:value_fun} shows the value of a \hyperref[def:psbe]{perfect Bayesian equilibrium} when $N=1$. Interestingly, the defender has the highest expected cost when the belief of compromise ($\mathbf{b}(1)$) is around $0.35$ rather than $0.5$.
\section{Problem Statement}\label{sec:problem_statement}
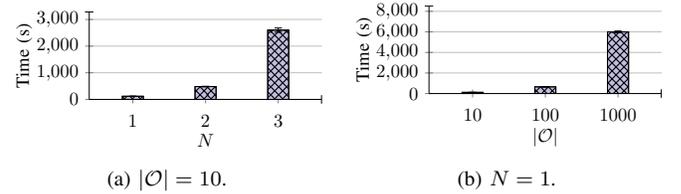
\begin{figure}
  \centering
  \begin{subfigure}[t]{0.49\columnwidth}
  \centering
  \scalebox{0.7}{
    \begin{tikzpicture}

\pgfplotstableread{
0.05368 0.000071
0.3324394 0.002386
2.582967 0.12
18.960885 1.64
}\datatable
%
%\pgfplotstableread{
%0.0048 0.0001
%0.0091 0.0002
%0.0121 0.0001
%0.08563 0.0001
%}\datatable

\pgfplotstableread{
118 5
482 11
2600 78
}\datatablee

\begin{axis}[
  spy using outlines={rectangle, magnification=3,
   width=3cm,height=2.5cm,connect spies},
   ybar,
    title style={align=center},
    ticks=both,
    xmin=0,
    xmax=2,
    ymax=2700,
    ymin=0,
    %ymin=10^(-3.5),
    %ymax=10^(3.5),
    %ymode=log,
%    log basis y={10},
    axis x line = bottom,
    axis y line = left,
    axis line style={-|},
    %nodes near coords = \rotatebox{90}{{\pgfmathprintnumber[fixed zerofill, precision=0]{\pgfplotspointmeta}}},
    %nodes near coords align={vertical},
    %every node near coord/.append style={font=\small, fill=white, yshift=3.3mm},
    enlarge y limits={lower, value=0.1},
    enlarge y limits={upper, value=0.22},
    ylabel=Time (s),
    xtick=data,
    xlabel=$N$,
    ymajorgrids,
    xticklabels={
        $1$,
        $2$,
        $3$
      },
    legend style={at={(0.42, -0.15)}, anchor=north, legend columns=2, draw=none},
    every axis legend/.append style={nodes={right}, inner sep = 0.2cm},
   x tick label style={align=center, yshift=-0.1cm},
    enlarge x limits=0.3,
    width=6cm,
    height=3.25cm,
    bar width=0.4cm
    ]

\addplot[draw=black,fill=Periwinkle!40,postaction={
        pattern=crosshatch
    }] plot [error bars/.cd, y dir=both, y explicit] table [x expr=\coordindex, y index=0, y error plus index=1, y error minus index=1] {\datatablee};
  \end{axis}
%\node[inner sep=0pt,align=center, scale=1, rotate=0, opacity=1] (obs) at (5.1,-0.75)
%{
%  $N$
%};
\end{tikzpicture}
  }
  \caption{$|\mathcal{O}|=10$.}\label{fig:hsvi_times_a}
\end{subfigure}
\hfill
  \begin{subfigure}[t]{0.49\columnwidth}
    \centering
  \scalebox{0.7}{
    \begin{tikzpicture}

\pgfplotstableread{
0.05368 0.000071
0.3324394 0.002386
2.582967 0.12
18.960885 1.64
}\datatable
%
%\pgfplotstableread{
%0.0048 0.0001
%0.0091 0.0002
%0.0121 0.0001
%0.08563 0.0001
%}\datatable

\pgfplotstableread{
118 5
650 17
6000 86
}\datatablee

\begin{axis}[
  spy using outlines={rectangle, magnification=3,
   width=3cm,height=2.5cm,connect spies},
   ybar,
    title style={align=center},
    ticks=both,
    xmin=0,
    xmax=2,
    ymax=7000,
    ymin=0,
    %ymin=10^(-3.5),
    %ymax=10^(3.5),
    %ymode=log,
%    log basis y={10},
    axis x line = bottom,
    axis y line = left,
    axis line style={-|},
    %nodes near coords = \rotatebox{90}{{\pgfmathprintnumber[fixed zerofill, precision=0]{\pgfplotspointmeta}}},
    %nodes near coords align={vertical},
    %every node near coord/.append style={font=\small, fill=white, yshift=3.3mm},
    enlarge y limits={lower, value=0.1},
    enlarge y limits={upper, value=0.22},
    ylabel=Time (s),
    xtick=data,
    xlabel=$|\mathcal{O}|$,
    ymajorgrids,
    xticklabels={
        $10$,
        $100$,
        $1000$
      },
    legend style={at={(0.42, -0.15)}, anchor=north, legend columns=2, draw=none},
    every axis legend/.append style={nodes={right}, inner sep = 0.2cm},
   x tick label style={align=center, yshift=-0.1cm},
    enlarge x limits=0.3,
    width=6cm,
    height=3.25cm,
    bar width=0.4cm
    ]

\addplot[draw=black,fill=Periwinkle!40,postaction={
        pattern=crosshatch
    }] plot [error bars/.cd, y dir=both, y explicit] table [x expr=\coordindex, y index=0, y error plus index=1, y error minus index=1] {\datatablee};
  \end{axis}
%\node[inner sep=0pt,align=center, scale=1, rotate=0, opacity=1] (obs) at (5.1,-0.75)
%{
%  $N$
%};
\end{tikzpicture}
  }
  \caption{$N=1$.}\label{fig:hsvi_times_b}
  \end{subfigure}
  \caption{Time required to compute a \hyperref[def:psbe]{perfect Bayesian equilibrium} of $\Gamma$ (\ref{eq:game_def}) with \hsvi \cite[Alg. 1]{horak_bosansky_hsvi} for different values of $N$ (\figref{fig:hsvi_times_a}) and $|\mathcal{O}|$ (\figref{fig:hsvi_times_b}); error bars indicate the $95\%$ confidence interval based on $20$ measurements; hyperparameters are listed in \appendixref{appendix:hyperparameters}.}
  \label{fig:hsvi_times}
\end{figure}
While the equilibrium defined above describes how the game \textit{ought to be played} by rational players, computing it is generally intractable, as illustrated in \figref{fig:hsvi_times}. Further, the equilibrium assumes a stationary game where players have correctly specified models, which is not the case in practice. To address these limitations, we relax the standard assumptions and consider a setting where the game is \textit{non-stationary} and players have \textit{misspecified} models, as described below.
%In particular, we consider the following problem.
\begin{problem}[Non-stationary game with misspecification]\label{main_problem}
\normalfont We consider a game $\Gamma_{\bm{\theta}_t}$ based on (\ref{eq:game_def}) that is parameterized by a time-dependent vector $\bm{\theta}_t$, which is hidden from the players. This vector can represent the transition function (\ref{eq:transitions}), the observation function (\ref{eq:obs_fun}), etc. (The game parameters of (\ref{eq:game_def}) that are not included in $\bm{\theta}_t$ are defined in \sectionref{sec:system_model} and known to both players.) Player $\mathrm{k}$ has a \textit{conjecture} of $\bm{\theta}_t$, denoted by $\overline{\bm{\theta}}^{(\mathrm{k})}_t \in \Theta_{\mathrm{k}}$, which is \textit{misspecified} if $\bm{\theta}_t \not \in \Theta_{\mathrm{k}}$. As $\bm{\theta}_t$ evolves, player $\mathrm{k}$ adapts its conjecture based on feedback $\mathbf{i}^{(\mathrm{k})}_t$ (\ref{eq:information_feedback}) and uses the conjecture to update its strategy $\pi_{\mathrm{k},t}$, starting from a \textit{base strategy} $\pi_{\mathrm{k},1}$. (Note that we do not make any assumption about the time evolution of $\bm{\theta}_t$.) Strategy updates are parameterized by a \textit{lookahead horizon} $\ell_{\mathrm{k}}$, which can be understood as a computational constraint. The defender conjectures $\ell_{\A}$ as $\overline{\ell}_{\mathrm{A}} \in \mathcal{L}$, which captures the defender's uncertainty about the attacker's computational capacity. We assume the attacker knows $\ell_\D$ and the defender's conjectures.
\end{problem}
We illustrate \probref{main_problem} using the following example.

\mybox{\textbf{Example: Moving target defense}.}{Black!5}{Black!2}{
The attacker has performed reconnaissance and knows the game model (\ref{eq:game_def}), i.e., $\Theta_{\A}=\{\bm{\theta}_1\}$. However, the configuration changes regularly $\bm{\theta}_1 \rightarrow \bm{\theta}_2 \rightarrow \hdots$ via a moving target defense scheme, leading to a misspecified model.
}
%\begin{example}[Backdoor]
%\normalfont
%\end{example}
%\begin{example}[Moving target defense]
%\normalfont
%\end{example}

Solving \probref{main_problem} leads to the following questions.
\begin{itemize}
\item[\faQuestionCircle] What is an effective method for a player to update its conjecture and its strategy?
\item[\faQuestionCircle] Do the sequences of conjectures converge?
\item[\faQuestionCircle] Once the parameters $\bm{\theta}_t$ remain constant, how can the steady state of $\Gamma_{\bm{\theta}_t}$ be characterized?
\end{itemize}

\section{Online Learning with Adaptive Conjectures}\label{sec:online_learning}
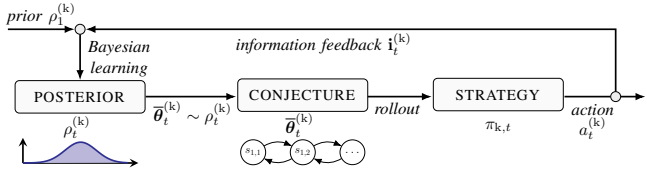
\begin{figure}[H]
  \centering
  \scalebox{0.8}{
    \begin{tikzpicture}

\node[scale=1] (kth_cr) at (5.9,0.46)
{
\begin{tikzpicture}

  \def\B{11};
  \def\Bs{3.0};
  \def\xmax{\B+3.2*\Bs};
  \def\ymin{{-0.1*gauss(\B,\B,\Bs)}};
  \def\h{0.08*gauss(\B,\B,\Bs)};
  \def\N{50}

  \begin{axis}[every axis plot post/.append style={
      mark=none,domain=0:20,
      samples=\N,smooth},
               xmin=0, xmax=20,
               ymin=0, ymax={1.1*gauss(\B,\B,\Bs)},
               axis lines=middle,
               axis line style=thick,
               enlargelimits=upper, % extend the axes a bit to the right and top
               ticks=none,
%               ylabel=$\mathbb{P}(\Gamma)$,
               every axis x label/.style={at={(current axis.right of origin)},anchor=north},
               width=3.5cm,
               height=2cm,
               clip=false
              ]

    \addplot[Blue,thick,name path=B] {gauss(x,\B,\Bs)};
    \path[name path=xaxis](0,0) -- (20,0);
    \addplot[Blue!25] fill between[of=xaxis and B];
  \end{axis}
  \node[inner sep=0pt,align=center, scale=0.8, color=black] (hacker) at (0.95,0.6) {
    $\rho^{(\mathrm{k})}_t$
  };

\node[scale=0.9] (box) at (0.95,1.1)
{
\begin{tikzpicture}
  \draw[rounded corners=0.5ex, fill=black!2] (0,0) rectangle node (m1){} (2.4,0.6);
  \node[inner sep=0pt,align=center, scale=1, color=black] (hacker) at (1.25,0.3) {
    \textsc{posterior}
  };
\end{tikzpicture}
};
\end{tikzpicture}
};

\node[scale=1] (kth_cr) at (9.6,0.42)
{
\begin{tikzpicture}

\node[scale=0.75] (box) at (2.75,1.35)
{
\begin{tikzpicture}
\node[draw,circle, minimum width=10mm, scale=0.5](s0) at (0.8,1.3) {\Large$s_{1,1}$};
\node[draw,circle, minimum width=10mm, scale=0.5](s1) at (1.9,1.3) {\Large$s_{1,2}$};
\node[draw,circle, minimum width=10mm, scale=0.55](s3) at (3,1.3) {\Large$\hdots$};

\draw[-{Latex[length=1.7mm]}, bend left] (s0) to (s1);
\draw[-{Latex[length=1.7mm]}, bend left] (s1) to (s0);
\draw[-{Latex[length=1.7mm]}, bend left] (s1) to (s3);
\draw[-{Latex[length=1.7mm]}, bend left] (s3) to (s1);
\end{tikzpicture}
};

  \node[inner sep=0pt,align=center, scale=0.8, color=black] (hacker) at (2.75,1.8) {
    $\overline{\bm{\theta}}^{(\mathrm{k})}_{t}$
  };

\node[scale=0.9] (box) at (2.75,2.3)
{
\begin{tikzpicture}
  \draw[rounded corners=0.5ex, fill=black!2] (0,0) rectangle node (m1){} (2.4,0.6);
  \node[inner sep=0pt,align=center, scale=1, color=black] (hacker) at (1.25,0.3) {
    \textsc{conjecture}
  };
\end{tikzpicture}
};
\end{tikzpicture}
};

\node[scale=1] (kth_cr) at (12.85,0.73)
{
\begin{tikzpicture}

  \node[inner sep=0pt,align=center, scale=0.8, color=black] (hacker) at (2.8,1.95) {
    $\pi_{\mathrm{k},t}$
  };
%  \node[scale=0.5](test) at (2.6,1.75) {
%    \begin{tikzpicture}
%\draw[-] (7,1) -- (6,1) -- (6,2);
%\draw[-] (6,2) -- (7,2) -- (7,1);
%%\draw[fill=Red] (6,2) -- (6,1) -- (7,1);
%\draw[-] (7,1) -- (7,1,-1) -- (7,2,-1) -- (7,2) -- cycle;
%\draw[-] (7,2) -- (6,2) -- (6,2,-1) -- (7,2,-1);
%\draw[-] (6.5,1.5) node {\huge $\pi$};
%    \end{tikzpicture}
%  };
\node[scale=0.9] (box) at (2.75,2.5)
{
\begin{tikzpicture}
  \draw[rounded corners=0.5ex, fill=black!2] (0,0) rectangle node (m1){} (2.4,0.6);
  \node[inner sep=0pt,align=center, scale=1, color=black] (hacker) at (1.25,0.3) {
    \textsc{strategy}
  };
\end{tikzpicture}
};
\end{tikzpicture}
};

%\node[scale=1] (kth_cr) at (20.1,0.4)
%{
%\begin{tikzpicture}
%
%\node[scale=1] (box) at (2.95,1.5)
%{
%\begin{tikzpicture}
%\node[scale=0.12](r1) at (0,0) {\router{}};
%%\node[rack switch, xshift=0.1cm,yshift=0.3cm, scale=0.6] at (0.17,-1) (sw1){};
%\node[server, scale=0.7](s2) at (-0.5,-1.14) {};
%\node[server, scale=0.7](s3) at (0.2,-1.14) {};
%\node[server, scale=0.7](s4) at (0.9,-1.14) {};
%
%\draw[-, color=black] (0,-0.18) to (0.0, -0.55);
%\draw[-, color=black] (-0.9,-0.5) to (1, -0.5);
%%\draw[-, color=black] (0,-0.77) to (0, -0.5);
%\draw[-, color=black] (-0.7,-0.5) to (-0.7, -0.7);
%\draw[-, color=black] (0,-0.5) to (0, -0.7);
%\draw[-, color=black] (0.7,-0.5) to (0.7, -0.7);
%\end{tikzpicture}
%};

%\node[scale=0.9] (box) at (2.8,2.7)
%{
%\begin{tikzpicture}
%  \draw[rounded corners=0.5ex, fill=black!2] (0,0) rectangle node (m1){} (3.5,0.6);
%  \node[inner sep=0pt,align=center, scale=1, color=black] (hacker) at (1.8,0.3) {
%    \textsc{infrastructure}
%  };
%\end{tikzpicture}
%};
%\end{tikzpicture}
%};
  \node[inner sep=0pt,align=center, scale=0.8, color=black] (hacker) at (7.82,0.55) {
    $\overline{\bm{\theta}}_t^{(\mathrm{k})} \sim \rho_t^{(\mathrm{k})}$
  };
  \node[inner sep=0pt,align=center, scale=0.8, color=black] (hacker) at (11.22,0.63) {
    \textit{rollout}
  };
  \node[inner sep=0pt,align=center, scale=0.8, color=black] (hacker) at (14.42,0.415) {
    \textit{action}\\
    $a_t^{(\mathrm{k})}$
  };
  \node[inner sep=0pt,align=center, scale=0.8, color=black] (hacker) at (5.31,2.2) {
    \textit{prior} $\rho_1^{(\mathrm{k})}$
  };
  \node[inner sep=0pt,align=center, scale=0.8, color=black] (hacker) at (6.55,1.46) {
    \textit{Bayesian}\\
    \textit{learning}
  };
  \node[inner sep=0pt,align=center, scale=0.8, color=black] (hacker) at (10,1.71) {
    \textit{information feedback} $\mathbf{i}^{(\mathrm{k})}_t$
  };
  \node[draw,circle, fill=gray2, scale=0.5] (feedback) at (14.8,0.83) {};
  \node[draw,circle, fill=gray2, scale=0.5](prior) at (5.9,1.95) {};
  \draw[-{Latex[length=1.7mm]}, thick] (6.99, 0.83) to (8.5, 0.83);
  \draw[-{Latex[length=1.7mm]}, thick] (10.68, 0.83) to (11.75, 0.83);
  \draw[-, thick] (13.95, 0.83) to (feedback);
  \draw[-{Latex[length=1.7mm]}, thick]  (feedback) to (15.3, 0.83);
  \draw[-{Latex[length=1.7mm]}, thick]  (feedback) to (14.8, 1.95) to (prior);
  \draw[-{Latex[length=1.7mm]}, thick] (4.7, 1.95) to (prior);
  \draw[-{Latex[length=1.7mm]}, thick] (prior) to (5.9, 1.1);
\end{tikzpicture}
  }
  \caption{\textbf{C}onjectural \textbf{O}nline \textbf{L}earning (\hyperref[alg:online_rollout]{\textsc{col}}); the figure illustrates a time step during which player $\mathrm{k}$ updates its conjecture $\overline{\bm{\theta}}_t^{(\mathrm{k})}$ and its strategy $\pi_{\mathrm{k},t}$.}
  \label{fig:pipeline}
\end{figure}

We address the above questions and develop \textbf{C}onjectural \textbf{O}nline \textbf{L}earning (\hyperref[alg:online_rollout]{\textsc{col}}), a game-theoretic method for \textit{online learning} in $\Gamma_{\bm{\theta}_t}$ (\probref{main_problem}). Using \hyperref[alg:online_rollout]{\textsc{col}}, each player iteratively adapts its conjecture through \textit{Bayesian learning} and updates its strategy through \textit{rollout}, which is a form of approximate dynamic programming; see \figref{fig:pipeline} \cite{bertsekas2021rollout}. The pseudocode of \hyperref[alg:online_rollout]{\textsc{col}} is listed in \myalgref{alg:online_rollout} and the main steps are described below.

At time $t$, player $\mathrm{k}$ updates its base strategy $\pi_{\mathrm{k},1}$ as follows\footnote{$J_{\mathrm{A}}$ depends on $s$, but this dependence is omitted for notational clarity.}.
\begin{align}
&\pi_{\mathrm{k}, t}(\mathbf{b}_t) \in \mathscr{R}(\overline{\bm{\theta}}^{(\mathrm{k})}_{t}, \mathbf{b}_t, \overline{J}_{\mathrm{k}}^{(\bm{\pi}_t)}, \ell_{\mathrm{k}}) \triangleq \argmin_{a_t^{(\mathrm{k})}, a^{(\mathrm{k})}_{t+1},\hdots,a^{(\mathrm{k})}_{t+\ell_{\mathrm{k}}-1}}\label{eq:rollout_operator}\\
&\mathbb{E}_{\bm{\pi}_t}\left[\sum_{j=t}^{t+\ell_{\mathrm{k}}-1}\gamma^{j-t}c_{\mathrm{k}}(S_j, A_j^{(\mathrm{D})}) + \gamma^{\ell_{\mathrm{k}}} \overline{J}_{\mathrm{k}}^{(\bm{\pi}_t)}(\mathbf{B}_{t+\ell_{\mathrm{k}}}) \mid \mathbf{b}_t\right],\nonumber
\end{align}
where $\ell_{\mathrm{k}}$ is the lookahead horizon, $\mathscr{R}$ is the rollout operator, $c_{\D}\triangleq c$ (\ref{eq:cost_fun}), $c_{\A} \triangleq -c$, $\bm{\pi}_t = (\pi_{\mathrm{k}, 1}, \overline{\pi}_{\mathrm{-k},t})$, $\overline{J}^{(\bm{\pi}_t)}_{\mathrm{k}}$ is the cost function induced by $\overline{\bm{\theta}}^{(\mathrm{k})}_t$ (line 19 in \myalgref{alg:online_rollout}), and $\overline{\pi}_{\mathrm{-k},t}$ is the conjectured strategy of the opponent (lines 14--17 in \myalgref{alg:online_rollout}).

\begin{figure}[H]
  \centering
  \begin{subfigure}[t]{\columnwidth}
  \centering
  \scalebox{0.8}{
    \begin{tikzpicture}

\pgfplotstableread{
3.13 0.11
39.08 3.9
467 7.2
6510 29
}\tentable

\pgfplotstableread{
1.23 0.1
4.68 0.13
16.85 0.49
66.85 1.05
}\onetable

\pgfplotstableread{
2.1 0.15
18.15 0.39
134 2.9
1747.97 13
}\fivetable

\pgfplotstableread{
27.64 1.2
355 9.4
1388 21
}\fivesamples

\begin{axis}[
   ybar=0.1cm,
    title style={align=center},
    ticks=both,
    %x=0.5cm,
    % ytick={0, 2000, 5000, 7000}
    %xmin=3,
    xmax=3,
    %ymax=1120,
    %ymin=10^(-3.5),
    %ymax=10^(3.5),
    ymode=log,
    log basis y={10},
    axis x line = bottom,
    axis y line = left,
    axis line style={-|},
    nodes near coords align={vertical}
    ylabel=Convergence time (min),
    xtick=data,
    ymajorgrids,
    xticklabels={
        $1$, $2$, $3$, $4$
      },
    ylabel=Time (s),
    legend style={at={(0.29, 1)}, anchor=north, legend columns=3, draw=none},
    every axis legend/.append style={nodes={right}, inner sep = 0.2cm},
   x tick label style={align=center, yshift=-0.1cm},
    enlarge x limits=0.15,
    width=11cm,
    height=3.25cm,
    bar width=0.4cm,
    ]

\addplot[draw=black,fill=Periwinkle!40,postaction={
        pattern=crosshatch
      }] plot [error bars/.cd, y dir=both, y explicit] table [x expr=\coordindex, y index=0, y error plus index=1, y error minus index=1] {\onetable};

\addplot[draw=black,fill=SeaGreen!30,postaction={
        pattern=north west lines
      }] plot [error bars/.cd, y dir=both, y explicit] table [x expr=\coordindex, y index=0, y error plus index=1, y error minus index=1] {\fivetable};
\addplot[draw=black,fill=Red,postaction={
        pattern=grid
      }] plot [error bars/.cd, y dir=both, y explicit] table [x expr=\coordindex, y index=0, y error plus index=1, y error minus index=1] {\tentable};
%%
%\addplot[draw=black,fill=SeaGreen!60,postaction={
%        pattern=north east lines
%      }] plot [error bars/.cd, y dir=both, y explicit] table [x expr=\coordindex, y index=0, y error plus index=1, y error minus index=1] {\datatableeeee};
%%
%\addplot[draw=black,fill=Red,postaction={
%        pattern=bricks
%    }] plot [error bars/.cd, y dir=both, y explicit] table [x expr=\coordindex, y index=0, y error plus index=1, y error minus index=1] {\datatableeeeee};

    \legend{$|\mathcal{O}|=1$, $|\mathcal{O}|=2$, $|\mathcal{O}|=3$}
  \end{axis}
\node[inner sep=0pt,align=center, scale=1, rotate=0, opacity=1] (obs) at (4.9,-0.65)
{
  $\ell_{\mathrm{k}}$
};
\end{tikzpicture}
  }
  \caption{Exact rollout.}\label{fig:rollout_times_a}
\end{subfigure}
\hfill
  \begin{subfigure}[t]{\columnwidth}
    \centering
  \scalebox{0.8}{
    \begin{tikzpicture}

\pgfplotstableread{
3.13 0.11
39.08 3.9
467 7.2
6510 29
}\tentable

\pgfplotstableread{
1.23 0.1
4.68 0.13
16.85 0.49
66.85 1.05
}\onetable

\pgfplotstableread{
2.1 0.15
18.15 0.39
134 2.9
1747.97 13
}\fivetable

\pgfplotstableread{
27.64 1.2
355 9.4
1388 21
10488 91
}\fivesamples

\begin{axis}[
   ybar=0.1cm,
    title style={align=center},
    ticks=both,
    %x=0.5cm,
    % ytick={0, 2000, 5000, 7000}
    %xmin=3,
    xmax=3,
    %ymax=1120,
    %ymin=10^(-3.5),
    %ymax=10^(3.5),
    ymode=log,
    log basis y={10},
    axis x line = bottom,
    axis y line = left,
    axis line style={-|},
    nodes near coords align={vertical}
    ylabel=Convergence time (min),
    xtick=data,
    ymajorgrids,
    xticklabels={
        $1$, $2$, $3$, $4$
      },
    ylabel=Time (s),
    legend style={at={(0.185, 0.94)}, anchor=north, legend columns=2, draw=none},
    every axis legend/.append style={nodes={right}, inner sep = 0.2cm},
   x tick label style={align=center, yshift=-0.1cm},
    enlarge x limits=0.15,
    width=11cm,
    height=3.25cm,
    bar width=0.4cm,
    ]

\addplot[draw=black,fill=Periwinkle!40,postaction={
        pattern=crosshatch
      }] plot [error bars/.cd, y dir=both, y explicit] table [x expr=\coordindex, y index=0, y error plus index=1, y error minus index=1] {\fivesamples};

%\addplot[draw=black,fill=SeaGreen!30,postaction={
%        pattern=north west lines
%      }] plot [error bars/.cd, y dir=both, y explicit] table [x expr=\coordindex, y index=0, y error plus index=1, y error minus index=1] {\fivetable};
%%%
%\addplot[draw=black,fill=Red,postaction={
%        pattern=grid
%      }] plot [error bars/.cd, y dir=both, y explicit] table [x expr=\coordindex, y index=0, y error plus index=1, y error minus index=1] {\tentable};
%%
%\addplot[draw=black,fill=SeaGreen!60,postaction={
%        pattern=north east lines
%      }] plot [error bars/.cd, y dir=both, y explicit] table [x expr=\coordindex, y index=0, y error plus index=1, y error minus index=1] {\datatableeeee};
%%
%\addplot[draw=black,fill=Red,postaction={
%        pattern=bricks
%    }] plot [error bars/.cd, y dir=both, y explicit] table [x expr=\coordindex, y index=0, y error plus index=1, y error minus index=1] {\datatableeeeee};
%    \legend{$|\mathcal{O}|=1$, $|\mathcal{O}|=2$, $|\mathcal{O}|=3$}
  \end{axis}
\node[inner sep=0pt,align=center, scale=1, rotate=0, opacity=1] (obs) at (4.9,-0.65)
{
  $\ell_{\mathrm{k}}$
};
\end{tikzpicture}
  }
  \caption{Monte-Carlo rollout with $|\mathcal{O}|=26178$.}\label{fig:rollout_times_b}
  \end{subfigure}
  \caption{Compute time of rollout (\ref{eq:rollout_operator}) for varying lookahead horizons $\ell_{\mathrm{k}}$ and observation space sizes $|\mathcal{O}|$; see \appendixref{appendix:hyperparameters} for hyperparameters.}
  \label{fig:rollout_times}
\end{figure}
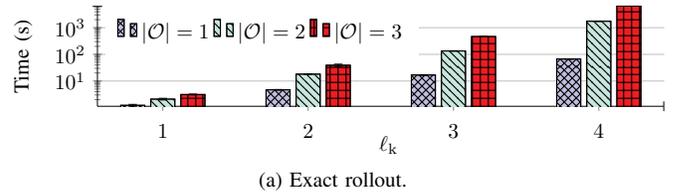
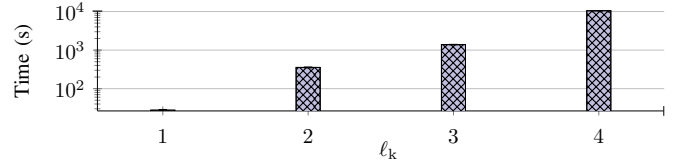

The Bellman equation in (\ref{eq:rollout_operator}) corresponds to one step of policy iteration with the base strategy as the starting point \cite[Eqs. 6.4.1-22]{puterman}. The effect of $\ell_{\mathrm{k}}>1$ is that the starting point is moved closer to the \hyperref[eq:best_responses]{best response} strategy through $\ell_{\mathrm{k}}-1$ value iterations \cite[Eq. 6.3.2-4]{puterman} \cite{bertsekas2021rollout}. Hence, the computational complexity of (\ref{eq:rollout_operator}) grows exponentially with $\ell_{\mathrm{k}}$, as shown in \figref{fig:rollout_times} \cite{bertsekas2021rollout}. To manage this complexity for large instantiations of $\Gamma_{\bm{\theta}_t}$, we estimate $\mathbb{E}$ in (\ref{eq:rollout_operator}) using Monte-Carlo samples \cite{bertsekas2021rollout}.

\begin{remark}
(\ref{eq:rollout_operator}) computes the next action as if the conjectures were true, i.e., the action is computed based on \textit{(enforced) certainty equivalence} \cite[p. 185]{bertsekas2021rollout}\cite[p. 232]{stochastic_systems_kumar}.
\end{remark}

\begin{figure}[H]
  \centering
  \scalebox{0.75}{
    \input{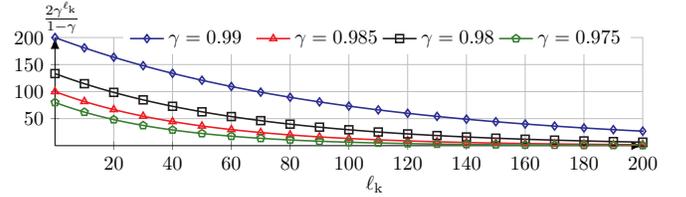}
  }
  \caption{Illustration of \theoremref{thm:cost_improvement_bound}; the x-axis indicates the lookahead horizon $\ell_{\mathrm{k}}$; the y-axis indicates the factor of the performance bound in (\ref{cost_conjecture_bound}); the curves relate to different discount factors $\gamma$.}
  \label{fig:rollout_bound}
\end{figure}

We know from dynamic programming that $\pi_{\mathrm{k},t}$ (\ref{eq:rollout_operator}) improves on the base strategy $\pi_{\mathrm{k},1}$ \cite[Prop. 1]{pomdp_rollout}. The extent of the improvement depends on the lookahead horizon and the accuracy of the conjectures as follows.
\begin{theorem}\label{thm:cost_improvement_bound}
The conjectured cost of player $\mathrm{k}$'s rollout strategy $\pi_{\mathrm{k}, t}$ satisfies
\begin{align}
\overline{J}_{\mathrm{k}}^{(\pi_{\mathrm{k}, t}, \overline{\pi}_{\mathrm{-k},t})}(\mathbf{b}) \leq \overline{J}_{\mathrm{k}}^{(\pi_{\mathrm{k}, 1}, \overline{\pi}_{\mathrm{-k},t})}(\mathbf{b}) && \forall \mathbf{b} \in \mathcal{B}. \tag{14a}\label{eq:improvement_conjecture}
\end{align}
Assuming $(\overline{\bm{\theta}}^{(\mathrm{k})}_t, \overline{\ell}_{-\mathrm{k}})$ predicts the game $\ell_{\mathrm{k}}$ steps ahead, then
\begin{align}
 \norm{\overline{J}_{\mathrm{k}}^{(\pi_{\mathrm{k}, t}, \overline{\pi}_{\mathrm{-k},t})}-J_{\mathrm{k}}^{\star}} \leq \frac{2\gamma^{\ell_{\mathrm{k}}}}{1-\gamma}\norm{\overline{J}_{\mathrm{k}}^{(\pi_{\mathrm{k}, 1}, \overline{\pi}_{\mathrm{-k},t})}-J_{\mathrm{k}}^{\star}},\tag{14b}\label{cost_conjecture_bound}
\end{align}
where $J_{\mathrm{k}}^{\star}$ is the optimal cost-to-go function when facing $\pi_{-\mathrm{k}, t}$ and $\norm{J}\triangleq \max_{x}|J(x)|$.
\end{theorem}
\setcounter{equation}{14}
\Theoremref{thm:cost_improvement_bound} states that the performance bound improves superlinearly when the lookahead horizon $\ell_{\mathrm{k}}$ increases or when the conjectured cost function $\overline{J}_{\mathrm{k}}$ moves closer to $J^{\star}_{\mathrm{k}}$ (\ref{eq:best_responses}), as shown in \figref{fig:rollout_bound}. In particular, (\ref{cost_conjecture_bound}) suggests that $\ell_{\mathrm{k}}$ controls the trade-off between computational cost and rollout performance; see \figref{fig:rollout_times}. We provide proof in \appendixref{app:cost_improvement_bound_proof}.

After computing (\ref{eq:rollout_operator}) and executing the corresponding action, player $\mathrm{k}$ receives the feedback $\mathbf{i}^{(\mathrm{k})}_t$ (\ref{eq:information_feedback}) and updates its conjectures as $\overline{\bm{\theta}}^{(\mathrm{k})}_{t} \sim \rho^{(\mathrm{k})}_{t}$ and $\overline{\ell}_{\A, t} \sim \mu_t$ (line 14 in \myalgref{alg:online_rollout}), where $\rho^{(\mathrm{k})}_{t}$ and $\mu_t$ are adapted through \textit{Bayesian learning} as
\begin{subequations}\label{eq:bayesian_learning}
\begin{align}
\rho^{(\mathrm{k})}_{t}(\overline{\bm{\theta}}_t^{(\mathrm{k})})&\triangleq \frac{\mathbb{P}[\mathbf{i}^{(\mathrm{k})}_{t}\mid\overline{\bm{\theta}}_t^{(\mathrm{k})}, \mathbf{b}_{t-1}]\rho^{(\mathrm{k})}(\overline{\bm{\theta}}_{t-1}^{(\mathrm{k})})}{\int_{\Theta_{\mathrm{k}}}\mathbb{P}[\mathbf{i}^{(\mathrm{k})}_{t}\mid\overline{\bm{\theta}}_t^{(\mathrm{k})}, \mathbf{b}_{t-1}]\rho^{(\mathrm{k})}_{t-1}(\mathrm{d}\overline{\bm{\theta}}_t^{(\mathrm{k})})} \label{eq:bayesian_estimator_game_1}\\
\mu_{t}(\overline{\ell}_{\A})&\triangleq \frac{\mathbb{P}[\mathbf{i}_t^{(\mathrm{D})}\mid  \overline{\ell}_{\A}, \mathbf{b}_{t-1}]\mu_{t-1}(\overline{\ell}_{\A})}{\sum_{\tilde{\ell}_{\A}\in \mathcal{L}}\mathbb{P}[\mathbf{i}_t^{(\mathrm{D})} \mid  \tilde{\ell}_{\A}, \mathbf{b}_{t-1}]\mu_{t-1}(\tilde{\ell}_{\A})}.\label{eq:bayesian-update}
\end{align}
\end{subequations}
These updates are well-defined under this assumption.
%For each player $\mathrm{k}$,
\begin{assumption}\label{assumption:bayes}
(\textit{i}) $\mathcal{L}$ is finite and $\Theta_{\mathrm{k}}$ is a compact subset of an Euclidean space; (\textit{ii}) $\rho^{(\mathrm{k})}_1$ and $\mu_1$ have full support; and (\textit{iii}) for all feasible $(\mathbf{i}^{(\mathrm{k})},\mathbf{b})$, there exists $\overline{\bm{\theta}} \in \Theta_{\mathrm{k}}$ and $\overline{\ell}_{\A} \in \mathcal{L}$ that assign positive probability to $\mathbf{i}^{(\mathrm{k})}$ in $\mathbf{b}$.
\end{assumption}
\begin{algorithm}
\scriptsize
  \SetNoFillComment
  \SetKwProg{myInput}{Input:}{}{}
  \SetKwProg{myOutput}{Output:}{}{}
  \SetKwProg{myalg}{Algorithm}{}{}
  \SetKwProg{myproc}{Procedure}{}{}
  \SetKw{KwTo}{inp}
  \SetKwFor{Forp}{in parallel for}{\string do}{}%
  \SetKwFor{Loop}{Loop}{}{EndLoop}
  \DontPrintSemicolon
  \SetKwBlock{DoParallel}{do in parallel}{end}
  \myInput{
    \upshape Initial belief $\mathbf{b}_1$, game model $\Gamma_{\bm{\theta}_1}$, \\
    $\quad\quad\quad\text{ }$base strategies $\bm{\pi}_{1}\triangleq (\pi_{\mathrm{D},1}, \pi_{\mathrm{A},1})$, priors $(\mu_1, \rho^{(\D)}_1, \rho^{(\A)}_1)$,\\
    $\quad\quad\quad\text{ }$discount factor $\gamma$, lookahead horizons $\ell_{\mathrm{D}},\ell_{\mathrm{A}}$.
  }{}
  \myOutput{
    \upshape A sequence of action profiles $\mathbf{a}_{1},\mathbf{a}_{2}, \hdots$.
  }{}
  \caption{\textbf{C}onjectural \textbf{O}nline \textbf{L}earning.}\label{alg:online_rollout}
  \myalg{}{
    \tcc{Initialization}
    $\mathbf{h}^{(\D)}_1 \leftarrow (\mathbf{b}_1), \mathbf{h}^{(\A)}_1 \leftarrow (\mathbf{b}_1)$,\;
    $s_1 \sim \mathbf{b}_1$\;
    $\overline{\pi}_{\mathrm{A}, 1} \leftarrow \pi_{\A, 1}, \overline{\pi}_{\mathrm{D}, 1} \leftarrow \pi_{\D, 1}$\;
    $a^{(\mathrm{D})}_{1} \sim \pi_{\mathrm{D},1}(\mathbf{b}_1), a^{(\mathrm{A})}_{1} \sim \pi_{\mathrm{A},1}(\mathbf{b}_1, s_1)$\;
    $s_{2} \sim f(\cdot \mid s_1, (a^{(\mathrm{D})}_{1}, a^{(\mathrm{A})}_{1}))$\;
    \For{$t=2,3,\hdots$}{
      \tcc{Defender learning}
      $o_t \sim z(\cdot \mid s_t)$\;
      $\mathbf{i}^{(\mathrm{D})}_{t} \leftarrow (o_{t}), \quad \mathbf{h}^{(\mathrm{D})}_{t} \leftarrow (\mathbf{h}^{(\mathrm{D})}_{t-1}, \mathbf{i}^{(\mathrm{D})}_{t}, a^{(\mathrm{D})}_{t-1})$\;
      Update $\rho^{(\D)}_t$ and $\mu_t$ (\ref{eq:bayesian_learning}) and set $\overline{\bm{\theta}}^{(\mathrm{D})}_{t} \sim \rho^{(\D)}_t$ and $\overline{\ell}_{\mathrm{A}, t} \sim \mu_t$\;
      $\mathbf{b}_{t} \leftarrow \mathds{B}(\mathbf{h}^{(\mathrm{D})}_{t}, \overline{\pi}_{\mathrm{A},t-1})$ \;
      Estimate $\overline{J}_{\mathrm{A}}^{(\pi_{\mathrm{D},t-1}, \pi_{\mathrm{A},1})}$ using $\Gamma_{\overline{\bm{\theta}}^{(\mathrm{D})}_{t}}$\;
      Compute $\overline{\pi}_{\mathrm{A}, t}(\mathbf{b}_t) \in \mathscr{R}(\overline{\bm{\theta}}^{(\mathrm{D})}_{t}, \mathbf{b}_t, \overline{J}_{\mathrm{A}}^{(\pi_{\mathrm{D},t-1}, \pi_{\mathrm{A},1})}, \overline{\ell}_{\mathrm{A}, t})$\;
      Estimate $\overline{J}_{\mathrm{D}}^{(\pi_{\mathrm{D},1}, \overline{\pi}_{\mathrm{A},t})}$ using $\Gamma_{\overline{\bm{\theta}}^{(\mathrm{D})}_{t}}$\;
      $\pi_{\mathrm{D},t}(\mathbf{b}_t) \in \mathscr{R}(\overline{\bm{\theta}}^{(\mathrm{D})}_{t}, \mathbf{b}_t, \overline{J}_{\mathrm{D}}^{(\pi_{\mathrm{D},1}, \overline{\pi}_{\mathrm{A},t})}, \ell_{\mathrm{D}})$\;
      \tcc{Attacker learning}
%      \If{$\mathrm{k} = \mathrm{A}$}{
      $\mathbf{i}^{(\mathrm{A})}_{t} \leftarrow (o_{t}, s_t, a^{(\mathrm{D})}_{t-1}), \quad \mathbf{h}^{(\mathrm{A})}_{t} \leftarrow (\mathbf{h}^{(\mathrm{A})}_{t-1}, \mathbf{i}^{(\mathrm{A})}_{t}, a^{(\mathrm{A})}_{t-1})$\;
      Update $\rho^{(\A)}_t$ using (\ref{eq:bayesian_estimator_game_1}) and set $\overline{\bm{\theta}}^{(\mathrm{A})}_{t} \sim \rho^{(\A)}_t$\;
        Estimate $\overline{J}_{\mathrm{A}}^{(\pi_{\mathrm{A},1}, \pi_{\mathrm{D},t})}$ using $\Gamma_{\overline{\bm{\theta}}^{(\mathrm{A})}_{t}}$\;
      $\pi_{\mathrm{A},t}(\mathbf{b}_t) \in \mathscr{R}(\overline{\bm{\theta}}^{(\mathrm{A})}_{t}, \mathbf{b}_t, \overline{J}_{\mathrm{A}}^{(\pi_{\mathrm{A},1}, \pi_{\mathrm{D},t})}, \ell_{\mathrm{A}})$\;
%    }
    $a^{(\mathrm{D})}_{t} \sim \pi_{\mathrm{D},t}(\mathbf{b}_t), \quad a^{(\mathrm{A})}_{t} \sim \pi_{\mathrm{A},t}(\mathbf{b}_t, s_t)$\;
    $s_{t+1} \sim f(\cdot \mid s_t, (a^{(\mathrm{D})}_{t}, a^{(\mathrm{A})}_{t}))$
    }
  }
  \normalsize
\end{algorithm}
\subsection{Convergence and Equilibrium Analysis}\label{sec:convergen_analysis}
When player $\mathrm{k}$ updates its conjectures through (\ref{eq:bayesian_learning}), the goal is to minimize the \textit{discrepancy} between the feedback distributions induced by the conjectures and the observed feedback (\ref{eq:information_feedback}). We define this discrepancy as
\begin{align}
K(\overline{\alpha},\nu)&\triangleq \mathbb{E}_{\mathbf{b}\sim \nu}\mathbb{E}_{\mathbf{I}^{(\mathrm{k})}}\left[\ln\left(\frac{\mathbb{P}[\mathbf{I}^{(\mathrm{k})} \mid \alpha, \mathbf{b}]}{\mathbb{P}[\mathbf{I}^{(\mathrm{k})} \mid \overline{\alpha}, \mathbf{b}]}\right)\mid  \alpha, \mathbf{b}\right],\label{eq:discrepancy}
\end{align}
where $\alpha \in \{\bm{\theta}, \ell_{\A}\}$ and $\nu \in \Delta(\mathcal{B})$ is an occupancy measure. We say that conjectures that minimize (\ref{eq:discrepancy}) are \textit{consistent} with $\nu$ \cite{kl_divergence}\footnote{We use the standard convention that $-\ln 0 = \infty$ and $0\ln 0 = 0$.}. Hence, the sets of consistent conjectures at time $t$ are
\begin{subequations}\label{eq:consistent_conjecture_sets}
\begin{align}
\overline{\bm{\theta}}^{(\mathrm{k})}_t &\in  \Theta_{\mathrm{k}}^\star(\nu_{t}) \triangleq \argmin_{\overline{\bm{\theta}}^{(\mathrm{k})}_{t}\in \Theta_{\mathrm{k}}}K(\overline{\bm{\theta}}^{(\mathrm{k})}_{t}, \nu_{t})\label{eq:consistent_model_conjecture}\\
\overline{\ell}_{\A,t} &\in  \mathcal{L}^\star(\nu_t) \triangleq \argmin_{\overline{\ell}_{\A,t}\in \mathcal{L}}K(\overline{\ell}_{\A,t}, \nu_t),\label{eq:consistent_conjecture}
\end{align}
\end{subequations}
where $\nu_t(\mathbf{b})\triangleq \frac{1}{t}\sum_{\tau=1}^{t}\mathbbm{1}_{\mathbf{b}=\mathbf{b}_{\tau}}$ is the empirical occupancy measure and $\bm{\pi}_{\mathbf{h}_t}$ is the empirical strategy profile at time $t$.

Intuitively, $\Theta_{\mathrm{k}}^\star$ and $\mathcal{L}^\star$ contain the conjectures that player $\mathrm{k}$ considers possible after observing feedback generated by $\nu_t$ and $\bm{\pi}_{\mathbf{h}_t}$ \cite{berk}. A desirable property of the conjecture distributions (\ref{eq:bayesian_learning}) is, therefore, that they concentrate on $\Theta_{\mathrm{k}}^\star$ and $\mathcal{L}^\star$ (\ref{eq:consistent_conjecture_sets}). This property is guaranteed asymptotically under the following conditions.
\begin{assumption}[Regularity]\label{assumption:regularity}
For fixed values of $\mathbf{i}^{(\mathrm{k})}$ and $\bm{\theta}$,
  \begin{enumerate}
  \item The mapping $\mathbf{b}\mapsto \ln\mathbb{P}[\mathbf{i}^{(\mathrm{k})} | \bm{\theta}, \mathbf{b}]$ is Lipschitz w.r.t. the Wasserstein-$1$ distance, and the Lipschitz constant is independent of $\mathbf{i}^{(\mathrm{k})}$ and $\bm{\theta}$.
  \item The mapping $\bm{\theta}\mapsto \ln\mathbb{P}[\mathbf{i}^{(\mathrm{k})} | \bm{\theta}, \mathbf{b}]$ is continuous and there exists an integrable function $g_{\mathbf{b}}(\mathbf{i}^{(\mathrm{k})})$ for all $\mathbf{b}\in \mathcal{B}$ such that $|\ln\frac{\mathbb{P}[\mathbf{i}^{(\mathrm{k})} \mid \bm{\theta}, \mathbf{b}]}{\mathbb{P}[\mathbf{i}^{(\mathrm{k})} \mid \overline{\bm{\theta}}, \mathbf{b}]}|\leq g_{\mathbf{b}}(\mathbf{i}^{(\mathrm{k})})$ for all $\overline{\bm{\theta}}\in \Theta_{\mathrm{k}}$.
  \end{enumerate}
\end{assumption}
\begin{theorem}\label{thm:conjecture_convergence}
Given Assumptions \ref{assumption:bayes}--\ref{assumption:regularity}, the following holds for any sequence $(\bm{\pi}_{\mathbf{h}_t}, \nu_t)_{t \geq 1}$ generated by \hyperref[alg:online_rollout]{\textsc{col}}.
\begin{align}
\lim_{t\rightarrow\infty}\sum_{\overline{\ell}_{\A} \in \mathcal{L}} \left(K(\overline{\ell}_{\A},\nu_{t})-K_{\mathcal{L}}^\star(\nu_{t})\right)\mu_{t+1}(\overline{\ell}_{\mathrm{A}})&=0\tag{A}
\end{align}
a.s.-$\mathbb{P}^{\mathscr{R}}$, and provided that $\bm{\theta}_t=\bm{\theta}_1$ for all $t$,
\begin{align}
\lim_{t\rightarrow\infty}\int_{\Theta_{\mathrm{k}}} \left(K(\overline{\bm{\theta}},\nu_{t})-K^\star_{\Theta_{\mathrm{k}}}(\nu_{t})\right)\rho^{(\mathrm{k})}_{t+1}(\mathrm{d}\overline{\bm{\theta}})&=0\tag{B}
\end{align}
a.s.-$\mathbb{P}^{\mathscr{R}}$, where $(K_{\mathcal{L}}^\star,K_{\Theta_{\mathrm{k}}}^\star)$ denote the minimal values of (\ref{eq:consistent_conjecture_sets}) and $\mathbb{P}^{\mathscr{R}}$ is a probability measure over the set of realizable histories $\underset{t\geq 1}{\Crossb}(\mathcal{H}_t^{(\D)} \times \mathcal{H}_t^{(\A)})$ that is induced by $(\bm{\pi}_{\mathbf{h}_t})_{t\geq 1}$.
\end{theorem}
%\begin{proof}
%See
%\end{proof}
\Theoremref{thm:conjecture_convergence} states that the conjectures produced by \hyperref[alg:online_rollout]{\textsc{col}} are asymptotically consistent (\ref{eq:consistent_conjecture_sets}) (see Appendices \ref{app:proof_conjecture_convergence_A}--\ref{app:proof_conjecture_convergence_B} for the proof). This consistency means that if the sequence $(\bm{\pi}_{\mathbf{h}_t}, \nu_t)_{t \geq 1}$ generated by \hyperref[alg:online_rollout]{\textsc{col}} converges, then it converges to an equilibrium of the following form.
\begin{definition}[Berk-Nash equilibrium, adapted from \protect{\cite{esponda21berk_mdp}}]\label{def:berk_nash}
$(\bm{\pi}, \nu) \in \Pi \times \Delta(\mathcal{B})$ is a Berk-Nash equilibrium of $\Gamma_{\bm{\theta}_t}$ (\probref{main_problem}, \hyperref[alg:online_rollout]{\textsc{col}}) iff there exist a $\rho^{(\mathrm{k})} \in \Delta(\Theta_{\mathrm{k}})$ for each player $\mathrm{k} \in \{\D, \A\}$ such that
\begin{enumerate}[(i),leftmargin=*]
\item \textsc{bounded rationality}. $\pi_{\mathrm{k}}$ is a \hyperref[eq:best_responses]{best response} against $\pi_{-\mathrm{k}}$ for any $\mathbf{b}$ given $(\nu, \rho^{(\mathrm{k})}, \rho^{(-\mathrm{k})})$, $(\ell_{\mathrm{k}},\ell_{-\mathrm{k}})$, and $\bm{\pi}_1$.
\item \textsc{consistency}. $\rho^{(\mathrm{k})} \in \Delta(\Theta_{\mathrm{k}}^{\star}(\nu))$.
\item \textsc{stationarity}. $(\bm{\pi}, \nu)$ is a limit point of some sequence $(\bm{\pi}_{\mathbf{h}_t}, \nu_t)_{t \geq 1}$ generated by \hyperref[alg:online_rollout]{\textsc{col}}, satisfying
\begin{align*}
\nu(\mathbf{b}^{\prime})= \int_{\mathcal{B}}\mathbb{E}_{A^{(\D)},O, \Gamma_{\bm{\hat{\theta}}}}\left[\delta_{\mathbf{b}^\prime}\big(\mathds{B}(\mathbf{b}, A^{(\D)}, O, \pi_{\A})\big)\right]\mathrm{d}\nu(\mathbf{b}),
\end{align*}
where $\mathds{B}$ is the belief operator defined in (\ref{eq_belief_operator}) and $\Gamma_{\bm{\hat{\theta}}}$ is parameterized by $\bm{\hat{\theta}} \triangleq \int_{\Theta_{\D}}\overline{\bm{\theta}}\mathrm{d}\rho^{(\D)}(\overline{\bm{\theta}})$.
\end{enumerate}
\end{definition}
\begin{corollary}[Berk-Nash is a fixed point of \textsc{col}]\label{cor:berk}
  $\quad$
\begin{enumerate}[(A)]
\item If the sequence $(\bm{\pi}_{\mathbf{h}_t}, \nu_t)_{t \geq 1}$ generated by \hyperref[alg:online_rollout]{\textsc{col}} converges to $(\bm{\pi}, \nu)$, then $(\bm{\pi}, \nu)$ is a \hyperref[def:berk_nash]{Berk-Nash equilibrium}.
\item Given a \hyperref[def:berk_nash]{Berk-Nash equilibrium} $(\bm{\pi}, \nu)$ and assuming
\begin{subequations}\label{eq:corollary_conditions}
\begin{align}
&\overline{\bm{\theta}}^{(\mathrm{k})} \in \Theta_{\mathrm{k}}^{\star}(\nu) \implies \mathbb{P}[\mathbf{I}^{(\mathrm{k})}| \overline{\bm{\theta}}^{(\mathrm{k})},\mathbf{b}]=\mathbb{P}[\mathbf{I}^{(\mathrm{k})}| \bm{\theta},\mathbf{b}] \label{eq:corollary_cond_1}\\
&\ell_{\D}=\ell_{\A}=\infty,\label{eq:corollary_cond_2}
\end{align}
\end{subequations}
then $(\bm{\pi}, \mathds{B})$ is a \hyperref[def:psbe]{perfect Bayesian equilibrium}.
\end{enumerate}
\end{corollary}
\begin{proof}
Condition (\textit{i}) in \defref{def:berk_nash} is ensured by rollout (\ref{eq:rollout_operator}); condition (\textit{ii}) is ensured asymptotically by \theoremref{thm:conjecture_convergence}; and condition (\textit{iii}) is a consequence of convergence. Hence, statement (A) holds. Now consider statement (B). Condition 2) in \defref{def:psbe} is satisfied by definition of $\mathds{B}$ (\ref{eq_belief_operator}). To see why condition 1) also must hold, note that assumption (\ref{eq:corollary_cond_1}) together with condition (\textit{ii}) of \defref{def:berk_nash} implies that $\mathbb{P}[\mathbf{I}^{(\mathrm{k})}| \bm{\theta},\mathbf{b}]=\mathbb{P}[\mathbf{I}^{(\mathrm{k})}| \overline{\bm{\theta}}^{(\mathrm{k})},\mathbf{b}]$ for any $\overline{\bm{\theta}}^{(\mathrm{k})} \sim \rho^{(\mathrm{k})}$. Consequently, it follows from assumption (\ref{eq:corollary_cond_2}) that $\pi_{\A} \in \mathscr{B}_{\A}(\pi_\D)$ and $\pi_{\D} \in \mathscr{B}_{\D}(\pi_\A)$ for any $\mathbf{b}_1$ (\ref{eq:best_responses}).
\end{proof}
\Corref{cor:berk} states that if the sequence $(\bm{\pi}_{\mathbf{h}_t}, \nu_t)_{t \geq 1}$ generated by \hyperref[alg:online_rollout]{\textsc{col}} converges, then it must converge to a \hyperref[def:berk_nash]{Berk-Nash equilibrium}. Further, under the conditions defined in (\ref{eq:corollary_conditions}), this equilibrium is also a \hyperref[def:psbe]{perfect Bayesian equilibrium}. The converse is not necessarily true, however, since the \hyperref[def:psbe]{perfect Bayesian equilibrium} does not enforce condition (\textit{iii}) of the \hyperref[def:berk_nash]{Berk-Nash equilibrium}. This condition requires that $\nu_t$ converges to a stationary distribution, which is not guaranteed to exist. We provide an example in \appendixref{appendix:berk_nash_example}. Whether a stationary distribution exists or not depends primarily on the parameter vector $\bm{\theta}_t$ and the observation function $z$ (\ref{eq:obs_fun}).
\section{Digital Twin and System Identification}
To implement and evaluate the method described above for the \apt use case (\sectionref{sec:use_case}), we estimate the parameters of $\Gamma$ (\ref{eq:game_def}) using a digital twin of the target infrastructure.

\subsection{Creating a Digital Twin of the Target Infrastructure}\label{sec:dt}
The configuration of the target infrastructure is listed in \appendixref{appendix:infrastructure_configuration}. The topology is shown in \figref{fig:use_case}. It consists of $N=64$ servers, some of which are vulnerable to \apt{}s.

We create a digital twin of the target infrastructure using an open-source emulation system, which is based on Linux containers and emulates network connectivity with Linux bridges \cite{csle_docs}; see \figref{fig:digital_twin}. The containers run software functions replicating important components of the target infrastructure, such as web servers, vulnerabilities, and the \snort \idps (ruleset v2.9.17.1). We implement network isolation and traffic shaping using network namespaces and the \netem module in the Linux kernel \cite{netem}. Resource allocation to containers, e.g., \cpu and memory, is enforced using \cgroups. We emulate the attacker using the actions in \tableref{tab:attacker_actions}. Similarly, we emulate the defender through hypervisor-based recovery of servers and blocking of \textsc{ip} addresses using a firewall \cite{4365686}. Software programs implementing these actions are available at \cite{csle_docs}.
\subsection{Estimating the Observation Distribution}\label{sec:identification}
Following the \apt use case described in \sectionref{sec:use_case}, we define the observation $o_{t}$ (\ref{eq:obs_fun}) to be the priority-weighted sum of the number of \ids alerts at time $t$. For the evaluation reported in this paper, we collect about $M=10^5$ measurements of $o_t$ in the digital twin (the measurements are available at \cite{csle_docs}, where $|\mathcal{O}|=26178$). Using these measurements, we estimate the observation distribution $z$ (\ref{eq:obs_fun}) with the empirical distribution $\widehat{z}$, where $\widehat{z} \overset{\text{a.s.}}{\rightarrow} z$ as $M \rightarrow \infty$ (Glivenko-Cantelli theorem).
%\begin{remark}
%The $2\times |\mathcal{O}|$ matrix with the rows $\widehat{z}(\cdot | 0)$ and $\widehat{z}(\cdot | 1)$ has about $342 \times 10^6$ second-order minors, out of which $99.9\%$ are non-negative. This suggests that the \textsc{tp}-2 assumption of Thm. \ref{thm:threshold_defender} can be made.
%\end{remark}
\section{Experimental Evaluation of \hyperref[alg:online_rollout]{\textsc{col}}}
\begin{figure}
  \centering
  \scalebox{1}{
    \input{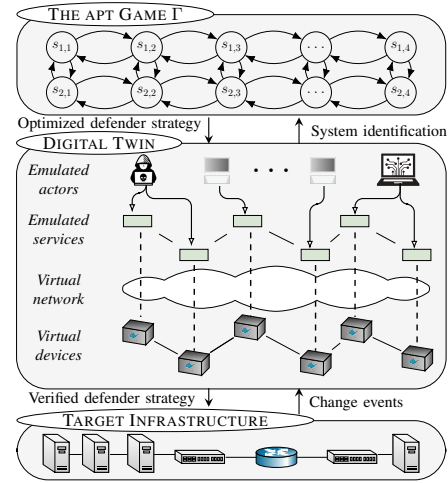}
  }
  \caption{The digital twin is a virtual replica of the target infrastructure; we use the twin for evaluation and data collection \cite{csle_docs}.}
  \label{fig:digital_twin}
\end{figure}

We implement \hyperref[alg:online_rollout]{\textsc{col}} (\figref{fig:pipeline}, \myalgref{alg:online_rollout}) in Python and evaluate it through simulation and emulation studies based on the digital twin (\figref{fig:digital_twin}). The source code of our implementation is available at \cite{csle_docs}. The simulation environment and the digital twin runs on a server with a $24$-core \textsc{intel} \textsc{xeon} \textsc{gold} \small $2.10$ GHz \normalsize \textsc{cpu} and $768$ \textsc{gb} \textsc{ram}. Hyperparameters for \hyperref[alg:online_rollout]{\textsc{col}} and $\Gamma$ (\ref{eq:game_def}) are listed in \appendixref{appendix:hyperparameters}.

\begin{table}
\centering
\begin{tabular}{lll} \toprule
  {\textit{Type}} & {\textit{Actions}} & {\textsc{mitre att\&ck} technique} \\ \midrule
  Reconnaissance  & \tcpp \syn scan, \udp scan & \textsc{t1046} service scanning\\
                  & \tcpp \xmas scan & \textsc{t1046} service scanning \\
                  & \vulscan & \textsc{t1595} active scanning \\
                  & ping-scan & \textsc{t1018} system discovery\\
%  & &\\
  Brute-force & \telnet, \ssh & \textsc{T1110} brute force\\
                  & \ftp, \cassandra & \textsc{T1110} brute force\\
                  &  \irc, \mongo, \mysql & \textsc{T1110} brute force\\
                  & \smtp, \postgres & \textsc{T1110} brute force\\
%                  &\\
  Exploit & \cve-2017-7494 & \textsc{t1210} service exploitation\\
                  &\cve-2015-3306 & \textsc{t1210} service exploitation\\
                  & \cve-2010-0426 & \textsc{T1068} privilege escalation\\
                  & \cve-2015-5602 & \textsc{T1068} privilege escalation\\
                  & \cve-2015-1427 & \textsc{t1210} service exploitation\\
                  & \cve-2014-6271 & \textsc{t1210} service exploitation\\
                  & \cve-2016-10033 & \textsc{t1210} service exploitation\\
                  & \textsc{sql} injection (\cwe-89) & \textsc{t1210} service exploitation \\
  \bottomrule\\
\end{tabular}
\caption{Attacker actions on the digital twin (\sectionref{sec:use_case}); when the attacker takes the stop action (\sectionref{sec:actions}), a randomly selected action is applied to every reachable server; actions are identified by the vulnerability identifiers in the Common Vulnerabilities and Exposures (\cve) database \cite{cve} and the Common Weakness Enumeration (\cwe) list \cite{cwe}; the actions are also linked to the corresponding attack techniques in \textsc{mitre att\&ck} \cite{strom2018mitre}.}\label{tab:attacker_actions}
\end{table}
\subsection{Evaluation Scenarios}\label{sec:eval_scenarios}
We define five scenarios to evaluate the performance properties of \hyperref[alg:online_rollout]{\textsc{col}} (\figref{fig:pipeline}, \myalgref{alg:online_rollout}). The first four scenarios are evaluated in simulation, and the fifth scenario is evaluated on the digital twin. Our aim in evaluating these scenarios is to assess a) the computational requirements of rollout (\ref{eq:rollout_operator}); b) the convergence rate of \hyperref[alg:online_rollout]{\textsc{col}} for different instantiations of $\Gamma_{\bm{\theta}_t}$ (\probref{main_problem}); and c), the benefit of \hyperref[alg:online_rollout]{\textsc{col}} compared to existing intrusion response systems.

Each evaluation scenario is based on an instantiation of \probref{main_problem} for the \textsc{apt} use case (\sectionref{sec:use_case}) with the model described in \sectionref{sec:system_model}. In such an instantiation, the defender and the attacker take actions at time steps $t=1,2,\hdots$. During each step, they perform one action each: either a (passive) \textit{continue} action or a \textit{stop action} (see \sectionref{sec:actions}). The defender's stop action corresponds to hypervisor-based recovery of servers (scenarios \ref{scenario1}--\ref{scenario4}) or blocking of \textsc{ip} addresses (scenario \ref{scenario5}) \cite{4365686}. The attacker's stop action is drawn randomly from \tableref{tab:attacker_actions}. After executing the actions, the observations are either sampled from the estimated observation distribution (scenarios \ref{scenario1}--\ref{scenario4}) or measured directly from the digital twin (scenario \ref{scenario5}). The main difference between the evaluation scenarios is how the attacker and the defender models are misspecified, as explained below.

\begin{figure}
  \centering
  \scalebox{0.75}{
    \input{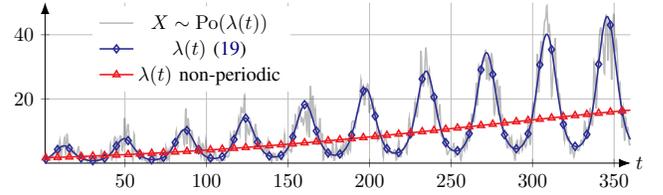}
  }
  \caption{The arrival rate function (\ref{eq:eptmp}) used in \scenarioref{scenario2}; the blue curve shows the arrival rate $\lambda(t)$; the red curve shows the trend of $\lambda(t)$ without the periodic effects; and the shaded black curve shows the number of arrivals.}
  \label{fig:arrivals}
\end{figure}

\begin{figure}
  \centering
  \scalebox{0.75}{
    \input{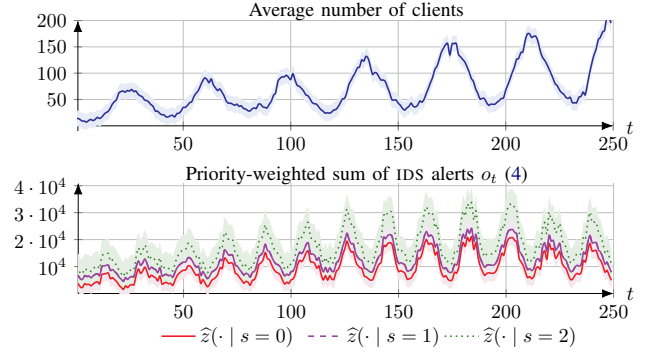}
  }
  \caption{Estimated distributions of the number of clients and the priority-weighted sum of \ids alerts $o_t$ (\ref{eq:obs_fun}) during different arrival rates $\lambda(t)$ (\ref{eq:eptmp}) based on the \textsc{apt} actions listed in \tableref{tab:attacker_actions} and the rate function shown in \figref{fig:arrivals}; the curves indicate mean values and the shaded areas indicate standard deviations from $3$ measurements; the measurements are available at \cite{csle_docs}.}
  \label{fig:alerts}
\end{figure}

\begin{scenario}[Defender is uncertain about $\ell_{\A}$]\label{scenario1}
\normalfont In this scenario, the game is stationary (i.e., $\bm{\theta}_t=\bm{\theta}_1$ for all $t$). $\bm{\theta}_1$ represents the complete game model (\ref{eq:game_def}) and is known to both players (i.e., $\rho_{1}^{(\mathrm{k})}(\bm{\theta}_1)=1$), but the defender is uncertain about the attacker's computational capacity $\ell_{\A}$, i.e., $\mu_1(\ell_{\A}) < 1$.
\end{scenario}
\begin{scenario}[Non-stationary $\bm{\theta}_t$]\label{scenario2}
\normalfont In this scenario, $\ell_{\A}$ is known to the defender, but the game is non-stationary, and $z$ (\ref{eq:obs_fun}) is parameterized by $\bm{\theta}_t$, which represents the number of clients. Hence, $\bm{\theta}_t$ changes whenever a client arrives or departs. Clients have exponential service times and arrive following a Poisson process with the following rate function (see Figs. \ref{fig:arrivals}--\ref{fig:alerts}) \cite{478761}
\begin{align}
\lambda(t) &= \exp\Biggl(\underbrace{\sum_{i=1}^{\mathrm{dim}(\bm{\psi})} \bm{\psi}_i t^i}_{\text{trend}} + \underbrace{\sum_{k=1}^{\mathrm{dim}(\bm{\chi})}\bm{\chi}_k \sin(\bm{\omega}_k t + \bm{\phi}_k)}_{\text{periodic}}\Biggr).\label{eq:eptmp}
\end{align}
See \appendixref{appendix:hyperparameters} for the parameter values.
\end{scenario}
\begin{scenario}[Misspecified model conjectures $\rho^{(\D)}_t,\rho^{(\A)}_t$]\label{scenario3}
\normalfont In this scenario, the game is stationary (i.e., $\bm{\theta}_t=\bm{\theta}_1$ for all $t$) and $\bm{\theta}_1$ represents the compromise probability $p_{\mathrm{A}}$ (\ref{eq:transitions}). Further, the attacker's computational capacity $\ell_{\A}$ is known to the defender, but both players are uncertain about $\bm{\theta}_1$ and have misspecified conjectures, i.e., $\bm{\theta}_1 \not\in \Theta_{\A} \cup \Theta_{\D}$.
\end{scenario}

\begin{scenario}[Defender is uncertain about $\ell_{\A}$ and $\bm{\theta}_t$]\label{scenario4}
\normalfont This scenario is the same as \scenarioref{scenario3}, except that the defender is uncertain about $\ell_{\A}$, i.e., $\mu_1(\ell_{\A}) < 1$.
\end{scenario}
\begin{scenario}[Comparison with the \snort \idps \cite{snort}]\label{scenario5}
\normalfont In this scenario, we compare \hyperref[alg:online_rollout]{\textsc{col}} with the \textsc{snort} \idps (ruleset v2.9.17.1). We use two baselines: \textsc{snort-high} and \textsc{snort-medium}, which block \textsc{ip} traffic that generates alerts with high and medium priority, respectively. The attacker follows the fixed strategy $\pi_{\A}(\mathsf{S}\mid \cdot)=1$ and spoofs its \textsc{ip} address. The observation $o_t$ (\ref{eq:obs_fun}) represents a \snort alert, where $o_t=0$ means no alert. The defender's response action (\sectionref{sec:actions}) corresponds to blocking the \textsc{ip} address that generated the alert. $\bm{\theta}_t$ parameterizes $z$ (\ref{eq:obs_fun}) and represents the distributions of alert priorities generated by the clients and the attacker. Specifically, $\Theta_{\mathrm{D}} = \{\bm{\theta}^{\prime}, \bm{\theta}^{\prime\prime}\}$ and
%\begin{subequations}\label{eq:scenario_5_model_change}
\begin{align}
\bm{\theta}^{\prime}\triangleq
\kbordermatrix{
    & \mathrm{C} & \mathrm{A}  \\
  \mathrm{N} & 0.85 & 0.4\\
  \mathrm{M} & 0.1 & 0.3 \\
  \mathrm{H} & 0.05 & 0.3 \\
  } \quad
\bm{\theta}^{\prime\prime}\triangleq
\kbordermatrix{
    & \mathrm{C} & \mathrm{A}  \\
  \mathrm{N} & 0.4 & 0.1\\
  \mathrm{M} & 0.3 & 0.1 \\
  \mathrm{H} & 0.3 & 0.8 \\
}, \label{eq:scenario_5_model_change}
\end{align}
where $\bm{\theta}_t=\bm{\theta}^{\prime}$ for $t < 50$ and $\bm{\theta}_{t}=\bm{\theta}^{\prime\prime}$ for $t \geq 50$. Here, $\mathrm{N}, \mathrm{M}$, and $\mathrm{H}$ refer to no alert, medium priority alert, and high priority alert, respectively. Similarly, $\mathrm{C}$ and $\mathrm{A}$ refer to the client and the attacker, respectively.
%The defender is uncertain about $\bm{\theta}_t$, i.e., $\Theta_{\mathrm{D}} = \{\bm{\theta}^{\prime}, \bm{\theta}^{\prime\prime}\}$ (\ref{eq:scenario_5_model_change}).
%  the attacker follows a fixed strategy that always attacks and we configure \snort with two
\end{scenario}

\begin{figure*}
  \centering
  \scalebox{0.7}{
    \input{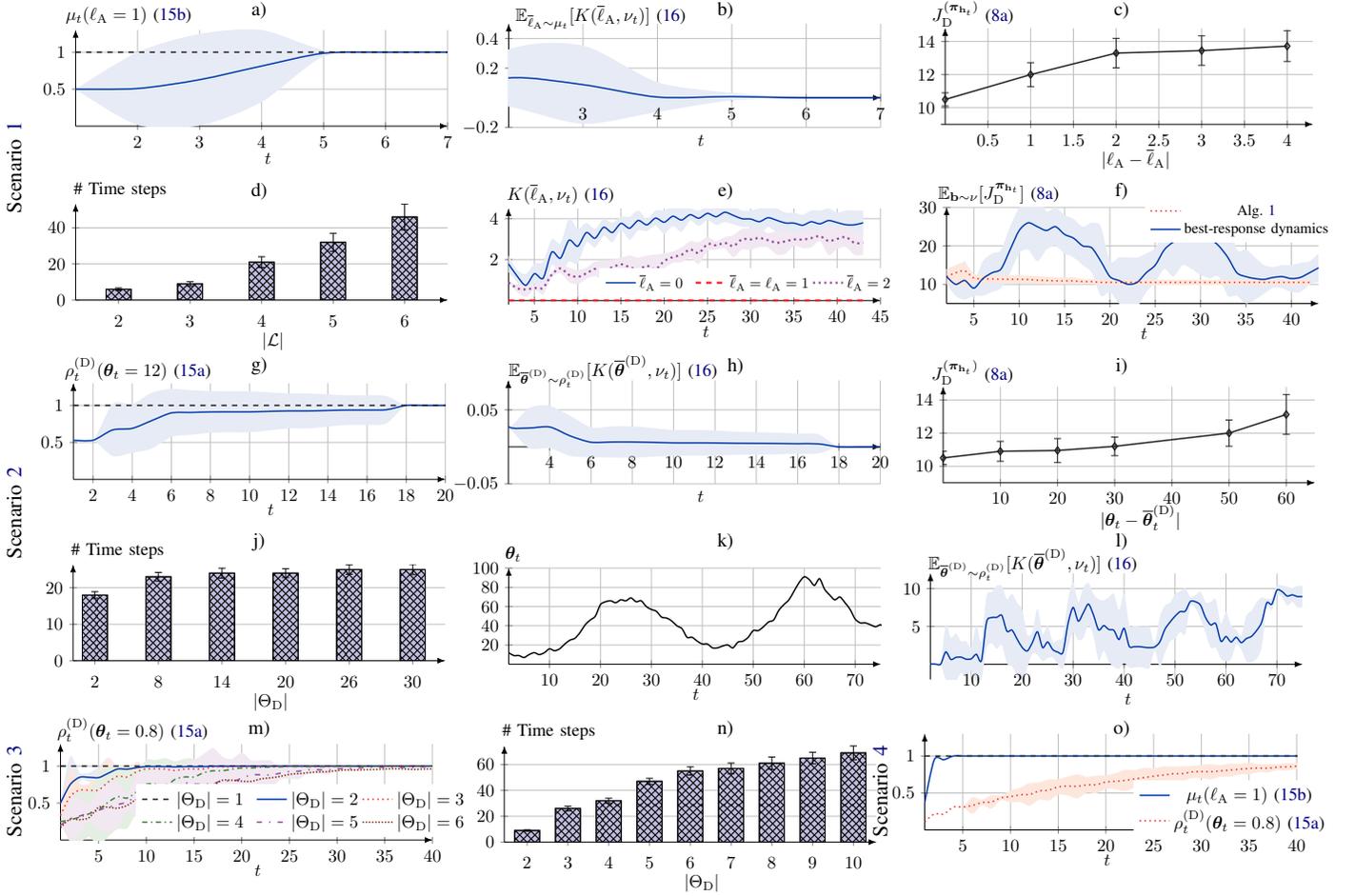}
  }
  \caption{Evaluation results; a)--f) relate to \scenarioref{scenario1}; g)--l) relate to \scenarioref{scenario2}; m)--n) relate to \scenarioref{scenario3}; and o) relate to \scenarioref{scenario4}; values indicate the mean; the shaded areas and the error bars indicate the 95\% confidence interval based on $20$ random seeds; hyperparameters are listed in \appendixref{appendix:hyperparameters}.}
  \label{fig:evaluation_results}
\end{figure*}

\begin{figure}
  \centering
  \scalebox{0.7}{
    \input{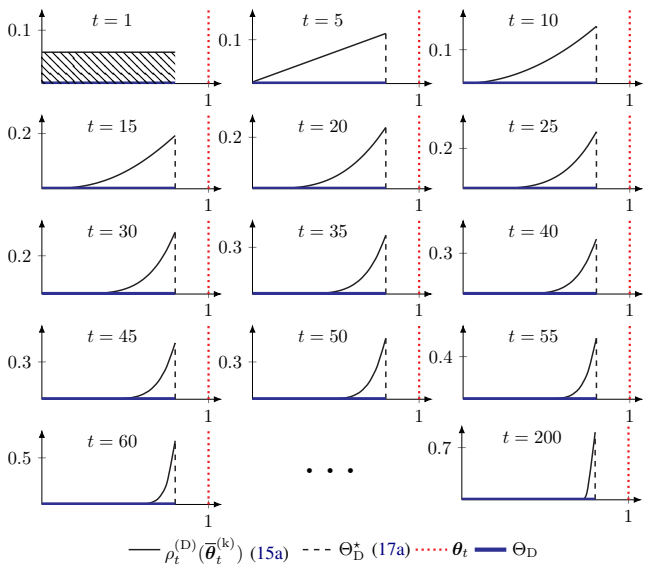}
  }
  \caption{Evolution of $\rho_t^{(\D)}$ (\ref{eq:bayesian_estimator_game_1}) when $\Theta_{\D} = \{0.0, 0.05,\hdots,0.8\}$ and $\bm{\theta}_t=1$ for all $t$ (\scenarioref{scenario3}); hyperparameters are listed in \appendixref{appendix:hyperparameters}.}
  \label{fig:posterior_evolution}
\end{figure}

\begin{figure}
  \centering
  \scalebox{0.76}{
    \input{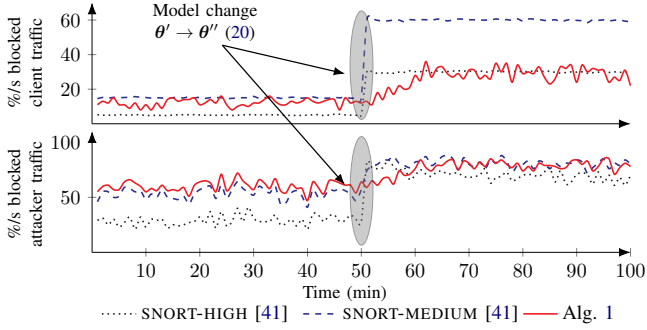}
  }
  \caption{\scenarioref{scenario5}: percentage of blocked network traffic in the digital twin.}
  \label{fig:ips_eval}
\end{figure}

% These parameters can be set based on domain knowledge or be obtained through system measurements. In fact, companies such as Google, Meta, and IBM, have documented procedures for estimating such parameters, see e.g., \cite{google_failure_model, facebook_faulure_model}.
\begin{remark}
In practice, the prior over $\bm{\theta}_1$ for the instantiations described above can be defined based on domain knowledge or obtained through system measurements. Companies like Google, Meta, and IBM have documented procedures for estimating such distributions \cite{google_failure_model}. Similarly, the prior over $\ell_{-\mathrm{k}}$ can be obtained from opponent modeling \cite{opponent_modeling}.
\end{remark}
\subsection{Evaluation Results (Figs. \ref{fig:evaluation_results}--\ref{fig:ips_eval}, \tableref{tab:convergence_comparison})}\label{sec:eval_results}

\vspace{2mm}

\noindent\textbf{\scenarioref{scenario1}.}
\Figref{fig:evaluation_results}.a--b show the evolution of the conjecture distribution $\mu_t$ (\ref{eq:bayesian-update}) and the discrepancy (\ref{eq:discrepancy}) of the conjecture $\overline{\ell}_{\A,t}$ when $\mathcal{L}=\{1,2\}$ and $\ell_{\A}=1$. We observe that $\mu_t$ converges and concentrates on the consistent conjecture (\ref{eq:consistent_conjecture_sets}) after $5$ time steps, as predicted by \theoremref{thm:conjecture_convergence}.A. \Figref{fig:evaluation_results}.d shows the rate of convergence for different $|\mathcal{L}|$, indicating that a larger $|\mathcal{L}|$ leads to a slower convergence. This is expected since a larger $|\mathcal{L}|$ means the defender has more uncertainty about $\ell_{\A}$.

\Figref{fig:evaluation_results}.c shows the expected cost of the defender as a function of $|\ell_\A - \overline{\ell}_{\A,t}|$, which quantifies the inaccuracy of the defender's conjecture $\overline{\ell}_{\A,t}$. We observe that the defender's cost is increasing with the inaccuracy of its conjecture. Next, \figref{fig:evaluation_results}.e shows the evolution of the discrepancy (\ref{eq:discrepancy}) of different conjectures. We observe that the discrepancies of the incorrect conjectures increase over time and that the discrepancy of the correct conjecture is $0$ (by definition).

Lastly, \figref{fig:evaluation_results}.f shows the expected cost of \hyperref[alg:online_rollout]{\textsc{col}} and the expected cost of \hyperref[eq:best_responses]{best response} reinforcement learning with \cem \cite{cem_rubinstein} (i.e., \textit{\hyperref[eq:best_responses]{best response} dynamics} \cite{NisaRougTardVazi07}). We note that the expected cost of reinforcement learning oscillates. Similar behavior of reinforcement learning has been observed in related work \cite{hammar_stadler_cnsm_20,self_play_cyclic}. The oscillation indicates that the players alternate between different \hyperref[eq:best_responses]{best responses} in a cycle. By contrast, the expected cost of \hyperref[alg:online_rollout]{\textsc{col}} is significantly more stable, and its behavior is consistent with convergence to a \hyperref[def:berk_nash]{Berk-Nash equilibrium}. The strategy oscillations induced by reinforcement learning lead to unpredictability, making it an impractical solution for operational systems. In comparison, the \hyperref[def:berk_nash]{Berk-Nash equilibrium} provides a robust and reliable strategy for the defender; see \figref{fig:evaluation_results}.f.

\vspace{2mm}

\noindent\textbf{\scenarioref{scenario2}.} Figures \ref{fig:evaluation_results}.g--h show the evolution of the defender's conjecture distribution $\rho^{(\D)}_t$ (\ref{eq:bayesian_estimator_game_1}) and the discrepancy (\ref{eq:discrepancy}) of the conjecture $\overline{\bm{\theta}}^{(\D)}_{t}$ when $\Theta_{\mathrm{D}}=\{12,9\}$ and $\bm{\theta}_t=12$ for all $t$. We observe that $\rho^{(\D)}_t$ converges and concentrates on the consistent conjecture (\ref{eq:consistent_conjecture_sets}) after $18$ time steps, as predicted by \theoremref{thm:conjecture_convergence}.B. \Figref{fig:evaluation_results}.j shows the rate of convergence for varying $|\Theta_{\D}|$. As expected, when the defender's uncertainty about $\bm{\theta}_t$ increases (i.e., when $|\Theta_{\D}|$ increases), the time it takes for the sequence of conjectures to converge increases.

\Figref{fig:evaluation_results}.i shows the defender's cost as a function of $|\bm{\theta}_t-\overline{\bm{\theta}}^{(\mathrm{D})}_{t}|$, which quantifies the inaccuracy of the defender's conjecture $\overline{\bm{\theta}}^{(\D)}_{t}$. We observe that the cost is increasing with the inaccuracy of the conjecture. Lastly, Figs. \ref{fig:evaluation_results}.k--l show the expected discrepancy (\ref{eq:discrepancy}) of the posterior (\ref{eq:bayesian_estimator_game_1}) when $\bm{\theta}_t$ is changing at every time step, whereby $\rho^{(\D)}_t$ does not converge. (Note that \theoremref{thm:conjecture_convergence} only applies when $\bm{\theta}_t$ remains fixed.)

\vspace{2mm}

\noindent\textbf{\scenarioref{scenario3}.} Figures \ref{fig:evaluation_results}.m--n show the time for the defender's conjecture $\overline{\bm{\theta}}_t^{(\D)}$ to converge for different sizes of $\Theta_{\D}$ when $\bm{\theta}_t$ is fixed. We observe that the time to converge increases with the size of $\Theta_{\D}$, which is expected since the size of $\Theta_{\D}$ represents the defender's degree of uncertainty about $\bm{\theta}_t$. \Figref{fig:posterior_evolution} shows the evolution of $\rho_t^{(\D)}$. We observe that $\rho_t^{(\D)}$ starts from a uniform distribution over $\Theta_{\D}$ and as $t\rightarrow \infty$, it concentrates on the set of consistent conjectures $\Theta_{\D}^{\star}$ (\ref{eq:consistent_model_conjecture}).

\vspace{2mm}

\noindent\textbf{\scenarioref{scenario4}.} \Figref{fig:evaluation_results}.o shows the evolution of the defender's conjecture distributions  $\mu_t$ and $\rho^{(\mathrm{D})}_t$ (\ref{eq:bayesian_learning}). We observe that both distributions converge, which is consistent with \theoremref{thm:conjecture_convergence}. The convergence of $\mu_t$ is significantly faster than that of $\rho^{(\mathrm{D})}_t$. We believe this difference is because $|\mathcal{L}| < |\Theta_{\D}|$, which means that the defender is more uncertain about $\ell_{\A}$ than about $\bm{\theta}_t$.

\vspace{2mm}

\noindent\textbf{\scenarioref{scenario5}.} \Figref{fig:ips_eval} shows the percentage of blocked attacker and client traffic when running the \snort \idps \cite{snort} and \hyperref[alg:online_rollout]{\textsc{col}} on the digital twin. We observe that both block some client traffic and fail to block some attacker traffic, which is expected considering the false \ids alarms generated by the clients. When comparing \hyperref[alg:online_rollout]{\textsc{col}} with \snort we find that a) \hyperref[alg:online_rollout]{\textsc{col}} and \textsc{ips-high} blocks the least \textit{client} traffic; and b) \textsc{ips-medium} and \hyperref[alg:online_rollout]{\textsc{col}} blocks the most \textit{attacker} traffic. This suggests to us that \hyperref[alg:online_rollout]{\textsc{col}} balances the trade-off between blocking clients and the attacker based on the cost function (\ref{eq:cost_fun}). Further, \hyperref[alg:online_rollout]{\textsc{col}} adapts when $\bm{\theta}_t$ changes.

\vspace{2mm}

\noindent\textbf{Comparison with state-of-the-art methods.} The convergence times of \hyperref[alg:online_rollout]{\textsc{col}} and methods used in prior work are listed in \tableref{tab:convergence_comparison}. While we observe that \hyperref[alg:online_rollout]{\textsc{col}} converges faster than the baselines, a direct comparison is not feasible for two reasons. First, the baselines do not consider the same solution concept as us, i.e., the \hyperref[def:berk_nash]{Berk-Nash equilibrium} (\defref{def:berk_nash}). Second, the baselines make different assumptions about the computational capacity and information available to the players. For example, fictitious play \cite{brown_fictious_play}, which is a popular method among prior work, assumes that players a) have correctly specified models and b) have unlimited computational capacity. Similarly, reinforcement learning methods like \ppo \cite[Alg. 1]{ppo} and \nfsp \cite[Alg. 9]{heinrich_thesis} are designed for offline rather than online learning.

\begin{table}
  \centering
\resizebox{1\columnwidth}{!}{%
\begin{tabular}{llllll} \toprule
  {\textit{Method}} & {\textit{Fixed point}} & {\textit{Time (min)}} \\ \midrule
  \hyperref[alg:online_rollout]{\textsc{col}}, $|\mathcal{L}|=2,|\Theta_{\mathrm{k}}|=1$ & \hyperref[def:berk_nash]{Berk-Nash equilibrium} & $6.7 \pm 0.7$ \\
  \hyperref[alg:online_rollout]{\textsc{col}}, $|\mathcal{L}|=3,|\Theta_{\mathrm{k}}|=1$ & \hyperref[def:berk_nash]{Berk-Nash equilibrium} & $11.4 \pm 0.9$ \\
  \hyperref[alg:online_rollout]{\textsc{col}}, $|\mathcal{L}|=4,|\Theta_{\mathrm{k}}|=1$ & \hyperref[def:berk_nash]{Berk-Nash equilibrium} & $39.1 \pm 1.3$\\
  \hyperref[alg:online_rollout]{\textsc{col}}, $|\mathcal{L}|=8,|\Theta_{\mathrm{k}}|=1$ & \hyperref[def:berk_nash]{Berk-Nash equilibrium} & $137.3 \pm 2.8$ \\
  \hyperref[alg:online_rollout]{\textsc{col}}, $|\mathcal{L}|=1,|\Theta_{\mathrm{k}}|=2$ & \hyperref[def:berk_nash]{Berk-Nash equilibrium} & $12.9 \pm 0.9$ \\
  \hyperref[alg:online_rollout]{\textsc{col}}, $|\mathcal{L}|=1,|\Theta_{\mathrm{k}}|=32$ & \hyperref[def:berk_nash]{Berk-Nash equilibrium} & $17.9 \pm 1.0$\\
  \hyperref[alg:online_rollout]{\textsc{col}}, $|\mathcal{L}|=1,|\Theta_{\mathrm{k}}|=192$ & \hyperref[def:berk_nash]{Berk-Nash equilibrium} & $29.9 \pm 1.2$ \\
  \hyperref[alg:online_rollout]{\textsc{col}}, $|\mathcal{L}|=4,|\Theta_{\mathrm{k}}|=192$ & \hyperref[def:berk_nash]{Berk-Nash equilibrium} & $194.6 \pm 3.7$ \\
  Best-response dynamics (\figref{fig:evaluation_results}.f)  & $\epsilon$-Nash equilibrium \cite[Eq. 1]{nash51} & \textsc{dnc}\\
  \hsvi \cite[Alg. 1]{horak_bosansky_hsvi} (\figref{fig:hsvi_times})  & $\epsilon$-Nash equilibrium \cite[Eq. 1]{nash51} & \textsc{dnc}\\
  \nfsp \cite[Alg. 9]{heinrich_thesis}  & $\epsilon$-Nash equilibrium \cite[Eq. 1]{nash51} & $\approx 919$ min\\
  Fictitious play \cite{brown_fictious_play} & $\epsilon$-Nash equilibrium \cite[Eq. 1]{nash51} & $\approx 4800$ min\\
  \ppo \cite[Alg. 1]{ppo}, static $\pi_{\A}$  & \hyperref[eq:best_responses]{Best response} & $9.2 \pm 0.4$ min\\
  \bottomrule\\
\end{tabular}
}
\caption{Comparison with baseline methods in terms of speed of convergence; \textsc{dnc} is short for ``does not converge''; ``$\approx$'' means that the algorithm nearly converges; numbers indicate the mean and the standard deviation from evaluations with $3$ random seeds; hyperparameters are listed in \appendixref{appendix:hyperparameters}.}\label{tab:convergence_comparison}
\end{table}
\subsection{Discussion of the Evaluation Results}\label{sec:discussion}
The experimental findings can be summarized as follows.
\begin{itemize}
\item[\faLightbulbO] The conjectures produced by \hyperref[alg:online_rollout]{\textsc{col}} converge to consistent conjectures once the model parameters $\bm{\theta}_t$ remain fixed (\ref{eq:consistent_conjecture_sets}) (Figs. \ref{fig:evaluation_results}.a,g,m,o, \theoremref{thm:conjecture_convergence}). The rate of convergence decreases as $|\mathcal{L}|$ and $|\Theta_{\mathrm{k}}|$ increase (Figs. \ref{fig:evaluation_results}.d,j,n).
\item[\faLightbulbO] The defender's cost increases as the distance between its conjecture and the true model increases (Figs. \ref{fig:evaluation_results}.c,i).
\item[\faLightbulbO] \hyperref[alg:online_rollout]{\textsc{col}} allows configuring the computational capacity and the uncertainty of each player $\mathrm{k}$ by tuning $\ell_{\mathrm{k}}$ (\figref{fig:rollout_times}) and $(\mathcal{L},\Theta_{\mathrm{k}})$, respectively (\figref{fig:evaluation_results}).
\item[\faLightbulbO] \hyperref[alg:online_rollout]{\textsc{col}} leads to effective strategies (\theoremref{thm:cost_improvement_bound}) that are more stable than those obtained through reinforcement learning (\figref{fig:evaluation_results}.f, \tableref{tab:convergence_comparison}).
\item[\faLightbulbO] \hyperref[alg:online_rollout]{\textsc{col}} outperforms the \snort \idps \cite{snort} in several key metrics; see \figref{fig:ips_eval}.
\item[\faLightbulbO] Computation of \hyperref[def:psbe]{perfect Bayesian equilibria} and computation of exact rollout strategies is intractable for any non-trivial instantiation of $\Gamma$ (\figref{fig:hsvi_times}, \figref{fig:rollout_times_a}).
\item[\faLightbulbO] Approximate \hyperref[eq:best_responses]{best response} and rollout strategies of $\Gamma$ can be efficiently computed using stochastic approximation (\figref{fig:rollout_times_b}, \figref{fig:best_response_learning}, \theoremref{thm:threshold_defender}).
\end{itemize}
The above findings suggest that \hyperref[alg:online_rollout]{\textsc{col}} can produce effective security strategies without relying on a correctly specified model of the environment. The practical implication of this result is that \hyperref[alg:online_rollout]{\textsc{col}} is suitable for dynamic \textsc{it} infrastructures with short update cycles, which aligns with current trends of virtualization and zero-touch management.
\section{Related Work}\label{sec:related_work}
Since the early 2000s, researchers have studied automated security through modeling attacks and response actions on an \textsc{it} infrastructure as a game between an attacker and a defender (see textbooks \cite{nework_security_alpcan,tambe,kamhoua2021game} and surveys \cite{r1_ref2,r5_ref3}). The game is modeled in different ways depending on the use case. Examples include: \textsc{apt} games \cite{flipit,dynamic_game_linan_zhu,r1_ref4,r3_ref2,r3_ref3}, honeypot placement games \cite{honeypot_game,DBLP:journals/compsec/HorakBTKK19}, resource allocation games \cite{9923774}, distributed denial-of-service games \cite{9328143,posg_cyber_deception_network_epidemic}, jamming games \cite{altman_jamming_1}, data corruption games \cite{r1_ref3}, moving target defense games \cite{r1_ref5, r1_ref2}, and intrusion response games \cite{stocahstic_games_security_indep_nodes_nguyen_alpcan_basar, hammar_stadler_cnsm_20,zhu_basar_dynamic_policy_ids_config,9096400,5270307,hammar_stadler_tnsm_23,kim_gamesec23}. These games are formulated using various models from the game-theoretic literature. For example: stochastic games (see e.g., \cite{stocahstic_games_security_indep_nodes_nguyen_alpcan_basar,zhu_basar_dynamic_policy_ids_config,altman_jamming_1,r2_ref3}), extensive-form games (see e.g., \cite{nework_security_alpcan}), Blotto games (see e.g., \cite{9923774}), hypergames (see e.g., \cite{9559403,8750848}), \posgs (see e.g., \cite{posg_cyber_deception_network_epidemic,hammar_stadler_cnsm_20}), Stackelberg games (see e.g., \cite{posg_cyber_deception_network_epidemic,5270307, r1_ref3}), differential games (see e.g., \cite{r2_ref4, r3_ref2}), Bayesian games (see e.g., \cite{9328143, nework_security_alpcan, r1_ref4,r2_ref2}), and evolutionary games (see e.g., \cite{9096400, r1_ref5}).

The main difference between this paper and the works referenced above is that we propose a method for online learning in non-stationary security games in which players have misspecified game models. By contrast, the referenced works assume that players have correctly specified models.

While the study of learning with misspecified models has attracted long-standing interest in economics \cite{berk_nash}, engineering \cite{kagel_mechanism_design}, and psychology \cite{rabin_psychology}, it remains unexplored in the security context. Related research in the security literature includes a) games with imperfect and incomplete information; b) games with bounded rationality; and c) model-free learning approaches. In the following subsections, we describe how these three research areas relate to this paper. The main differences are listed in \tableref{tab:related_work}.
\begin{table*}
  \centering
\resizebox{1\textwidth}{!}{%
\begin{tabular}{llllll} \toprule
  {\textit{Paper}} & {\textit{Use case}} & {\textit{Method}} & {\textit{Equilibrium}} & {\textit{Evaluation}} & {\textit{Game type}} \\ \midrule
  \cite{5270307} Zonouz, 2009 & Intrusion response  & Dynamic programming & - & Testbed & Stationary \posg \\
  \cite{zhu_basar_dynamic_policy_ids_config} Zhu, 2009 & Intrusion detection  & Q-learning & Nash & Simulation & Stationary stochastic dynamic game \\
  \cite{10.1145/2764468.2764478} Balcan, 2015 & Resource allocation  & No-regret learning & - & Analytical & Stationary repeated Stackelberg game \\
  \cite{Lisy_Davis_Bowling_2016} Lis{\'{y}}, 2016 & Resource allocation  & No-regret learning & Nash & Analytical & Stationary \textsc{nfgss} \\
  \cite{8691466} Chen, 2019 & Risk management & Proximal optimization & Gestalt Nash & Simulation & Stationary one-stage game \\
  \cite{9144263} Sanjab, 2020 & Drone operation & Prospect theory & Stackelberg & Simulation & Stationary network interdiction game \\
  \cite{8750848} Bakker, 2020 & Intrusion response & Analytical & Hyper Nash & Simulation & Stationary repeated hypergame \\
  \cite{behavioral_gt_1} Abdallah, 2020 & Power grid & Analytical & Nash & Simulation & Stationary behavioral game \\
  \cite{HUANG2020101660} Huang, 2020 & \apt & Optimization & Perfect Bayes Nash & Simulation & Stationary \posg \\
  \cite{ZHAO2020106878} Zhao, 2020 & Intrusion response & Analytical & Nash & Simulation & Non-stationary Markov game \\
  \cite{9328143} Aydeger, 2021 & \textsc{ddos} & Optimization & Perfect Bayes Nash & Testbed & Stationary signaling game \\
  \cite{9559403} Wan, 2022 & \apt & Analytical & Hyper Nash & Simulation & Stationary hypergame \\
  \cite{bounded_rational_stackelberg_1} Gabrys, 2023 & Cyber deception & Analytical & Stackelberg & Simulation & Stationary Stackelberg game \\
  \cite{hammar_stadler_tnsm_23} Hammar, 2023 & Intrusion response & Fictitious play & Nash & Testbed & Stationary \posg \\
  \cite{behavioral_gt_2} Mavridis, 2023 & Cyber-physical security & Stochastic approximation & Nash & Simulation & Stationary differential game \\
  \cite{ge_li_zhu_infocomm_workshop} Ge, 2023 & Zero-trust & Meta-learning & - & Simulation & Stationary \pomdp \\
  \textbf{This paper, 2025} & \apt & Bayesian learning \& rollout & \hyperref[def:berk_nash]{Berk-Nash} & Testbed \& simulation & Non-stationary and misspecified \posg \\
  \bottomrule\\
\end{tabular}
}
\caption{Comparison between this paper and related work.}\label{tab:related_work}
\end{table*}
\subsection{Security Games with Imperfect and Incomplete Information}
Security games with imperfect and incomplete information include \textsc{posg}s \cite{posg_cyber_deception_network_epidemic,hammar_stadler_cnsm_20,hammar_stadler_tnsm_23} and Bayesian games \cite{9328143, nework_security_alpcan, r1_ref4,r2_ref2}. These games capture scenarios where players have private knowledge represented by \textit{types} or \textit{observations}. Still, the players' perceptions about how the game works are identical (as defined by a common prior); the only thing distinguishing players is the information each has received \cite{harsanyi_2, horak_solving_one_sided_posgs}. Conversely, our model allows to capture \textit{misspecification}, where players have incorrect \textit{conjectures} about the game's structure and the opponents' strategies. Such misspecification encapsulates players' subjective perception of how the game works and can include both game elements \cite{esponda21berk_mdp} and other players' strategies \cite{berk_nash}. Moreover, players can disagree on the very form of the game. For instance, one player may represent the game with a scalar, whereas another represents it with a high-dimensional vector (see \probref{main_problem}). Such disagreements are captured by the \hyperref[def:berk_nash]{Berk-Nash equilibrium} but not the \hyperref[def:psbe]{perfect Bayesian equilibrium}.

\subsection{Security Games with Bounded Rationality}
The concept of bounded rationality was introduced by Herbert A. Simon in the 1950s as a critique of the assumption of perfect rationality in classical game theory \cite{simon_original_br}. Research on security games with bounded rationality includes \cite{8691466,9144263,bounded_rational_stackelberg_1,8362263,9559403,8750848,behavioral_gt_1,behavioral_gt_2, r1_ref5,r2_ref3}. These works differ from this paper in three main ways. First, they do not consider model misspecification as we do in this paper. Second, they study different types of games, i.e., one-stage games \cite{8691466,8362263}; network interdiction games \cite{9144263}; differential games \cite{behavioral_gt_2}; Stackelberg games \cite{bounded_rational_stackelberg_1}; behavioral games \cite{behavioral_gt_2}; evolutionary games \cite{r1_ref5}, and hypergames \cite{9559403,8750848}. Third, they study different types of equilibria, i.e., Nash equilibria \cite{behavioral_gt_1}; Gestalt Nash equilibria \cite{8691466,8362263}; Stackelberg equilibria \cite{9144263, bounded_rational_stackelberg_1}; evolutionary stable equilibria \cite{r1_ref5}, and hyper Nash equilibria \cite{9559403,8750848}. By contrast, we study \hyperref[def:berk_nash]{Berk-Nash equilibria} \cite[Def. 1]{berk_nash} and a \posg where players have misspecified models. The benefit of our game model and equilibrium concept is that they better capture the \apt use case (\sectionref{sec:use_case}).

\subsection{Learning in Security Games}
Prior work that studies learning in security games includes \cite{10.1145/2764468.2764478,Lisy_Davis_Bowling_2016,ge_li_zhu_infocomm_workshop,tao_info,HUANG2020101660,zhu_basar_dynamic_policy_ids_config,hammar_stadler_cnsm_20,hammar_stadler_tnsm,hammar_stadler_tnsm_23,kim_gamesec23}. This paper differs from these works in two main ways. First, we design a novel way to update strategies using rollout with a conjectured model. This contrasts with all of the referenced works which consider other types of strategy updates, e.g., offline learning \cite{hammar_stadler_tnsm_23,kim_gamesec23,HUANG2020101660}, meta-learning \cite{ge_li_zhu_infocomm_workshop}, model-free learning \cite{zhu_basar_dynamic_policy_ids_config,8603817,hammar_stadler_cnsm_20,hammar_stadler_tnsm,yin2023zeroshot}, and learning with perfect rationality \cite{10.1145/2764468.2764478,Lisy_Davis_Bowling_2016}. The advantage of our approach is that it allows us to model players with limited computational capacity and varying degrees of misspecification. Second, we design a convergent Bayesian mechanism for online learning. In comparison, most of the prior work considers other types of learning mechanisms, e.g., fictitious play \cite{hammar_stadler_tnsm_23,kim_gamesec23}, reinforcement learning \cite{zhu_basar_dynamic_policy_ids_config,8603817,hammar_stadler_cnsm_20,hammar_stadler_tnsm,yin2023zeroshot}, online gradient descent \cite{r1_ref1}, and no-regret learning \cite{10.1145/2764468.2764478,Lisy_Davis_Bowling_2016}.
\section{Conclusion and Future Work}\label{sec:conclusions}
This paper presents \textbf{C}onjectural \textbf{O}nline \textbf{L}earning (\hyperref[alg:online_rollout]{\textsc{col}}), a new game-theoretic method for online learning of security strategies that applies to dynamic \textsc{it} environments where the attacker and defender are uncertain about the environment and the opponent's strategy. We formulate the interaction between an attacker and a defender as a non-stationary game where each player has a probabilistic conjecture about the game model, which may be misspecified in the sense that the true model has probability $0$. Both players iteratively adapt their conjecture using Bayesian learning and update their strategy using rollout. We prove that the conjectures converge to best fits (\theoremref{thm:conjecture_convergence}), and we provide a bound on the performance improvement that rollout enables with a conjectured model (\theoremref{thm:cost_improvement_bound}). To characterize the steady state of the game, we propose a novel equilibrium concept based on the \hyperref[def:berk_nash]{Berk-Nash equilibrium}, which represents a stable point where each player acts optimally given its conjecture (\defref{def:berk_nash}). We present \hyperref[alg:online_rollout]{\textsc{col}} through an \apt use case (\sectionref{sec:use_case}). Evaluations on a testbed show that \hyperref[alg:online_rollout]{\textsc{col}} produces effective security strategies that adapt to a changing environment (\figref{fig:evaluation_results}). It also leads to faster convergence than current reinforcement learning techniques and outperforms the \snort \idps \cite{snort} (\tableref{tab:convergence_comparison} and \figref{fig:ips_eval}).

This paper opens up several directions for future work. One direction is to adapt \hyperref[alg:online_rollout]{\textsc{col}} to additional use cases. In particular, this paper considers an attacker with inside information -- in future work, we plan to investigate use cases with less capable attackers. Another direction of future work is to complement our asymptotic results (\theoremref{thm:conjecture_convergence}) with finite sample bounds and to relate our definition of the \hyperref[def:berk_nash]{Berk-Nash equilibrium} with other equilibrium concepts. A third direction of future work is to instantiate our game with an observation function (\ref{eq:obs_fun}) that is based on dynamic information flow tracking \cite{217638} and to compare rollout with regret minimization \cite{r5_ref2}.

\section{Acknowledgments}
The authors would like to thank Branislav Bosansk{\'{y}} for sharing the code of the \hsvi algorithm.

\appendices
\section{Proof of \Theoremref{thm:threshold_defender}}\label{app:thm_threshold_defender}
Given $(\pi_{\mathrm{A}},\pi_{\mathrm{D}})$, the \hyperref[eq:best_responses]{best response} strategies $(\tilde{\pi}_{\mathrm{D}},\tilde{\pi}_{\mathrm{A}})$ are optimal strategies in two \pomdp{}s $\mathcal{M}_{\D}^{P}$ and $\mathcal{M}_{\A}^{P}$. Hence, it suffices to show that there exist optimal strategies $\pi_{\mathrm{D}}^{\star}$ and $\pi_{\mathrm{A}}^{\star})$ in $\mathcal{M}_\D^{P}$ and $\mathcal{M}_{\A}^{P}$ that satisfy (\ref{eq:defender_threshold_strategy}) and (\ref{eq:attacker_threshold_strategy}), respectively. Towards this proof, we state the following five lemmas.

\begin{lemma}\label{lemma:stopping_prob}
$\mathcal{M}_\D^{\mathrm{P}}$ can be formulated as a repeated optimal stopping problem.
\end{lemma}
\begin{proof}
By definition, an optimal strategy $\pi_{\mathrm{D}}^{\star}$ in $\mathcal{M}_\D^{P}$ satisfies
\begin{align}
&\pi_{\mathrm{D}}^{\star} \in \argmin_{\pi_{\mathrm{D}} \in \Pi_{\mathrm{D}}} \mathbb{E}_{(\pi_{\mathrm{D}},\pi_{\mathrm{A}})}\left[\sum_{t=1}^{\infty}\gamma^{t-1}c(S_t, A_t^{(\mathrm{D})}) \mid S_1 = 0 \right]\nonumber\\
  &\numeq{a} \argmin_{\pi_{\mathrm{D}} \in \Pi_{\mathrm{D}}} \Biggl[\mathbb{E}_{(\pi_{\mathrm{D}},\pi_{\mathrm{A}})}\left[\sum_{t=1}^{\tau_1}\gamma^{t-1}c(S_t, A_t^{(\mathrm{D})}) \mid S_1=0 \right] + \nonumber\\
  &\quad\quad\mathbb{E}_{(\pi_{\mathrm{D}},\pi_{\mathrm{A}})}\left[\sum_{t=\tau_1+1}^{\tau_2}\gamma^{t-1}c(S_t, A_t^{(\mathrm{D})}) \mid S_{\tau_1} = 0 \right] + \hdots \Biggr]\nonumber\\
  &= \argmin_{\pi_{\mathrm{D}} \in \Pi_{\mathrm{D}}} \Biggl[\gamma^{\tau_1}J_{\mathrm{D}}(\mathbf{e}_1) \label{eq:repeated_stopping}\\
  &\quad\quad + \mathbb{E}_{(\pi_{\mathrm{D}},\pi_{\mathrm{A}})}\left[\sum_{t=1}^{\tau_1}\gamma^{t-1}c(S_t, A_t^{(\mathrm{D})}) \mid S_1=0 \right]\Biggr], \nonumber
\end{align}
where $\tau_1, \tau_2, \hdots$ are the stopping times; $J_{\mathrm{D}}$ is the cost-to-go function in $\mathcal{M}_\D^{\mathrm{P}}$ (\ref{eq:objective_1}); $\mathbf{e}_1=(1,0)$; and (a) follows by linearity of $\mathbb{E}$. Since $\mathbb{E}_{\pi_{\mathrm{D}}}[J_{\mathrm{D}}(\mathbf{e}_1)]$ can be seen as a fixed recovery cost, (\ref{eq:repeated_stopping}) defines an optimal stopping problem.
\end{proof}
\begin{lemma}\label{lemma:concave_value}
The optimal cost-to-go function $J_{\mathrm{D}}^{\star}(\mathbf{b})$ in $\mathcal{M}_\D^{\mathrm{P}}$ is piecewise linear and concave with respect to $\mathbf{b} \in \mathcal{B}$.
\end{lemma}
\begin{proof}
Edward J. Sondik originally proved this property \cite[Thm. 2]{smallwood_1}. A more accessible proof can be found in \cite[Thms. 7.6.1--7.6.2]{krishnamurthy_2016}. For brevity, we omit it here.
\end{proof}
\begin{lemma}\label{lemma:convex_stopping_set}
Let $\mathscr{S} \subseteq \mathcal{B}$ denote the subset of the belief space in $\mathcal{M}_\D^{\mathrm{P}}$ where it is optimal to stop, i.e., $\mathbf{b} \in \mathscr{S} \iff \pi_{\D}^{\star}(\mathbf{b}) = \mathsf{S}$. If $N=1$, then $\mathscr{S}$ is a convex set \cite[Thm. 2.2.1]{krishnamurthy_2016}.
\end{lemma}
\begin{proof}
As $N=1$, we have that $\mathcal{B}=[0,1]$ and $\mathbf{b}$ is uniquely determined by $\mathbf{b}(1)$. For ease of notation, we use $\mathbf{b}$ as a shorthand for $\mathbf{b}(1)$. To show that $\mathscr{S}$ is convex, we need to show that $\mathbf{b}^{\prime},\mathbf{b}^{\prime\prime}\in \mathscr{S} \implies \lambda \mathbf{b}^{\prime} + (1-\lambda)\mathbf{b}^{\prime\prime} \in \mathscr{S}$ for any $\lambda \in [0,1]$. Since $J_\D^{\star}(\mathbf{b})$ is concave (\lemmaref{lemma:concave_value}), we have
\begin{align*}
J_{\mathrm{D}}^{\star}(\lambda \mathbf{b}^{\prime} + (1-\lambda)\mathbf{b}^{\prime\prime}) &\geq \lambda J_{\mathrm{D}}^{\star}(\mathbf{b}^{\prime}) + (1-\lambda)J_{\mathrm{D}}^{\star}(\mathbf{b}^{\prime\prime}).
\end{align*}
As $\mathbf{b}^{\prime},\mathbf{b}^{\prime\prime} \in \mathscr{S}$, the optimal action given $\mathbf{b}^{\prime}$ or $\mathbf{b}^{\prime\prime}$ is $\mathsf{S}$. Thus $J_{\mathrm{D}}^{\star}(\mathbf{b}^{\prime})=Q_{\D}^{\star}(\mathbf{b}^{\prime},\mathsf{S})$ and $J_{\mathrm{D}}^{\star}(\mathbf{b}^{\prime\prime})=Q^{\star}(\mathbf{b}^{\prime\prime},\mathsf{S})$, where $Q_{\D}^{\star}$ is the optimal Q-function in $\mathcal{M}_\D^{\mathrm{P}}$. Since $Q_{\D}^{\star}(\mathbf{b},\mathsf{S})=\mathbb{E}_{S}[c(S, \mathsf{S}) \mid \mathbf{b}] = \mathbf{b}c(1,\mathsf{S}) + (1-\mathbf{b})c(0,\mathsf{S})$, we obtain
\begin{align*}
&J_{\mathrm{D}}^{\star}(\lambda \mathbf{b}^{\prime} + (1-\lambda)\mathbf{b}^{\prime\prime}) \geq \lambda J_{\mathrm{D}}^{\star}(\mathbf{b}^{\prime}) + (1-\lambda)J_{\mathrm{D}}^{\star}(\mathbf{b}^{\prime\prime})\\
&= \lambda Q_{\D}^{\star}(\mathbf{b}^{\prime}, \mathsf{S}) + (1-\lambda)Q_{\D}^{\star}(\mathbf{b}^{\prime\prime}, \mathsf{S})\\
&= \lambda\big(\mathbf{b}^{\prime}c(1,\mathsf{S}) + (1-\mathbf{b}^{\prime})c(0,\mathsf{S})\big) + \\
&\quad\quad(1-\lambda)\big(\mathbf{b}^{\prime\prime}c(1,\mathsf{S}) + (1-\mathbf{b}^{\prime\prime})c(0,\mathsf{S})\big)\\
&= (\lambda \mathbf{b}^{\prime} + (1-\lambda)\mathbf{b}^{\prime\prime})c(1,\mathsf{S}) \\
&\quad\quad + (\lambda (1-\mathbf{b}^{\prime}) + (1-\lambda)(1-\mathbf{b}^{\prime\prime}))c(0,\mathsf{S})\\
&= Q_{\D}^{\star}\big(\lambda \mathbf{b}^{\prime} + (1-\lambda)\mathbf{b}^{\prime\prime}, \mathsf{S}\big)\\
&\geq J_{\mathrm{D}}^{\star}\big(\lambda \mathbf{b}^{\prime} + (1-\lambda)\mathbf{b}^{\prime\prime}\big) \quad\quad (J_{\mathrm{D}}^{\star}(\mathbf{b})=\min_{a^{(\mathrm{D})}}Q_{\D}^{\star}(\mathbf{b},a^{(\mathrm{D})}))\\
&\implies Q_{\D}^{\star}(\lambda \mathbf{b}^{\prime} + (1-\lambda)\mathbf{b}^{\prime\prime},\mathsf{S}) = J_{\mathrm{D}}^{\star}(\lambda \mathbf{b}^{\prime} + (1-\lambda)\mathbf{b}^{\prime\prime})\\
&\implies \lambda \mathbf{b}^{\prime} + (1-\lambda)\mathbf{b}^{\prime\prime} \in \mathscr{S} \quad\quad \forall \lambda \in [0,1].
\end{align*}
%where (a) follows because $J_{\mathrm{D}}^{\star}$ is optimal.
%This shows that if $\mathbf{b}^{\prime},\mathbf{b}^{\prime\prime} \in \mathscr{S}$ then $Q_{\D}^{\star}(\lambda \mathbf{b}^{\prime} + (1-\lambda)\mathbf{b}^{\prime\prime}, \mathsf{S}) = J_{\mathrm{D}}^{\star}(\lambda \mathbf{b}^{\prime} + (1-\lambda)\mathbf{b}^{\prime\prime})$, which means that $\lambda \mathbf{b}^{\prime} + (1-\lambda)\mathbf{b}^{\prime\prime} \in \mathscr{S}$ for any $\lambda \in [0,1]$. Hence, $\mathscr{S}$ is convex \cite[Thm. 2.2.1]{krishnamurthy_2016}.
\end{proof}

\begin{lemma}\label{lemma:thm_2_b_1}
If $s_t=N$, then $a^{(\mathrm{A})}_t=\mathsf{C}$ is a \hyperref[eq:best_responses]{best response} for any $\mathbf{b}_t \in \mathcal{B}$ and $\pi_{\D} \in \Pi_{\D}$.
\end{lemma}
\begin{proof}
Let $J^{\star}_\A$ and $Q_{\A}^{\star}$ be the optimal cost-to-go function and $Q$-function in $\mathcal{M}_\A^{P}$. Assume by contradiction that $a^{(\mathrm{A})}_t=\mathsf{C}$ is not a \hyperref[eq:best_responses]{best response}. Then the expected cost of $a^{(\mathrm{A})}_t=\mathsf{S}$ must be lower than that of $a^{(\mathrm{A})}_t=\mathsf{C}$, i.e.,
\begin{align*}
&Q_{\A}^{\star}((\mathbf{b}_t,N),\mathsf{S}) < Q_{\A}^{\star}((\mathbf{b}_t,N),\mathsf{C}) \numimp{a} \\
  &\mathbb{E}_{A_t^{(\mathrm{D})},\mathbf{B}_{t+1}}[-c(s_t, A^{(\D)}_t) + \gamma J^{\star}_\A(\mathbf{B}_{t+1}) \mid s_t=N,a_t^{(\A)}=\mathsf{S}] <  \\
  &\mathbb{E}_{A_t^{(\mathrm{D})},\mathbf{B}_{t+1}}[-c(s_t, A^{(\D)}_t) +\gamma  J^{\star}_\A(\mathbf{B}_{t+1}) \mid s_t=N,a_t^{(\A)}=\mathsf{C}]\numimp{b} \\
  &\mathbb{E}_{A_t^{(\mathrm{D})},S_{t+1}}\left[\sum_{o \in \mathcal{O}}z(o \mid S_{t+1})J^{\star}_\A(\mathds{B}(\mathbf{b}_{t}, A^{(\mathrm{D})}_{t}, o, \pi_{\mathrm{A}}))\mid N,\mathsf{S}\right]\\
  &<\mathbb{E}_{A_t^{(\mathrm{D})}, S_{t+1}}\left[\sum_{o \in \mathcal{O}}z(o\mid S_{t+1})J^{\star}_\A(\mathds{B}(\mathbf{b}_{t}, A^{(\mathrm{D})}_{t}, o, \pi_{\mathrm{A}}))| N,\mathsf{C}\right]\\
  &\numimp{c} 0 < 0 \quad\quad \text{(contradiction)},
\end{align*}
where $\pi_\A$ is the attacker strategy assumed by $\pi_\D$ and $\mathds{B}$ is defined in (\ref{eq_belief_operator}). Step (a) follows from Bellman's optimality equation \cite[Eq. 1]{bellman1957markovian}; (b) follows because $c$ (\ref{eq:cost_fun}) is independent of $a^{(\A)}_t$; and (c) follows because both $a^{(\mathrm{A})}_t=\mathsf{C}$ and $a^{(\mathrm{A})}_t=\mathsf{S}$ lead to the same state when $s_t=N$ (\ref{eq:transitions}).
\end{proof}
\begin{lemma}\label{lemma:thm_2_b_2}
If $\pi_\D(\mathsf{S}\mid \mathbf{b}_t)=1$ for some $\mathbf{b}_t \in \mathcal{B}$, then $a^{(\mathrm{A})}_t=\mathsf{C}$ is a \hyperref[eq:best_responses]{best response}.
\end{lemma}
\begin{proof}
Assume by contradiction that $a^{(\mathrm{A})}_t=\mathsf{C}$ is not a \hyperref[eq:best_responses]{best response}. Then the expected cost of $a^{(\mathrm{A})}_t=\mathsf{S}$ must be lower than that of $a^{(\mathrm{A})}_t=\mathsf{C}$, i.e.,
\begin{align*}
&Q_{\A}^{\star}((\mathbf{b}_t,s_t),\mathsf{S}) < Q_{\A}^{\star}((\mathbf{b}_t,s_t),\mathsf{C}) \numimp{a} J^{\star}_\A(\mathbf{e}_1) < J^{\star}_\A(\mathbf{e}_1)\\
&\numimp{b} 0 < 0 \quad\quad \text{(contradiction)},
\end{align*}
where (a) follows because $c$ (\ref{eq:cost_fun}) is independent of $a^{(\mathrm{A})}_t$ and because $\pi_\D(\mathsf{S} \mid \mathbf{b}_t)=1 \implies \mathbf{b}_{t+1}=\mathbf{e}_1$ (\ref{eq:transitions}).
\end{proof}
\subsection{Proof of \Theoremref{thm:threshold_defender}.A}
\lemmaref{lemma:convex_stopping_set} and the assumption that $N=1$ means that $\mathscr{S}=[\alpha^{\star}, \kappa]$, where $0 \leq \alpha^{\star} \leq \kappa \leq 1$. Thus, it suffices to show that $\kappa=1$. Bellman's optimality equation \cite[Eq. 1]{bellman1957markovian} implies that
\begin{align*}
&\pi^{\star}_{\mathrm{D}}(\mathbf{e}_{2}) \in \argmin_{a \in \{\mathsf{S}, \mathsf{C}\}}\left[\overbrace{c(1, \mathsf{S}) + \gamma J^{\star}_{\mathrm{D}}(\mathbf{e}_1)}^{\mathsf{S}}, \overbrace{c(1, \mathsf{C}) + \gamma J_{\mathrm{D}}^{\star}(\mathbf{e}_{2})}^{\mathsf{C}}\right]\\
&\numeq{a}\argmin_{a \in \{\mathsf{S}, \mathsf{C}\}}\bigg[c(1, \mathsf{S}) + \gamma J^{\star}_{\mathrm{D}}(\mathbf{e}_1), \\
  &\quad\quad \gamma^{\tau-1} c(1, \mathsf{S}) + \gamma^{\tau}J^{\star}_{\mathrm{D}}(\mathbf{e}_1) + \sum_{t=1}^{\tau-1}\gamma^{t-1}c(1, \mathsf{C})\bigg]\\
  &=\argmin_{a_t^{(\mathrm{D})} \in \{\mathsf{S}, \mathsf{C}\}}\bigg[q-r + \gamma J^{\star}_{\mathrm{D}}(\mathbf{e}_1), \\
&\quad\quad \gamma^{\tau-1}(q-r) + \gamma^{\tau}J^{\star}_{\mathrm{D}}(\mathbf{e}_1) + \left(\frac{1-\gamma^{\tau-1}}{1-\gamma}\right)1^p\bigg] \numeq{b} \{\mathsf{S}\},
\end{align*}
where $\tau \geq 1$ denotes the stopping time; $\mathbf{e}_{2}=(0,1)$; $\mathbf{e}_{1}=(1,0)$; and $J^{\star}_{\mathrm{D}}$ is the optimal cost-to-go function. Step (a) follows because $s=1$ is an absorbing state until the stop. Step (b) follows from (\ref{eq:cost_fun}) and the fact that $1 > q-r$, which implies that the cost per time-step is upper bounded by $c(N, \mathsf{C})=N^p=1^p=1 > q-r$. Thus, $\mathscr{S}=[\alpha^{\star},1]$. \qed
\subsection{Proof of \Theoremref{thm:threshold_defender}.B}
Since $N=1$, we have that $\mathcal{B}=[0,1]$ and $\mathbf{b}$ is uniquely determined by $\mathbf{b}(1)$. For ease of notation, we use $\mathbf{b}$ as a shorthand for $\mathbf{b}(1)$. Given \lemmaref{lemma:thm_2_b_1}, it suffices to consider the case when $s_t=0$. From \lemmaref{lemma:thm_2_b_2} and the assumption that $\pi_\D$ satisfies (\ref{eq:defender_threshold_strategy}), we know that $a_t^{(\A)}=\mathsf{S}$ is a \hyperref[eq:best_responses]{best response} iff $\mathbf{b}_t \in [0,\alpha^{\star})$, where $\alpha^{\star} \leq 1$. It further follows from \lemmaref{lemma:thm_2_b_2} that $\alpha^{\star}=0 \implies \beta^{\star}=0$. Thus \theoremref{thm:threshold_defender}.B holds when $\alpha^{\star}=0$. Now consider $\alpha^{\star}>0$. We know from \lemmaref{lemma:concave_value} and \lemmaref{lemma:convex_stopping_set} that the stopping set $\mathscr{S}_{\mathrm{A}} \subset \mathcal{B}$ for the attacker is a convex subset of $[0,\alpha^{\star})$. Since $0 \in \mathscr{S}_{\mathrm{A}}$ by assumption, it follows that $\mathscr{S}_{\mathrm{A}}=[0,\beta^{\star}]$ for some $\beta^{\star}$. \qed

\section{Proof of \Theoremref{thm:cost_improvement_bound}}\label{app:cost_improvement_bound_proof}
As (\ref{eq:rollout_operator}) implements one step of the policy iteration algorithm \cite[Eqs. 6.4.1-22]{puterman}, (\ref{eq:improvement_conjecture}) follows from standard results in dynamic programming theory, see e.g., \cite[Prop. 1]{pomdp_rollout}.

To prove (\ref{cost_conjecture_bound}), we adapt the proof in \cite[Prop 5.1.1]{bertsekas2019reinforcement}. Let $\tilde{\bm{\pi}}\triangleq (\pi_{\mathrm{k}, t}, \overline{\pi}_{\mathrm{-k},t})$ and $\bm{\pi}\triangleq (\pi_{\mathrm{k}, 1}, \overline{\pi}_{\mathrm{-k},t})$. Then define
\begin{align*}
(T_{\mathrm{k},\bm{\pi}}J_{\mathrm{k}})(\mathbf{b}_t) &\triangleq \mathbb{E}_{\bm{\pi}}[c(S_t, A^{(\mathrm{D})}_t) + \gamma J_{\mathrm{k}}(\mathbf{B}_{t+1}) \mid \bm{\pi}] && \forall \mathbf{b} \in \mathcal{B}.
\end{align*}
Since $J_{\mathrm{k}}^{(\bm{\pi})}$ is a fixed point of $T_{\mathrm{k},\bm{\pi}}$, i.e., $T_{\mathrm{k},\bm{\pi}}J_{\mathrm{k}}^{(\bm{\pi})}=J_{\mathrm{k}}^{(\bm{\pi})}$, and since $T^k_{\mathrm{k},\bm{\pi}}$ is a contraction mapping \cite[Prop. 6.2.4]{puterman}, we have that $\lim_{j\rightarrow \infty}T^j_{\mathrm{k},\bm{\pi}}J=J_{\mathrm{k}}^{(\bm{\pi})}$ for any $J$ \cite[Thm. 6.4.6, Cor. 6.4.7]{puterman}. As a consequence
\begin{align}
\norm{\overline{J}_{\mathrm{k}}^{(\tilde{\bm{\pi}})}-J_{\mathrm{k}}^{\star}} &= \lim_{j \rightarrow \infty}\norm{T^j_{\mathrm{k},\tilde{\bm{\pi}}}J_{\mathrm{k}}^{\star}-J_{\mathrm{k}}^{\star}}.\label{eq:cost_bound_1}
\end{align}
By repeated application of the triangle inequality:
\begin{align*}
  \norm{T^j_{\mathrm{k},\tilde{\bm{\pi}}}J_{\mathrm{k}}^{\star}-J_{\mathrm{k}}^{\star}} &\leq \norm{T^{j}_{\mathrm{k},\tilde{\bm{\pi}}}J_{\mathrm{k}}^{\star}-T^{j-1}_{\mathrm{k},\tilde{\bm{\pi}}}J_{\mathrm{k}}^{\star}} + \norm{T^{j-1}_{\mathrm{k},\tilde{\bm{\pi}}}J_{\mathrm{k}}^{\star}-J_{\mathrm{k}}^{\star}}\\
  &\leq \hdots \leq \sum_{m=1}^{j}\norm{T^{m}_{\mathrm{k},\tilde{\bm{\pi}}}J_{\mathrm{k}}^{\star}-T^{m-1}_{\mathrm{k},\tilde{\bm{\pi}}}J_{\mathrm{k}}^{\star}}.
\end{align*}
Since $T_{\mathrm{k},\tilde{\bm{\pi}}}$ is a contraction mapping with modulus $\gamma < 1$,
\begin{align*}
\norm{T_{\mathrm{k},\tilde{\bm{\pi}}}(T_{\mathrm{k},\tilde{\bm{\pi}}}J_{\mathrm{k}}^{\star})-T_{\mathrm{k},\tilde{\bm{\pi}}}J_{\mathrm{k}}^{\star}} &\leq \gamma\norm{T_{\mathrm{k},\tilde{\bm{\pi}}}J_{\mathrm{k}}^{\star}-J_{\mathrm{k}}^{\star}}\\
\implies \norm{T^j_{\mathrm{k},\tilde{\bm{\pi}}}J_{\mathrm{k}}^{\star}-T^{j-1}_{\mathrm{k},\tilde{\bm{\pi}}}J_{\mathrm{k}}^{\star}} &\leq \gamma^{j-1}\norm{T_{\mathrm{k},\tilde{\bm{\pi}}}J_{\mathrm{k}}^{\star}-J_{\mathrm{k}}^{\star}}.
\end{align*}
The above inequality means that
\begin{align*}
\sum_{m=1}^{j}\norm{T^{m}_{\mathrm{k},\tilde{\bm{\pi}}}J_{\mathrm{k}}^{\star}-T^{m-1}_{\mathrm{k},\tilde{\bm{\pi}}}J_{\mathrm{k}}^{\star}} &\leq \sum_{m=1}^{j}\gamma^{m-1}\norm{T_{\mathrm{k},\tilde{\bm{\pi}}}J_{\mathrm{k}}^{\star}-J_{\mathrm{k}}^{\star}}.
\end{align*}
Since limits preserve non-strict inequalities,
\begin{align}
  \lim_{j \rightarrow \infty}\norm{T^j_{\mathrm{k},\tilde{\bm{\pi}}}J_{\mathrm{k}}^{\star}-J_{\mathrm{k}}^{\star}} &\leq \lim_{j \rightarrow \infty}\sum_{m=1}^{j}\gamma^{m-1}\norm{T_{\mathrm{k},\tilde{\bm{\pi}}}J_{\mathrm{k}}^{\star}-J_{\mathrm{k}}^{\star}}\nonumber\\
                                                                                                           &= \frac{\norm{T_{\mathrm{k},\tilde{\bm{\pi}}}J_{\mathrm{k}}^{\star}-J_{\mathrm{k}}^{\star}}}{1-\gamma}.\label{eq:cost_bound_2}
\end{align}
Now let $\widehat{J}^{(\tilde{\bm{\pi}})}_{\mathrm{k}} \triangleq T^{\overline{\pi}_{\mathrm{-k}, t}}_{\mathrm{k},\ell_{\mathrm{k}}}\overline{J}_{\mathrm{k}}^{(\tilde{\bm{\pi}})}$, where $T^{\overline{\pi}_{\mathrm{-k}, t}}_{\mathrm{k},\ell_{\mathrm{k}}}$ is defined in (\ref{eq:rollout_operator}). Note that, since \theoremref{thm:cost_improvement_bound} assumes correct conjectures, $T_{\mathrm{k},\tilde{\bm{\pi}}}\widehat{J}^{(\tilde{\bm{\pi}})}_{\mathrm{k}}=T^{\overline{\pi}_{\mathrm{-k}, t}}_{\mathrm{k},1}\widehat{J}^{(\tilde{\bm{\pi}})}_{\mathrm{k}}$ and $T^{\overline{\pi}_{\mathrm{-k}, t}}_{\mathrm{k},1}J^{\star}_{\mathrm{k}}=J^{\star}_{\mathrm{k}}$ by definition.

Applying the triangle inequality to the numerator in (\ref{eq:cost_bound_2}),
\begin{align}
&\norm{T_{\mathrm{k},\tilde{\bm{\pi}}}J_{\mathrm{k}}^{\star}-J_{\mathrm{k}}^{\star}} \leq  \norm{T_{\mathrm{k},\tilde{\bm{\pi}}}J_{\mathrm{k}}^{\star}- T_{\mathrm{k},\tilde{\bm{\pi}}}\widehat{J}^{(\tilde{\bm{\pi}})}_{\mathrm{k}}}+ \nonumber\\
  &\norm{T_{\mathrm{k},\tilde{\bm{\pi}}}\widehat{J}^{(\tilde{\bm{\pi}})}_{\mathrm{k}} - T^{\overline{\pi}_{\mathrm{-k}, t}}_{\mathrm{k},1}\widehat{J}^{(\tilde{\bm{\pi}})}_{\mathrm{k}}} + \norm{T^{\overline{\pi}_{\mathrm{-k}, t}}_{\mathrm{k},1}\widehat{J}^{(\tilde{\bm{\pi}})}_{\mathrm{k}} - J_{\mathrm{k}}^{\star}}\nonumber\\
  &= \underbrace{\norm{T_{\mathrm{k},\tilde{\bm{\pi}}}J_{\mathrm{k}}^{\star}- T_{\mathrm{k},\tilde{\bm{\pi}}}\widehat{J}^{(\tilde{\bm{\pi}})}_{\mathrm{k}}}}_{\leq \gamma \norm{\widehat{J}^{(\tilde{\bm{\pi}})}_{\mathrm{k}}-J_{\mathrm{k}}^{\star}}} + \underbrace{\norm{T^{\overline{\pi}_{\mathrm{-k}, t}}_{\mathrm{k},1}\widehat{J}^{(\tilde{\bm{\pi}})}_{\mathrm{k}} - T^{\overline{\pi}_{\mathrm{-k}, t}}_{\mathrm{k},1}J_{\mathrm{k}}^{\star}}}_{\leq \gamma\norm{\widehat{J}^{(\tilde{\bm{\pi}})}_{\mathrm{k}}-J_{\mathrm{k}}^{\star}}}\nonumber\\
  &\leq 2\gamma \norm{\widehat{J}^{(\tilde{\bm{\pi}})}_{\mathrm{k}}-J_{\mathrm{k}}^{\star}}=2\gamma \norm{T^{\overline{\pi}_{\mathrm{-k}, t}}_{\mathrm{k},\ell_{\mathrm{k}}}\overline{J}_{\mathrm{k}}^{(\tilde{\bm{\pi}})}- T^{\overline{\pi}_{\mathrm{-k}, t}}_{\mathrm{k},\ell_{\mathrm{k}}}J_{\mathrm{k}}^{\star}}\nonumber\\
  &\leq 2\gamma^{\ell_{\mathrm{k}}}\norm{\overline{J}_{\mathrm{k}}^{(\bm{\pi})}-J_{\mathrm{k}}^{\star}}. \label{eq:cost_bound_3}
\end{align}
Combining (\ref{eq:cost_bound_1})--(\ref{eq:cost_bound_3}) gives (\ref{cost_conjecture_bound}). \qed

\section{Proof of \Theoremref{thm:conjecture_convergence}.A}\label{app:proof_conjecture_convergence_A}
The idea behind the proof is to express $\mu_t$ (\ref{eq:bayesian-update}) in terms of log-likelihood ratios, which can be shown to converge using the martingale convergence theorem \cite[Thm. 6.4.3]{Ash:2000uj}. Towards this proof, we prove the following two lemmas.
\begin{lemma}
Any sequence $(\bm{\pi}_{\mathbf{h}_t})_{t\geq 1}$ generated by \hyperref[alg:online_rollout]{\textsc{col}} induces a well-defined probability measure $\mathbb{P}^{\mathscr{R}}$ over the set of realizable histories $\mathbf{h}_t \in \underset{t\geq 1}{\Crossb}(\mathcal{H}_t^{(\D)} \times \mathcal{H}_t^{(\A)})$.
\end{lemma}
\begin{proof}
  Since the sample space of the random vectors $(\mathbf{I}_t^{(\D)}, \mathbf{I}_t^{(\A)})$ (\ref{eq:information_feedback}) is finite (and measurable) for each $t$, the space of realizable histories $\mathbf{h}_t \in \mathcal{H}^{(\D)}_{t} \times \mathcal{H}^{(\A)}_{t}$ is countable. By the extension theorem of Ionescu Tulcea, it thus follows that a measure $\mathbb{P}^{\mathscr{R}}$ over $\mathcal{H}^{(\D)}_{t}\times \mathcal{H}^{(\A)}_{t}$ exists for all $t\geq 1$ \cite{Tulcea49}.
\end{proof}
\begin{lemma}
\label{lem:lln}
For any $\overline{\ell}_{\A}\in \mathcal{L}$ and any sequence $(\nu_t, \bm{\pi}_{\mathbf{h}_t})_{t \geq 1}$ generated by \hyperref[alg:online_rollout]{\textsc{col}}, the following holds a.s.-$\mathbb{P}^{\mathscr{R}}$ as $t \rightarrow \infty$
\begin{align*}
\bigg|\underbrace{t^{-1}\sum_{\tau=1}^t \ln \frac{\mathbb{P}[\mathbf{i}^{(\D)}_{\tau+1} \mid \ell_{\A}, \mathbf{b}_\tau]}{\mathbb{P}[\mathbf{i}^{(\D)}_{\tau+1} \mid \overline{\ell}_{\A}, \mathbf{b}_\tau]}}_{\triangleq Z_{t+1}(\overline{\ell}_{\A})}-K(\overline{\ell}_{\A}, \nu_{t})\bigg|=0.
\end{align*}
\end{lemma}
\begin{proof}
By definition of $Z_{t}$ and $\nu_t$,
\begin{align*}
&Z_{t+1}(\overline{\ell}_{\A}) = \sum_{\mathbf{b} \in \mathcal{B}}\sum_{\tau=1}^{t}t^{-1}\mathbbm{1}_{\{\mathbf{b}\}}(\mathbf{b}_{\tau})\ln \frac{\mathbb{P}[\mathbf{i}^{(\D)}_{\tau+1} \mid \ell_{\A}, \mathbf{b}_\tau]}{\mathbb{P}[\mathbf{i}^{(\D)}_{\tau+1} \mid \overline{\ell}_{\A}, \mathbf{b}_\tau]}\\
&=\mathbb{E}_{\mathbf{b}\sim \nu_t}\left[\sum_{\tau=1}^{t}\frac{\ln \mathbb{P}[\mathbf{i}^{(\D)}_{\tau+1} \mid \ell_{\A}, \mathbf{b}]}{t} - \sum_{\tau=1}^{t}\frac{\ln \mathbb{P}[\mathbf{i}^{(\D)}_{\tau+1} \mid \overline{\ell}_{\A}, \mathbf{b}]}{t}\right],
\end{align*}
where we use $\sum_{\mathbf{b} \in \mathcal{B}}\mathbbm{1}_{\mathbf{b}_{\tau}=\mathbf{b}}=1$. This means that if the left sum above converges to $\mathbb{E}_{\mathbf{I}^{(\D)}}[\ln \mathbb{P}[\mathbf{I}^{(\D)} | \ell_{\A}, \mathbf{b}]]$ and the right sum converges to $\mathbb{E}_{\mathbf{I}^{(\D)}}[\ln \mathbb{P}[\mathbf{I}^{(\D)} | \overline{\ell}_{\A}, \mathbf{b}]]$, we obtain $Z_t \xrightarrow[t \to \infty]{} K(\overline{\ell}_{\A}, \nu_{t})$, which yields the desired result. As these two proofs are almost identical, we only provide the first proof here.

Let $X_{\tau} \triangleq \ln\mathbb{P}[\mathbf{i}^{(\D)}_{\tau+1} | \ell_\A, \mathbf{b}_{\tau}] - \mathbb{E}_{\mathbf{I}^{(\D)}}\left[\ln\mathbb{P}[\mathbf{I}^{(\D)} | \ell_\A, \mathbf{b}_{\tau}]\right]$. We will show that $(X_{\tau})_{\tau \geq 1}$ is a martingale difference sequence (\mds). To show this, we need to prove that (\textit{i}) $\mathbb{E}[X_{\tau} | \mathbf{h}_{\tau-1}]=0$; and (\textit{ii}) $\mathbb{E}[|X_{\tau}|] < \infty$. We start with (\textit{i}),
\begin{align*}
&\mathbb{E}[X_{\tau} | \mathbf{h}_{\tau-1}]\\
&=\mathbb{E}_{\mathbf{I}^{(\D)}}\left[\ln\mathbb{P}[\mathbf{I}^{(\D)} |\ell_\A, \mathbf{b}_{\tau}] - \mathbb{E}_{\mathbf{I}^{(\D)}}\left[\ln\mathbb{P}[\mathbf{I}^{(\D)} | \ell_\A, \mathbf{b}_{\tau}]\right] | \mathbf{h}_{\tau-1}\right]\\
&\numeq{a}\mathbb{E}_{\mathbf{I}^{(\D)}}\left[\ln\mathbb{P}[\mathbf{I}^{(\D)} | \ell_\A, \mathbf{b}_{\tau}]\right] -\mathbb{E}_{\mathbf{I}^{(\D)}}\left[\ln\mathbb{P}[\mathbf{I}^{(\D)} | \ell_\A, \mathbf{b}_{\tau}]\right]=0,
\end{align*}
where (a) follows because $\mathbf{I}^{(\D)}$ is conditionally independent of $\mathbf{h}_{\tau-1}$ given $\mathbf{b}_{\tau}$ (\ref{eq:belief_upd}).

To prove (\textit{ii}) we will write $X_{\tau}$ as an expression of the form $(\ln \mathbb{P}[\varphi])^2\mathbb{P}[\varphi]$, which is bounded by $1$\footnote{We use the standard convention that $(\ln 0)^20=0$.}. Towards this end, we start by applying Jensen's inequality to obtain
\begin{align}
\mathbb{E}[|X_{\tau}|]=(\mathbb{E}[|X_{\tau}|]^2)^{1/2}\leq (\mathbb{E}[X_{\tau}^2])^{1/2},\label{eq:jensen}
\end{align}
which means that it suffices to bound $\mathbb{E}[X_{\tau}^2]$.

Next, we write $\ln\mathbb{P}[\mathbf{i}^{(\D)}_{\tau+1} |\ell_\A, \mathbf{b}_{\tau}]$ as
\begin{align*}
&\sum_{\mathbf{I}^{(\D)}}\mathbbm{1}_{\{\mathbf{I}^{(\D)}\}}(\mathbf{i}^{(\D)}_{\tau+1})\frac{\mathbb{P}[\mathbf{I}^{(\D)} \mid \ell_\A, \mathbf{b}_{\tau}]}{\mathbb{P}[\mathbf{I}^{(\D)} \mid \ell_\A, \mathbf{b}_{\tau}]}\ln\mathbb{P}[\mathbf{I}^{(\D)} \mid \ell_\A, \mathbf{b}_{\tau}]\\
&=\frac{\mathbb{E}_{\mathbf{I}^{(\D)}}\left[\mathbbm{1}_{\{\mathbf{I}^{(\D)}\}}(\mathbf{i}^{(\D)}_{\tau+1})\ln\mathbb{P}[\mathbf{I}^{(\D)} \mid \ell_\A, \mathbf{b}_{\tau}]\right]}{\mathbb{P}[\mathbf{i}_{\tau+1}^{(\D)} \mid \ell_\A, \mathbf{b}_{\tau}]},
\end{align*}
which means that we can write $X_{\tau}$ as
\begin{align}
\frac{\mathbb{E}_{\mathbf{I}^{(\D)}}\left[\kappa \left(\mathbbm{1}_{\{\mathbf{I}^{(\D)}\}}(\mathbf{i}^{(\D)}_{\tau+1}) - \mathbb{P}[\mathbf{i}_{\tau+1}^{(\D)} | \ell_\A, \mathbf{b}_{\tau}]\right) \mid \ell_\A, \mathbf{b}_{\tau}\right]}{\mathbb{P}[\mathbf{i}_{\tau+1}^{(\D)} | \ell_\A, \mathbf{b}_{\tau}]},\label{eq:x_tau_1}
\end{align}
where $\kappa\triangleq \ln\mathbb{P}[\mathbf{I}^{(\D)} | \ell_\A, \mathbf{b}_{\tau}]$.

Since we focus on realizable histories, we have that $\mathbb{P}[\mathbf{i}_{\tau+1}^{(\D)} | \ell_\A, \mathbf{b}_{\tau}] \in (0,1]$. Hence, it is safe to suppress the denominator in (\ref{eq:x_tau_1}). Consequently, the square of (\ref{eq:x_tau_1}) can be bounded by the Cauchy-Schwarz inequality as
\begin{align*}
\leq \mathbb{E}_{\mathbf{I}^{(\D)}}\Big[\kappa^2 \big(\underbrace{\mathbbm{1}_{\{\mathbf{I}^{(\D)}\}}(\mathbf{i}^{(\D)}_{\tau}+1) - \mathbb{P}[\mathbf{i}_{\tau+1}^{(\D)} | \ell_\A, \mathbf{b}_{\tau}]}_{\triangleq \chi}\big)^2 | \ell_\A, \mathbf{b}_{\tau}\Big].
\end{align*}
Next, since $\kappa \geq 0$ and $\chi \in [0,1]$, we obtain that
\begin{align*}
X^2_{\tau} \leq \mathbb{E}_{\mathbf{I}^{(\D)}}\left[\kappa^2 \chi^2 | \ell_\A, \mathbf{b}_{\tau}\right] \leq \mathbb{E}_{\mathbf{I}^{(\D)}}\left[\kappa^2 | \ell_\A, \mathbf{b}_{\tau}\right],
\end{align*}
which means that
\begin{align*}
  \mathbb{E}[X_{\tau}^2] &\leq \mathbb{E}_{\mathbf{I}^{(\D)}}\left[\mathbb{P}[\mathbf{I}^{(\D)} | \ell_\A, \mathbf{b}_{\tau}]\ln^2\mathbb{P}[\mathbf{I}^{(\D)} | \ell_\A, \mathbf{b}_{\tau}] | \ell_\A, \mathbf{b}_{\tau}\right] \leq 1\\
  &\numimp{a} \mathbb{E}[|X_{\tau}|] \leq 1,
\end{align*}
where (a) follows from (\ref{eq:jensen}). Therefore, $(X_{\tau})_{\tau \geq 1}$ is an \mds. Consequently, the sequence $Y_{t} \triangleq \sum_{\tau=1}^{t}\frac{X_{\tau}}{\tau}$ is a martingale. By the martingale convergence theorem, $(Y_{\tau})_{\tau \geq 1}$ converges to a finite and integrable random variable a.s.-$\mathbb{P}^{\mathscr{R}}$ \cite[Thm. 6.4.3]{Ash:2000uj}. This convergence means that we can invoke Kronecker's lemma, which states that $\lim_{t \rightarrow \infty}t^{-1}\sum_{\tau=1}^tX_{\tau}=0$ a.s.-$\mathbb{P}^{\mathscr{R}}$ \cite[pp. 105]{pollard_2001}. As a consequence, the following holds a.s.-$\mathbb{P}^{\mathscr{R}}$ as $t \rightarrow \infty$
\begin{align*}
\mathbb{E}_{\mathbf{b} \sim \nu_t}\sum_{\tau=1}^t\frac{\ln\mathbb{P}[\mathbf{i}^{(\D)}_{\tau+1} | \ell_\A, \mathbf{b}]}{t} = \mathbb{E}_{\mathbf{b} \sim \nu_t}\mathbb{E}_{\mathbf{I}^{(\D)}}\left[\ln\mathbb{P}[\mathbf{I}^{(\D)} | \ell_\A, \mathbf{b}]\right].
\end{align*}
\end{proof}
\subsection{Proof of \Theoremref{thm:conjecture_convergence}.A}
To streamline analysis, we treat $\mu_t$ as a probability measure over $\mathcal{L}$ and use integral language.
From Bayes rule and the Markov property of $\mathbb{P}[\mathbf{I}^{(\D)}\mid \overline{\ell}_{\A},\mathbf{b}_t]$, we have that
\begin{align}
\mu_{t+1}(\overline{\ell}_{\A}) &= \frac{\mathbb{P}[\overline{\ell}_{\A}]\mathbb{P}[\mathbf{i}^{(\D)}_2,\hdots,\mathbf{i}^{(\D)}_{t+1} \mid \overline{\ell}_{\A}, \mathbf{h}^{(\mathrm{D})}_t]}{\mathbb{P}[\mathbf{i}^{(\D)}_2,\hdots,\mathbf{i}^{(\D)}_{t+1} \mid \mathbf{h}^{(\mathrm{D})}_t]}\nonumber\\
                            &\numeq{a} \frac{\mu_1(\overline{\ell}_{\A})\prod_{\tau=1}^t\mathbb{P}[\mathbf{i}^{(\D)}_{\tau+1} \mid \overline{\ell}_{\A},\mathbf{b}_{\tau}]}{\int_{\mathcal{L}}\mu_1(\mathrm{d}\overline{\ell}_{\A})\prod_{\tau=1}^t\mathbb{P}[\mathbf{i}^{(\D)}_{\tau+1} \mid \overline{\ell}_{\A},\mathbf{b}_{\tau}]}\nonumber\\
                            &=\frac{\mu_1(\overline{\ell}_{\A})\exp\left(\ln\left(\prod_{\tau=1}^t\frac{\mathbb{P}[\mathbf{i}^{(\D)}_{\tau+1} \mid \overline{\ell}_{\A},\mathbf{b}_{\tau}]}{\mathbb{P}[\mathbf{i}^{(\D)}_{\tau+1} \mid \ell_{\A},\mathbf{b}_{\tau}]}\right)\right)}{\int_{\mathcal{L}}\mu_1(\mathrm{d}\overline{\ell}_{\A})\exp\left(\ln\left(\prod_{\tau=1}^t\frac{\mathbb{P}[\mathbf{i}^{(\D)}_{\tau+1} \mid \overline{\ell}_{\A},\mathbf{b}_{\tau}]}{\mathbb{P}[\mathbf{i}^{(\D)}_{\tau+1} \mid \ell_{\A},\mathbf{b}_{\tau}]}\right)\right)}\nonumber\\
&=\frac{\mu_1(\overline{\ell}_{\A})\exp\left(-tZ_{t+1}(\overline{\ell}_\A)\right)}{\int_{\mathcal{L}}\mu_1(\mathrm{d}\overline{\ell}_{\A})\exp\left(-tZ_{t+1}(\overline{\ell}_\A)\right)},\label{eq:thm4a_first}
\end{align}
where $Z_t$ is defined in \lemmaref{lem:lln}. Step (a) above is well-defined by \assumptionref{assumption:bayes} and follows because $\mathbf{I}^{(\mathrm{D})}$ is conditionally independent of $\mathbf{h}^{(\mathrm{D})}_{t-1}$ given $\mathbf{b}_{t-1}$.

Using the expression in (\ref{eq:thm4a_first}), we obtain that
\begin{align}
&\mathbb{E}_{\overline{\ell}_{\A} \sim \mu_{t+1}}\Big[\overbrace{K(\overline{\ell}_{\A},  \nu_{t})-K_{\mathcal{L}}^\star(\nu_{t})}^{\triangleq \Delta K(\overline{\ell}_{\A}, \nu_{t})}\Big] \label{eq:sigma_proof}\\
&= \frac{\int_{\mathcal{L}}\Delta K(\overline{\ell}_{\A}, \nu_{t})\mu_1(\mathrm{d}\overline{\ell}_{\A})\exp\left(-tZ_{t+1}(\overline{\ell}_\A)\right)}{\int_{\mathcal{L}}\mu_1(\mathrm{d}\overline{\ell}_{\A})\exp\left(-tZ_{t+1}(\overline{\ell}_\A)\right)}\nonumber \\
&=\frac{\int_{\mathcal{L}}\overbrace{\Delta K(\overline{\ell}_{\A}, \nu_{t})\mu_1(\mathrm{d}\overline{\ell}_{\A})\exp\left(-t(Z_{t+1}(\overline{\ell}_\A)-K^{\star}_{\mathcal{L}}(\nu_{t+1}))\right)}^{\triangleq \sigma}}{\int_{\mathcal{L}}\mu_1(\mathrm{d}\overline{\ell}_{\A})\exp\left(-t(Z_{t+1}(\overline{\ell}_\A)-K^{\star}_{\mathcal{L}}(\nu_{t+1}))\right)}.\nonumber
\end{align}
By defining $\mathcal{L}_{\epsilon} \triangleq \{\overline{\ell}_A \mid \Delta K(\overline{\ell}_{\A}, \nu_{t+1}) \geq \epsilon\}$, we can write the numerator in (\ref{eq:sigma_proof}) as
\begin{align}
&\int_{\mathcal{L}\setminus \mathcal{L}_{\epsilon}}\sigma + \int_{\mathcal{L}_{\epsilon}}\sigma \leq \epsilon + \int_{\mathcal{L}\setminus \mathcal{L}_{\epsilon}}\sigma.\label{eq:thm4_eps_bound}
\end{align}
Given this bound, it suffices to show that
\begin{align*}
\lim_{t\rightarrow \infty} \frac{\int_{\mathcal{L}_{\epsilon}}\sigma}{\int_{\mathcal{L}}\mu_1(\mathrm{d}\overline{\ell}_{\A})\exp\left(-t(Z_{t+1}(\overline{\ell}_\A)-K^{\star}_{\mathcal{L}}(\nu_{t+1}))\right)} = 0,
\end{align*}
for arbitrarily small $\epsilon > 0$. Towards this proof, we note that the exponent in $\sigma$ (\ref{eq:sigma_proof}) can be written as
\begin{align}
&-t(Z_{t+1}(\overline{\ell}_\A)-K^{\star}_{\mathcal{L}}(\nu_{t+1})) \nonumber\\
&=-t(Z_{t+1}(\overline{\ell}_\A)-K^{\star}_{\mathcal{L}}(\nu_{t+1})) + K(\overline{\ell}_{A},\nu_{t+1}) -K(\overline{\ell}_{A},\nu_{t+1})\nonumber\\
&=-t(\Delta K(\overline{\ell}_{\A}, \nu_{t+1}) + Z_{t+1}(\overline{\ell}_\A) -K(\overline{\ell}_{A},\nu_{t+1})). \label{eq:numerator_wrangling_thm_4}
\end{align}
Next, we recall from \lemmaref{lem:lln} that for any $\epsilon > 0$, there exists $\eta > 0$ and $t_{\eta}\geq 1$ such that, for all $t \geq t_{\eta}$ and $\overline{\ell}_{\A} \in \mathcal{L}$, $|Z_t(\overline{\ell}_{\A} - K(\overline{\ell}_\A, \nu_{t+1}))| < \eta$. ($t_{\eta}$ is uniform as $|\mathcal{L}| < \infty$, \assumptionref{assumption:bayes}.) This fact together with (\ref{eq:numerator_wrangling_thm_4}) implies that
\begin{align}
&\frac{\int_{\mathcal{L}_{\epsilon}}\Delta K(\overline{\ell}_{\A}, \nu_{t+1})\mu_1(\mathrm{d}\overline{\ell}_{\A})\exp\left(-t(Z_{t+1}(\overline{\ell}_\A)-K^{\star}_{\mathcal{L}}(\nu_{t+1}))\right)}{\int_{\mathcal{L}}\mu_1(\mathrm{d}\overline{\ell}_{\A})\exp\left(-t(Z_{t+1}(\overline{\ell}_\A)-K^{\star}_{\mathcal{L}}(\nu_{t+1}))\right)}\nonumber\\
&\leq \frac{\int_{\mathcal{L}_{\epsilon}}\Delta K(\overline{\ell}_{\A}, \nu_{t+1})\mu_1(\mathrm{d}\overline{\ell}_{\A})\exp\left(-t(\Delta K(\overline{\ell}_\A, \nu_{t+1}) -\eta)\right)}{\int_{\mathcal{L}}\mu_1(\mathrm{d}\overline{\ell}_{\A})\exp\left(-t(\Delta K(\overline{\ell}_\A, \nu_{t+1}) + \eta)\right)}\nonumber\\
&= e^{2t\eta}\frac{\int_{\mathcal{L}_{\epsilon}}\Delta K(\overline{\ell}_{\A}, \nu_{t+1})\mu_1(\mathrm{d}\overline{\ell}_{\A})e^{-t\Delta K(\overline{\ell}_\A, \nu_{t+1})}}{\int_{\mathcal{L}}\mu_1(\mathrm{d}\overline{\ell}_{\A})e^{-t\Delta K(\overline{\ell}_\A, \nu_{t+1})}} & \tag{*}\label{eq:thm4_a_bound_pf1}
\end{align}
for all $t \geq t_{\eta}$.

Now, consider the numerator in (\ref{eq:thm4_a_bound_pf1}). Since $xe^{-tx}$ is decreasing for all $x>t^{-1}$ and since $\Delta K(\overline{\ell}_{\A}, \nu_{t+1}) \geq \epsilon$ for all $\overline{\ell}_{\A} \in \mathcal{L}_{\epsilon}$, we have that
\begin{align*}
\int_{\mathcal{L}_{\epsilon}}\Delta K(\overline{\ell}_{\A}, \nu_{t+1})\mu_1(\mathrm{d}\overline{\ell}_{\A})e^{-t\Delta K(\overline{\ell}_\A, \nu_{t+1})} \leq \epsilon e^{-t\epsilon}
\end{align*}
for any $t \geq \max [t_{\eta}, \frac{1}{\epsilon}]$.

Next, consider the denominator in (\ref{eq:thm4_a_bound_pf1}). By definition, $\exists \overline{\ell}_{\A} \in \mathcal{L}$ such that $K(\overline{\ell}_{\A},  \nu_{t})=K_{\mathcal{L}}^\star(\nu_{t}) \forall t$. This fact, together with the assumption that $\mu_1$ has full support (\assumptionref{assumption:bayes}), means that the denominator is a positive constant, which we denote by $k$. As a result, $(*) \leq e^{2t\eta}\epsilon e^{-t\epsilon}k^{-1}$. Let $\eta=\frac{\epsilon}{4}$. Then $e^{2t\eta}\epsilon e^{-t\epsilon}k^{-1}=e^{\frac{-t\epsilon}{2}}\epsilon k^{-1}$, which converges to $0$ as $t \rightarrow \infty$. Consequently, $\lim_{t \rightarrow \infty}\mathbb{E}_{\overline{\ell}_{\A}\sim \mu_t}[\Delta K(\overline{\ell}_{\A}, \nu_{t})] = 0$ a.s.-$\mathbb{P}^{\mathscr{R}}$.

\section{Proof of \Theoremref{thm:conjecture_convergence}.B}\label{app:proof_conjecture_convergence_B}
The proof of \theoremref{thm:conjecture_convergence}.B follows the same procedure as that of \theoremref{thm:conjecture_convergence}.A, with the difference that $\Theta_{\mathrm{k}}$ is allowed to be non-finite, whereas $\mathcal{L}$ in \theoremref{thm:conjecture_convergence}.A is finite (\assumptionref{assumption:bayes}).

Define $\Theta^{+}_{\mathrm{k},\epsilon}\triangleq \{\overline{\bm{\theta}} | \Delta K(\overline{\bm{\theta}}, \nu_{t})\geq \epsilon\}$ and $\Theta_{\mathrm{k},\frac{\epsilon}{2}}^{-}\triangleq \{\overline{\bm{\theta}} | \Delta K(\overline{\bm{\theta}}, \nu_{t})\leq \frac{\epsilon}{2}\}$. It then follows from (\ref{eq:thm4_eps_bound}) that
\begin{align}
  &\int_{\Theta_{\mathrm{k}}} \Big(\overbrace{K(\overline{\bm{\theta}},  \nu_{t})-K_{\Theta_{\mathrm{k}}}^\star(\nu_{t})}^{\Delta K(\overline{\bm{\theta}}, \nu_{t})}\Big)\rho^{(\mathrm{k})}_{t+1}(\mathrm{d}\overline{\bm{\theta}}) \leq \epsilon + \label{eq:thm4b_11}\\
  & \underbrace{\frac{\int_{\Theta^{+}_{\mathrm{k},\epsilon}}\Delta K(\overline{\bm{\theta}}, \nu_{t})\exp\left(-t\left(Z_t(\overline{\bm{\theta}})-K_{\Theta_{\mathrm{k}}}^\star(\nu_{t})\right)\right)\rho^{(\mathrm{k})}_1(\mathrm{d}\overline{\bm{\theta}})}{\int_{\Theta_{\mathrm{k},\frac{\epsilon}{2}}^{-}}\exp\left(-t\left(Z_t(\overline{\bm{\theta}})-K_{\Theta_{\mathrm{k}}}^\star(\nu_{t})\right)\right)\rho^{(\mathrm{k})}_1(\mathrm{d}\overline{\bm{\theta}})}}_{\triangleq \text{\ding{73}}},\nonumber
\end{align}
where \ding{73} is well-defined by Assumptions \ref{assumption:bayes}--\ref{assumption:regularity}.

(\ref{eq:thm4b_11}) implies that it suffices to prove that $\text{\ding{73}}\xrightarrow[]{t\rightarrow\infty} 0$ for arbitrarily small $\epsilon$. Applying \lemmaref{lem:lln} and (\ref{eq:thm4_a_bound_pf1}), we obtain
\begin{align}
  \text{\ding{73}}&\numleq{a} e^{2t\eta} \frac{\int_{\Theta^{+}_{\mathrm{k}, \epsilon}}\Delta K(\overline{\bm{\theta}}, \nu_{t})e^{-t\Delta K(\overline{\bm{\theta}}, \nu_{t})}\rho^{(\mathrm{k})}_1(\mathrm{d}\overline{\bm{\theta}})}{\int_{\Theta_{\mathrm{k}, \frac{\epsilon}{2}}^{-}}e^{-t\Delta K(\overline{\bm{\theta}}, \nu_{t})}\rho^{(\mathrm{k})}_1(\mathrm{d}\overline{\bm{\theta}})}\nonumber\\
                   &\numleq{b} e^{2t\eta} \frac{\epsilon e^{-t\epsilon}}{\int_{\Theta_{\mathrm{k}, \frac{\epsilon}{2}}^{-}}e^{-t\Delta K(\overline{\bm{\theta}}, \nu_{t})}\rho^{(\mathrm{k})}_1(\mathrm{d}\overline{\bm{\theta}})}\nonumber \\
  &\numleq{c} e^{2t\eta} \frac{\epsilon e^{-t\epsilon}}{e^{-t\frac{\epsilon}{2}}\int_{\Theta_{\mathrm{k}, \frac{\epsilon}{2}}^{-}}\rho^{(\mathrm{k})}_1(\mathrm{d}\overline{\bm{\theta}})} = e^{2t\eta} \frac{\epsilon e^{-t\frac{\epsilon}{2}}}{\rho^{(\mathrm{k})}_1(\Theta_{\mathrm{k}, \frac{\epsilon}{2}}^{-})}\label{eq:thm4_proof_eps_32}
\end{align}
for all $t\geq \max[t_\eta, \epsilon^{-1}]$, where (a) follows from (\ref{eq:thm4_a_bound_pf1}); (b) follows because $xe^{-tx}$ is decreasing for all $x>t^{-1}$; and c) follows because $e^{-t\Delta K(\overline{\bm{\theta}}, \nu_{t})}\geq e^{-t\frac{\epsilon}{2}}$.

Let $\eta=\epsilon/8$. Then the numerator in the final expression above becomes $\epsilon e^{-t\frac{\epsilon}{4}}$, which converges to $0$ as $t \rightarrow \infty$. Thus, what remains to show is that the denominator is positive in the limit, i.e., $\lim_{t\rightarrow\infty}\rho^{(\mathrm{k})}_1(\Theta_{\mathrm{k}, \frac{\epsilon}{2}}^{-})>0$. We prove this statement by establishing uniform continuity of $\Delta K(\overline{\bm{\theta}},\nu)$. Towards this end, we prove the following two lemmas.
\begin{lemma}
\label{lemma:compact-wasserstein}
$\mathcal{B}$ (\ref{eq:belief_upd}) is a compact subset of $\mathbb{R}^{|\mathcal{S}|}$ with the Euclidean metric $d$ and $\Delta(\mathcal{B})$ a compact metric space with the Wasserstein-$p$ distance $W_p$ ($p \geq 1$).
\end{lemma}
\begin{proof}
Since $\mathcal{S}$ is finite, $\mathcal{B}=\Delta(\mathcal{S})$ is a compact subset of $\mathbb{R}^{|\mathcal{S}|}$ and $(\mathcal{B}, d)$ is a Polish space. To prove that $(\Delta(\mathcal{B}), W_p)$ is compact we will show that every sequence $(\nu_n)_{n=1}^\infty\subset \Delta(\mathcal{B})$ admits a subsequence converging to some limit point in $\Delta(\mathcal{B})$. Since $\mathcal{B}$ is compact and $\nu_n(\mathcal{B})=1$, this collection of measures is tight as $\exists \mathcal{C}\subseteq \mathcal{B}$ such that $\nu_n(\mathcal{C})=1>1-\epsilon$ for any $\epsilon>0$ and $\nu_n$. Therefore, $(\nu_n)_{n=1}^\infty$ admits a limit point $\nu^{\star}\in \Delta(\mathcal{B})$ w.r.t the topology of weak convergence (Prokhorov's theorem \cite[Ch. 1,\S 5]{billing}). We will show that $\nu^{\star}$ is also a limit point under $W_p$. By Skorokhod's representation theorem \cite[p. 70]{billing}, there exists a sequence of $\mathcal{B}$-valued random variables $\{V_1,\ldots, V_n,\ldots, V^{\star} \}$ such that $V_n$ has the probability law $\nu_n$ and $V_n$ converges to $V^{\star}$ almost surely as $n \rightarrow \infty$. By the dominance convergence theorem and the facts that $\mathcal{B}$ is compact and $d$ is continuous, $\lim_{n\rightarrow \infty}\mathbb{E}[d(V_n, V^{\star})^p]= 0$. Consequently, for any coupling $\xi$ between $\nu_n$ and $\nu^{\star}$,
\begin{align*}
\lim_{n\rightarrow\infty}\left(\int d(x,y)^p d\xi(x,y)\right)^{\frac{1}{p}}=0.
\end{align*}
Since $W_p(\nu_n, \nu^{\star})$ is the infimum of the left-hand side above (by definition), $\lim_{n\rightarrow\infty}W_p(\nu_n, \nu^{\star})=0$. Hence, every sequence $(\nu_n)_{n=1}^\infty\subset \Delta(\mathcal{B})$ admits a subsequence converging to some limit point in $\Delta(\mathcal{B})$ under $W_p$. Thus, $(\Delta(\mathcal{B}), W_p)$ is compact.
\end{proof}
\begin{lemma}\label{lemma:KL-continuity}
$\Delta K(\overline{\bm{\theta}}, \nu)\triangleq K(\overline{\bm{\theta}},\nu)-K_{\Theta_{\mathrm{k}}}^\star(\nu)$ is a continuous map from $(\Theta_{\mathrm{k}}, d)\times (\Delta(\mathcal{B}), {W}_1)$ to $\mathbb{R}$, where $d$ and $W_1$ denote the Euclidean and the Wasserstein-$1$ distance, respectively.
\end{lemma}
\begin{proof}
We start by showing that $K(\overline{\bm{\theta}}, \nu)$ is continuous by proving that for any convergent sequence $(\overline{\bm{\theta}}_n, \nu_n) \xrightarrow[n \to \infty]{} (\overline{\bm{\theta}}, \nu)$, the difference $|K(\overline{\bm{\theta}}_n, \nu_n)-K(\overline{\bm{\theta}}, \nu)|$ converges to $0$. This difference can be bounded by the triangle inequality as
\begin{align*}
&|K(\overline{\bm{\theta}}_n, \nu_n)-K(\overline{\bm{\theta}}, \nu)|\leq \\
&\underbrace{|K(\overline{\bm{\theta}}_n, \nu_n)-K(\overline{\bm{\theta}}_n, \nu)|}_{\triangleq \text{\ding{172}}}+\underbrace{|K(\overline{\bm{\theta}}_n, \nu)-K(\overline{\bm{\theta}}, \nu)|}_{\triangleq \text{\ding{173}}}.
\end{align*}
Consider the left expression above (\ding{172}). (\ref{eq:discrepancy}) implies that
\begin{align*}
\text{\ding{172}}=&\left|\int_\mathcal{B}\mathbb{E}_{\mathbf{I}^{(\mathrm{k})}}\left[\ln\left(\frac{\mathbb{P}[\mathbf{I}^{(\mathrm{k})} \mid \bm{\theta}, \mathbf{b}]}{\mathbb{P}[\mathbf{I}^{(\mathrm{k})} \mid \overline{\bm{\theta}}_n, \mathbf{b}]}\right)\right]\nu_n(\mathrm{d}\mathbf{b} )\right.\\
&\left. -\int_\mathcal{B}\mathbb{E}_{\mathbf{I}^{(\mathrm{k})}}\left[\ln\left(\frac{\mathbb{P}[\mathbf{I}^{(\mathrm{k})} \mid \bm{\theta}, \mathbf{b}]}{\mathbb{P}[\mathbf{I}^{(\mathrm{k})} \mid \overline{\bm{\theta}}_n, \mathbf{b}]}\right)\right]\nu(\mathrm{d}\mathbf{b})\right|,
\end{align*}
which is an integral probability metric (\textsc{ipm}) with the testing function $f(\mathbf{b})\triangleq \mathbb{E}_{\mathbf{I}^{(\mathrm{k})}}\left[\ln\left(\frac{\mathbb{P}[\mathbf{I}^{(\mathrm{k})} \mid \bm{\theta}, \mathbf{b}]}{\mathbb{P}[\mathbf{I}^{(\mathrm{k})} \mid \overline{\bm{\theta}}_n, \mathbf{b}]}\right)\right]$. This function is assumed to be Lipschitz continuous (\assumptionref{assumption:regularity}.1). As the function can be rescaled, we can, without loss of generality, assume the Lipschitz constant to be $1$. Since the Wasserstein distance is equivalent to the \textsc{ipm} w.r.t the class of 1-Lipschitz functions, \ding{172} is upper-bounded by $W_p(\nu_n,\nu)$. Hence, as $\nu_n$ converges to $\nu$ in $W_p$, \ding{172} converges to $0$. Therefore, $\nu \mapsto K(\overline{\bm{\theta}},\nu)$ is continuous. We now show that \ding{173} converges. Using \assumptionref{assumption:regularity}.2 and the dominated convergence theorem, we obtain that
\begin{align*}
\lim_{n\rightarrow\infty}&\mathbb{E}_{\mathbf{b}\sim \nu}\mathbb{E}_{\mathbf{I}^{(\mathrm{k})}}\left[\ln\left(\frac{\mathbb{P}[\mathbf{I}^{(\mathrm{k})} \mid \bm{\theta}, \mathbf{b}]}{\mathbb{P}[\mathbf{I}^{(\mathrm{k})} \mid \overline{\bm{\theta}}_n, \mathbf{b}]}\right)\right]\\
=& \mathbb{E}_{\mathbf{b}\sim \nu}\mathbb{E}_{\mathbf{I}^{(\mathrm{k})}}\left[\ln\left(\frac{\mathbb{P}[\mathbf{I}^{(\mathrm{k})} \mid \bm{\theta}, \mathbf{b}]}{\mathbb{P}[\mathbf{I}^{(\mathrm{k})} \mid \overline{\bm{\theta}}, \mathbf{b}]}\right)\right],
  \end{align*}
which implies that \ding{173} converges to $0$ as $n\rightarrow\infty$. Consequently, $\overline{\bm{\theta}}\mapsto K(\overline{\bm{\theta}},\nu)$ is continuous. Finally, since $\Theta_{\mathrm{k}}$ is compact (\assumptionref{assumption:bayes}) and $K(\overline{\bm{\theta}},\nu)$ is continuous in both $\overline{\bm{\theta}}$ and $\nu$, we can apply Berge's maximum theorem to $K(\overline{\bm{\theta}},\nu)$ w.r.t. $\overline{\bm{\theta}}$. This theorem states that the mapping $\nu\mapsto K_{\Theta_{\mathrm{k}}}^\star(\nu)$ is continuous (\ref{eq:consistent_model_conjecture}). Since continuity is preserved under subtraction, it follows that $\Delta K$ is also continuous \cite[Thm. 17.31]{aliprantis06}.
\end{proof}
\subsection{Proof of \Theoremref{thm:conjecture_convergence}.B}
Lemmas \ref{lemma:compact-wasserstein}--\ref{lemma:KL-continuity} and the compactness of $\Theta_{\mathrm{k}}$ (\assumptionref{assumption:bayes}) imply uniform continuity of $\Delta K(\overline{\bm{\theta}}, \nu)$ and that $|\Theta^\star_{\mathrm{k}}(\nu)|>0$ \cite[Thm. 17.31]{aliprantis06}. As a consequence, for each $\overline{\bm{\theta}}_\nu\in \Theta^\star_{\mathrm{k}}(\nu)$, $\overline{\bm{\theta}}' \in \Theta_{\mathrm{k}}$, and $\nu',\nu \in \Delta(\mathcal{B})$, $\exists \delta_m$ such that $d(\overline{\bm{\theta}}_\nu, \overline{\bm{\theta}}')< \delta_m$, $W_1(\nu, \nu')< \delta_m$, and $d(\Delta K(\overline{\bm{\theta}}',\nu'), \Delta K(\overline{\bm{\theta}}_{\nu},\nu))\leq m \numimp{a} \Delta K(\overline{\bm{\theta}}',\nu') < m$ for each $m > 0$, where (a) follows because $\overline{\bm{\theta}}_\nu\in \Theta^\star_{\mathrm{k}}(\nu) \implies \Delta K(\overline{\bm{\theta}}_\nu, \nu)=0$. Define the ball $B(\nu, \delta_m)\triangleq \{\nu' \mid  W_1(\nu,\nu')< \delta_m, \nu' \in \Delta(\mathcal{B})\}$. It follows that, for any $\nu\in \Delta(\mathcal{B})$ and $\nu'\in B(\nu, \delta_m)$,
\begin{align*}
\underbrace{\{\overline{\bm{\theta}}'| d(\overline{\bm{\theta}}',\overline{\bm{\theta}}_\nu)< \delta_m\}}_{\Theta_{\nu}(\delta_m)} \subset \underbrace{\{\overline{\bm{\theta}}'| \Delta K(\overline{\bm{\theta}}',\nu')\leq m \}}_{\Theta_{\nu^{\prime}}(m)}.
\end{align*}
Thus, for any $\nu$ and $\nu'\in B(\nu, \delta_m)$,
\begin{align*}
\rho_1^{(\mathrm{k})}(\Theta_{\nu^{\prime}}(m))&\geq \rho_1^{(\mathrm{k})}(\Theta_{\nu}(\delta_m))\numgeq{a} 0,
\end{align*}
where (a) follows because $\rho_1^{(\mathrm{k})}$ has full support (\assumptionref{assumption:bayes}).

Since $\Delta(\mathcal{B})$ is compact (\lemmaref{lemma:compact-wasserstein}), the set $\{B(\nu,\delta_m)\}_{\nu\in \Delta(\mathcal{B})}$ forms an open cover for a compact space, which means that there exists a finite subcover $\{B(\nu_i,\delta_m)\}_{i=1}^M$. As a conseqeuence, each $\nu'\in \Delta(\mathcal{B})$ belongs to some Wasserstein ball $B(\nu_i,\delta_m)$. Let $r \triangleq \min_{i}\rho^{(\mathrm{k})}_1(\Theta_{\nu_i}(\delta_m)) > 0$. We then have that
\begin{align*}
\rho_1^{(\mathrm{k})}(\Theta_{\nu^{\prime}}(m))&\geq \rho_1^{(\mathrm{k})}(\Theta_{\nu_i}(\delta_m)) \geq r.
\end{align*}
Now recall the denominator $\rho^{(\mathrm{k})}_1(\Theta_{\mathrm{k}, \frac{\epsilon}{2}}^{-})$ in (\ref{eq:thm4_proof_eps_32}). Let $m=\frac{\epsilon}{2}$. Then $\rho_1^{(\mathrm{k})}(\Theta_{\mathrm{k}, \frac{\epsilon}{2}}^{-})\geq r>0$ for any $\epsilon > 0$. Hence, $\lim_{t\rightarrow \infty} \frac{\epsilon e^{-t\frac{\epsilon}{4}}}{\rho^{(\mathrm{k})}_1(\Theta_{\mathrm{k}, \frac{\epsilon}{2}}^{-})}=0$.\qed
\section{Example Derivation of a \hyperref[def:berk_nash]{Berk-Nash Equilibrium}}\label{appendix:berk_nash_example}
We use the following example to illustrate the steps required to find a \hyperref[def:berk_nash]{Berk-Nash equilibrium}.
\begin{example}\label{example:1}
\normalfont Consider \probref{main_problem} with $N=1$, $p_{\mathrm{A}}=1$, $\mathcal{O}=\{0,1\}$, $z_{\bm{\theta}_t}(\cdot | 0) = \operatorname{Ber}(p)$, $z_{\bm{\theta}_t}(1 | 1) = \operatorname{Ber}(q)$, $\mathbf{b}_1(1)=0$, and $c$ being defined as in \figref{fig:cost_function}. Let the rollout parameters be ($\ell_{\mathrm{A}}=0$, $\ell_{\mathrm{D}}=1$) and let $\bm{\pi}_1$ be threshold strategies with $\beta \geq 0$ and $\alpha \leq 1$ (\theoremref{thm:threshold_defender}). Finally, let $\mathcal{L}=\{\ell_{\mathrm{A}}\}$, $\Theta_{\mathrm{A}}=\{\bm{\theta}_1\}$, and $\Theta_{\mathrm{D}}=\{\overline{\bm{\theta}}_{a}, \overline{\bm{\theta}}_b\}$, where $z_{\overline{\bm{\theta}}_{a}}(0 | 0)=z_{\overline{\bm{\theta}}_{a}}(1 | 1)=z_{\overline{\bm{\theta}}_{b}}(1 | 0)=z_{\overline{\bm{\theta}}_{b}}(0 | 1)=1$.
\end{example}
First note that the definition of $z_{\overline{\bm{\theta}}_{a}}$, $z_{\overline{\bm{\theta}}_{b}}$, and $\mathbf{b}_1$ imply that $\mathbf{b}_t(1) \in \{0,1\}$ for all $t$, which simplifies the following derivation. For ease of notation, we write $\mathbf{b}$ instead of $\mathbf{b}(1)$. To derive a \hyperref[def:berk_nash]{\textsc{bne}}, we start with condition (\textit{i}) in \defref{def:berk_nash}. Since $\ell_{\A}=0$, it suffices to consider $\pi_{\D}$. By the principle of optimality
\begin{align*}
\pi_\D(\mathbf{b}) \in \argmin_{a \in \mathcal{A}_{\D}}\mathbb{E}_{S, \mathbf{B}^{\prime}}\left[c(S,a) + \gamma \overline{J}^{(\bm{\pi}_1)}_{\D, \overline{\bm{\theta}}}(\mathbf{B}^{\prime}) \mid \mathbf{b}, \bm{\pi}_1\right].
\end{align*}
Let $\mathbf{P}_{\overline{\bm{\theta}},\bm{\pi}_1}$ and $\mathbf{c}_{\bm{\pi}_1}$ be the belief transition matrix and the vector of expected stage costs induced by $(\overline{\bm{\theta}}, \bm{\pi}_1)$, respectively.  From the definition of $\Theta_{\mathrm{D}}$ we obtain that
\begin{align*}
  \mathbf{P}_{\overline{\bm{\theta}}_a,\bm{\pi}_1} \numeq{a}
  \begin{bmatrix}
    1-q & q\\
    1 & 0
  \end{bmatrix},
\text{ }
  \mathbf{P}_{\overline{\bm{\theta}}_b,\bm{\pi}_1} \numeq{b}
  \begin{bmatrix}
    1-p & p\\
    1 & 0
  \end{bmatrix},
\text{ }
\mathbf{c}_{\bm{\pi}_1} \numeq{c}
  \begin{bmatrix}
    0\\
    -1
  \end{bmatrix},
\end{align*}
where (a)--(c) follow because $\alpha \in (0,1] \implies \pi_{\D}(1)=\mathsf{S}, \pi_{\D}(0)=\mathsf{C}$.

By definition, $\overline{J}^{(\bm{\pi}_1)}_{\D, \overline{\bm{\theta}}} = (\mathbf{1}_{2}-\gamma\mathbf{P}_{\overline{\bm{\theta}},\bm{\pi}_1})^{-1}\mathbf{c}_{\bm{\pi}_1}$, where $\mathbf{1}_{2}$ is the $2 \times 2$ identity matrix. Therefore,
\begin{align*}
  \overline{J}^{(\bm{\pi}_1)}_{\D,\overline{\bm{\theta}}_a} &= (\mathbf{1}_{2}-\gamma\mathbf{P}_{\overline{\bm{\theta}}_a,\bm{\pi}_1})^{-1}\mathbf{c}_{\bm{\pi}_1} \\
  &=  \left(\mathbf{1}_{2} -\gamma
  \begin{bmatrix}
    1-q & q\\
    1 & 0
  \end{bmatrix}
        \right)^{-1}
  \begin{bmatrix}
    0\\
    -1
  \end{bmatrix}\\
                                                          &=
                                                            \begin{bmatrix}
    1-\gamma+ \gamma q & -\gamma q\\
    -\gamma & 1
  \end{bmatrix}^{-1}
  \begin{bmatrix}
    0\\
    -1
  \end{bmatrix}\\
  &=  \begin{bmatrix}
    \frac{-1}{(\gamma-1)(1+\gamma q)} & \frac{-\gamma q}{(\gamma-1)(1+\gamma q)}\\
    \frac{-\gamma}{(\gamma-1)(1+\gamma q)} & \frac{\gamma-1 - \gamma q}{(\gamma-1)(1+\gamma q)}
  \end{bmatrix}
  \begin{bmatrix}
    0\\
    -1
  \end{bmatrix}\\
&=\frac{1}{(\gamma-1)(1+\gamma q)}
  \begin{bmatrix}
    \gamma q\\
    1+ \gamma (q-1)
  \end{bmatrix}.
\end{align*}
Similarly,
\begin{align*}
  \overline{J}^{(\bm{\pi}_1)}_{\D,\overline{\bm{\theta}}_b} &= (\mathbf{1}_{2}-\gamma\mathbf{P}_{\overline{\bm{\theta}}_b,\bm{\pi}_1})^{-1}\mathbf{c}_{\bm{\pi}_1} \\
  &=  \left(\mathbf{1}_{2} -\gamma
  \begin{bmatrix}
    1-p & p\\
    1 & 0
  \end{bmatrix}
        \right)^{-1}
  \begin{bmatrix}
    0\\
    -1
  \end{bmatrix}\\
&=\frac{1}{(\gamma-1)(1+\gamma p)}
  \begin{bmatrix}
    \gamma p\\
    1+ \gamma (p-1)
  \end{bmatrix}.
\end{align*}
Hence, to meet condition (\textit{i}), the defender's rollout strategy must satisfy $\pi_{\D}(0)=\mathsf{C}$ and $\pi_{\D}(1)=\mathsf{S}$, which is ensured by (\ref{eq:rollout_operator}). As a consequence, $\bm{\pi}=\bm{\pi}_1$ in any \hyperref[def:berk_nash]{\textsc{bne}}.

Now consider condition (\textit{ii}); (\ref{eq:discrepancy}) can be written as
\begin{align*}
  K(\overline{\bm{\theta}},\nu)&= \mathbb{E}_{\mathbf{b}\sim \nu}\mathbb{E}_{\mathbf{I}^{(\mathrm{D})}}\left[\ln\left(\frac{\mathbb{P}[\mathbf{I}^{(\mathrm{D})} \mid \bm{\theta}, \mathbf{b}]}{\mathbb{P}[\mathbf{I}^{(\mathrm{D})} \mid \overline{\bm{\theta}}, \mathbf{b}]}\right)\mid  \bm{\theta}, \mathbf{b}\right]\\
                               &=-\sum_{\mathbf{b} \in \{0,1\}}\nu(\mathbf{b})\sum_{o \in \{0,1\}}z_{\bm{\theta}}(o\mid \mathbf{b})\ln z_{\overline{\bm{\theta}}}(o\mid \mathbf{b}) +\operatorname{const.}
\end{align*}
Minimizing the above expression with respect to $\bm{\overline{\theta}}$ yields $\Theta^{\star}_\D=\{\overline{\bm{\theta}}_a\}$ if $(p=0,q=1)$. Conversely, $\Theta^{\star}_\D=\{\overline{\bm{\theta}}_{b}\}$ if $(p=1,q=0)$. Otherwise, $\Theta^{\star}_{\mathrm{D}}(\nu)=\{\overline{\bm{\theta}}_a,\overline{\bm{\theta}}_{b}\}$.

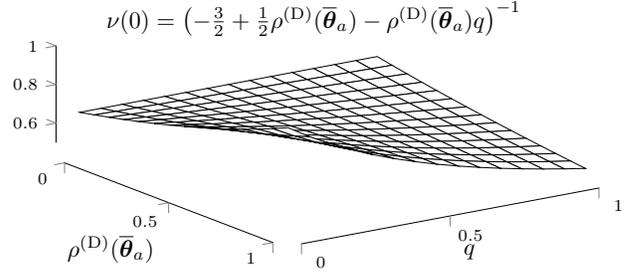
\begin{figure}
  \centering
  \scalebox{0.9}{
    \begin{tikzpicture}
\node[scale=1] (plot12) at (0,0)
{
\begin{tikzpicture}
\pgfplotsset{
%  /tikz/normal shift/.code 2 args = {%
%    \pgftransformshift{%
%        \pgfpointscale{#2}{\pgfplotspointouternormalvectorofticklabelaxis{#1}}%
%    }%
%  },%
%  %
%  shift/.style = {
%    tick align        = outside,
%    scaled ticks      = false,
%    enlargelimits     = false,
%    ticklabel shift   = {#1},
%    axis lines*       = left,
%    xtick style       = {normal shift={x}{#1}},
%    ytick style       = {normal shift={y}{#1}},
%    x axis line style = {normal shift={x}{#1}},
%    y axis line style = {normal shift={y}{#1}},
%  },
%  shift/.default = 10pt,
%  shift3d/.style = {
%    shift=#1,
%    ztick style       = {normal shift={z}{#1}},
%    z axis line style = {normal shift={z}{#1}},
%  },
%  shift3d/.default = 10pt,
%  every axis/.append style = {thick},tick style = {thick,black},
  %
  % #1 = x, y, or z
  % #2 = the shift value
  /tikz/normal shift/.code 2 args = {%
    \pgftransformshift{%
        \pgfpointscale{#2}{\pgfplotspointouternormalvectorofticklabelaxis{#1}}%
    }%
  },%
  range3frame/.style = {
    tick align        = outside,
    scaled ticks      = false,
    enlargelimits     = false,
    ticklabel shift   = {10pt},
    axis lines*       = left,
    line cap          = round,
    clip              = false,
    xtick style       = {normal shift={x}{10pt}},
    ytick style       = {normal shift={y}{10pt}},
    ztick style       = {normal shift={z}{10pt}},
    x axis line style = {normal shift={x}{10pt}},
    y axis line style = {normal shift={y}{10pt}},
    z axis line style = {normal shift={z}{10pt}},
  }
}
\begin{axis} [
range3frame, view={55}{45},
    width =9cm,
    height = 5cm,
    xtick = {0,0.5, 1},
    ytick = {0,0.5, 1},
%    ytick style={draw=none},
%    xtick style={draw=none},
%    ztick style={draw=none},
%    xlabel = $\rho_{\mathrm{D}}(\overline{\bm{\theta}}_a)$, ylabel = $q$,
    ticklabel style = {font = \scriptsize}
%    shift
]
%\addplot3 [mesh, domain=0:1, samples=30] { -1/(-1.5 + 0.5*x -x*y)};
\addplot3 [surf, colormap/blackwhite, fill=white, point meta=0, domain=0:1, samples=15] { -1/(-1.5 + 0.5*x -x*y)};

%\addplot3 [surf, colormap/hot2, domain = -2:2, samples = 50]
%      { x/exp(x^2+y^2) };
\end{axis}
\end{tikzpicture}
};
\node[inner sep=0pt,align=center, scale=1, rotate=0, opacity=1] (obs) at (0,2)
{
  $\nu(0)=\left(-\frac{3}{2} + \frac{1}{2}\rho^{(\mathrm{D})}(\overline{\bm{\theta}}_a) - \rho^{(\mathrm{D})}(\overline{\bm{\theta}}_a)q\right)^{-1}$
};
\node[inner sep=0pt,align=center, scale=1, rotate=0, opacity=1] (obs) at (-3,-1.4)
{
$\rho^{(\mathrm{D})}(\overline{\bm{\theta}}_a)$
};
\node[inner sep=0pt,align=center, scale=1, rotate=0, opacity=1] (obs) at (2.3,-1.45)
{
$q$
};
\end{tikzpicture}
  }
  \caption{\hyperref[def:berk_nash]{Berk-Nash equilibria} of \exampleref{example:1} when $p=\frac{1}{2}$.}
  \label{fig:berk_nash}
\end{figure}

Lastly, condition (\textit{iii}) is satisfied iff $\mathbf{P}^T_{\overline{\bm{\theta}},\bm{\pi}_1}\nu = \nu$. Since $\mathbf{P}^T_{\overline{\bm{\theta}},\bm{\pi}_1} = \rho^{(\mathrm{D})}(\overline{\bm{\theta}}_a)\mathbf{P}^T_{\overline{\bm{\theta}}_a,\bm{\pi}_1} + (1-\rho^{(\mathrm{D})}(\overline{\bm{\theta}}_a))\mathbf{P}^T_{\overline{\bm{\theta}}_b,\bm{\pi}_1}$, solving this equation gives
\begin{align}
\nu(0) = -\left(-1-p + \rho^{(\D)}(\overline{\bm{\theta}}_a)p -\rho^{(\D)}(\overline{\bm{\theta}}_a)q\right)^{-1},\label{eq:stationary_cond_berk}
\end{align}
which means the \hyperref[def:berk_nash]{\textsc{bne}} is not unique and may not exist; see \figref{fig:berk_nash} on the next page. For example, if $p=1$ and $q=0$, then (\ref{eq:stationary_cond_berk}) requires that $\rho^{(\D)}(\overline{\bm{\theta}}_a)=1$, but this means that $\rho^{(\D)} \not \in \Delta(\Theta^{\star}(\nu)$, which violates condition (\textit{ii}).
\section{Hyperparameters}\label{appendix:hyperparameters}
\begin{table}
\centering
\resizebox{1\columnwidth}{!}{
\begin{tabular}{ll} \toprule
  \textbf{Figures and Tables} & {\textbf{Values}} \\
    \hline
    \figref{fig:hsvi_times} & $\mathcal{O}=\{0,\hdots,9\}$, $p_{\mathrm{A}}=0.1, \gamma=0.99$ \\
                           & $z(\cdot \mid 0) = \mathrm{BetaBin}(n=10,\alpha=0.7,\beta=3)$\\
                           & $z(\cdot \mid 1) = \mathrm{BetaBin}(n=10,\alpha=1,\beta=0.7$\\
    \figref{fig:value_fun} & $\mathcal{O}=\{0,\hdots,9\}$, $p_{\mathrm{A}}=0.1$, $N=1$, $\gamma=0.99$\\
                           & $z(\cdot \mid 0) = \mathrm{BetaBin}(n=10,\alpha=0.7,\beta=3)$\\
                           & $z(\cdot \mid 1) = \mathrm{BetaBin}(n=10,\alpha=1,\beta=0.7$\\
    \figref{fig:rollout_times} & $N=10$, $p_{\mathrm{A}}=0.1, \gamma=0.99$ \\
                   & $\pi_{\D, 1}(\mathsf{S} \mid \mathbf{b}_t) = 1 \iff \mathbb{P}[S_t \geq 1 \mid \mathbf{b}_t] \geq 0.75$\\
                   & $\pi_{\A,1}(\mathsf{S} \mid \mathbf{b}_t,s_t) = 0.5$\\
                   & Cost function of base strategy estimated \\
                   &using $100$ MC samples w. horizon $50$\\
    \figref{fig:evaluation_results}.a--e & $\ell_{\mathrm{A}}=\ell_{\mathrm{D}}=1, \mathcal{L}=\{1,2\}, p_{\mathrm{A}}=1$\\
                   & $\pi_{\D, 1}(\mathsf{S} \mid \mathbf{b}_t) = 1 \iff \mathbb{P}[S_t \geq 1 \mid \mathbf{b}_t] \geq 0.75$\\
                   & $\pi_{\A,1}(\mathsf{S} \mid \mathbf{b}_t,s_t) = 0.05$\\
                   &$N=10$, $\mathcal{O}, z$ (\figref{fig:alerts}) \\
                   & using $100$ MC samples w. horizon $50$\\
    \figref{fig:best_response_learning}, \figref{fig:evaluation_results}.f & $\mathcal{O}, z$ (\figref{fig:alerts}), $p_{\mathrm{A}}=0.1, \gamma=0.99$ \\
                   & $\mathcal{L}={0,1,2}$, $N=10$\\
                   & $\pi_{\D, 1}(\mathsf{S} \mid \mathbf{b}_t) = 1 \iff \mathbb{P}[S_t \geq 1 \mid \mathbf{b}_t] \geq 0.75$\\
                   & $\pi_{\A,1}(\mathsf{S} \mid \mathbf{b}_t,s_t) = 0.05$\\
                   & \hyperref[eq:best_responses]{Best response} computation: \cem \cite{cem_rubinstein}\\
                   & \hyperref[eq:best_responses]{Best responses} parameterized following \theoremref{thm:threshold_defender}\\
    Figs. \ref{fig:evaluation_results}.g--o, \ref{fig:posterior_evolution} & $\ell_{\mathrm{A}}=\ell_{\mathrm{D}}=1, p_{\mathrm{A}}=1, \gamma=0.99$\\
                   & $\pi_{\D, 1}(\mathsf{S} \mid \mathbf{b}_t) = 1 \iff \mathbb{P}[S_t \geq 1 \mid \mathbf{b}_t] \geq 0.75$\\
                   & $\pi_{\A,1}(\mathsf{S} \mid \mathbf{b}_t,s_t) = 0.05$\\
                   &$N=10$, $\mathcal{O}, z$ (\figref{fig:alerts}) \\
                   & using $100$ MC samples w. horizon $50$\\
    \figref{fig:evaluation_results}.k & $\Theta_{\D} = \{0,\hdots, 200\}$\\
    \tableref{tab:convergence_comparison} & $\ell_{\mathrm{A}}=\ell_{\mathrm{D}}=1, p_{\mathrm{A}}=1, \gamma=0.99$\\
                   & $\pi_{\D, 1}(\mathsf{S} \mid \mathbf{b}_t) = 1 \iff \mathbb{P}[S_t \geq 1 \mid \mathbf{b}_t] \geq 0.75$\\
                   & $\pi_{\A,1}(\mathsf{S} \mid \mathbf{b}_t,s_t) = 0.05$\\
                   &$N=10$, $\mathcal{O}, z$ (\figref{fig:alerts}, $t=10$) \\
                              & using $100$ MC samples w. horizon $50$\\
    \figref{fig:arrivals} & $\bm{\psi} = (\frac{1}{2}, 10^{-2}, -10^{5})$, $\bm{\chi} = (1.0593)$ \\
                              &$\bm{\phi}=(-0.5193)$, $\bm{\omega} = (0.054\pi)$\\
                              &time-step: $30$s, service time: $\operatorname{Exp}(\mu=4)t$\\
    \figref{fig:ips_eval} & $\mathcal{O}$ (\figref{fig:alerts}), $\pi_{\A,1}(\mathsf{S} \mid \cdot) =1$, $\ell_{\D}=1$\\
    Confidence intervals & computed using the Student-t distribution\\
    Base strategies $\pi_{\D,1}\pi_{\A,1}$ & Approximate threshold \hyperref[eq:best_responses]{best responses} \\
                                          & against randomized opponents (\figref{fig:best_response_learning})\\
    Priors $\mu_1,\rho_1^{(\mathrm{D})},\rho_1^{(\mathrm{A})}$ & uniform\\
    Cost function (\ref{eq:cost_fun}), \figref{fig:cost_function} & $p=5/4$, $q=1$, $r=2$\\
  {\textbf{\hsvi Parameter}} &  \\
  \hline
    $\epsilon$ & $0.1$\\
  {\textbf{Cross-entropy method \cite{cem_rubinstein}}} &   \\
  \hline
    $\lambda$ (fraction of samples to keep) & $0.15,100$\\
    $K$ population size & $100$\\
    $M$ number of samples for each evaluation & $50$ \\
  {\textbf{\ppo \cite[Alg. 1]{ppo} parameters}} &   \\
  \hline
  lr $\alpha$, batch, \# layers, \# neurons, clip $\epsilon$ & $10^{-5}$, $4\cdot 10^{3}t$, $4$, $64$, $0.2$,\\
    GAE $\lambda$, ent-coef, activation & $0.95$, $10^{-4}$, ReLU \\
  {\textbf{\nfsp \cite[Alg. 9]{heinrich_thesis} parameters}} &  \\
  \hline
  lr RL, lr SL, batch, \# layers,\# neurons, $\mathcal{M}_{RL}$ & $10^{-2}$, $5\cdot 10^{-3}$, $64$, $2$,$128$, $2\times 10^{5}$ \\
  $\mathcal{M}_{SL}$,$\epsilon$, $\epsilon$-decay, $\eta$ & $2\times 10^{6}$, $0.06$, $0.001$, $0.1$\\
    \bottomrule\\
\end{tabular}
}
\caption{Hyperparameters.}\label{tab:hyperparams}
\end{table}
The hyperparameters used for the evaluation in this paper are listed in \tableref{tab:hyperparams} and were obtained through grid search.
\section{Configuration of the Infrastructure in \figref{fig:use_case}}\label{appendix:infrastructure_configuration}
The configuration of the target infrastructure (\figref{fig:use_case}) is available in \tableref{tab:target_infra_config}.

\begin{table*}
\centering
\resizebox{1\linewidth}{!}{%
\begin{tabular}{llllll} \toprule
  {\textit{ID(s)}} & {\textit{Type}} & {\textit{Operating system}} & {\textit{Zone}} & {\textit{Services}} & {\textit{Vulnerabilities}} \\ \midrule
  $1$ & Gateway & \ubuntu 20 & - & \snort (ruleset v2.9.17.1), \ssh, \openflow v1.3, \ryu \sdn controller & -\\
  $2$ & Gateway & \ubuntu 20 & \dmz & \snort (ruleset v2.9.17.1), \ssh, \ovs v2.16, \openflow v1.3 & -\\
  $28$ & Gateway & \ubuntu 20 & \rnd & \snort (ruleset v2.9.17.1), \ssh, \ovs v2.16, \openflow v1.3 & -\\
  $3$,$12$ & Switch & \ubuntu 22 & \dmz & \ssh, \openflow v1.3 , \ovs v2.16& -\\
  $21,22$ & Switch & \ubuntu 22 & - & \ssh, \openflow v1.3, \ovs v2.16 & -\\
  $23$ & Switch & \ubuntu 22 & \admin & \ssh, \openflow v1.3, \ovs v2.16 & -\\
  $29$-$48$ & Switch & \ubuntu 22 & \rnd & \ssh, \openflow v1.3, \ovs v2.16 & -\\
  $13$-$16$ & Honeypot & \ubuntu 20 & \dmz & \ssh, \snmp, \postgres, \ntp & -\\
  $17$-$20$ & Honeypot & \ubuntu 20 & \dmz & \ssh, \irc, \snmp, \ssh, \postgres & -\\
  $4$ & App node & \ubuntu 20 & \dmz & \http, \dns, \ssh & \cwe-1391\\
  $5$, $6$ & App node & \ubuntu 20 & \dmz & \ssh, \snmp, \postgres, \ntp & -\\
  $7$ & App node & \ubuntu 20 & \dmz & \http, \telnet, \ssh & \cwe-1391\\
  $8$ & App node & \debian \jessie & \dmz & \ftp, \ssh, \apache 2,\snmp & \cve-2015-3306\\
  $9$,$10$ & App node & \ubuntu 20 & \dmz & \ntp, \irc, \snmp, \ssh, \postgres & -\\
  $11$ & App node & \debian \jessie & \dmz & \apache 2, \smtp, \ssh & \cve-2016-10033\\
  $24$ & Admin system & \ubuntu 20 & \admin & \http, \dns, \ssh & \cwe-1391\\
  $25$ & Admin system & \ubuntu 20 & \admin & \ftp, \mongo, \smtp, \tomcat, \ts 3, \ssh & -\\
  $26$ & Admin system & \ubuntu 20 & \admin & \ssh, \snmp, \postgres, \ntp & -\\
  $27$ & Admin system & \ubuntu 20 & \admin & \ftp, \mongo, \smtp, \tomcat, \ts 3, \ssh & \cwe-1391\\
  $49$-$59$ & Compute node & \ubuntu 20 & \rnd & \spark, \hdfs & -\\
  $60$ & Compute node & \debian \wheezy & \rnd & \spark, \hdfs, \apache 2,\snmp, \ssh & \cve-2014-6271\\
  $61$ & Compute node & \debian 9.2 & \rnd & \irc, \apache 2, \ssh & \cwe-89\\
  $62$ & Compute node & \debian \jessie & \rnd & \spark, \hdfs, \ts 3, \tomcat, \ssh & \cve-2010-0426\\
  $63$ & Compute node & \debian \jessie & \rnd & \ssh, \spark, \hdfs & \cve-2015-5602\\
  $64$ & Compute node & \debian \jessie & \rnd & \samba, \ntp, \ssh, \spark, \hdfs & \cve-2017-7494\\
  \bottomrule\\
\end{tabular}
}
\caption{Configuration of the target infrastructure shown in \figref{fig:use_case}; vulnerabilities in specific software products are identified by the vulnerability identifiers in the Common Vulnerabilities and Exposures (\cve) database \cite{cve}; vulnerabilities that are not described in the \cve database are categorized according to the types of the vulnerabilities they exploit based on the Common Weakness Enumeration (\cwe) list \cite{cwe}.}\label{tab:target_infra_config}
\end{table*}

\bibliographystyle{IEEEtran}
\bibliography{references,url}

% Generated by IEEEtran.bst, version: 1.14 (2015/08/26)
\begin{thebibliography}{100}
\providecommand{\url}[1]{#1}
\csname url@samestyle\endcsname
\providecommand{\newblock}{\relax}
\providecommand{\bibinfo}[2]{#2}
\providecommand{\BIBentrySTDinterwordspacing}{\spaceskip=0pt\relax}
\providecommand{\BIBentryALTinterwordstretchfactor}{4}
\providecommand{\BIBentryALTinterwordspacing}{\spaceskip=\fontdimen2\font plus
\BIBentryALTinterwordstretchfactor\fontdimen3\font minus
  \fontdimen4\font\relax}
\providecommand{\BIBforeignlanguage}[2]{{%
\expandafter\ifx\csname l@#1\endcsname\relax
\typeout{** WARNING: IEEEtran.bst: No hyphenation pattern has been}%
\typeout{** loaded for the language `#1'. Using the pattern for}%
\typeout{** the default language instead.}%
\else
\language=\csname l@#1\endcsname
\fi
#2}}
\providecommand{\BIBdecl}{\relax}
\BIBdecl

\bibitem{game_t_sec_survey}
M.~H. Manshaei, Q.~Zhu, T.~Alpcan, T.~Basar, and J.-P. Hubaux, ``Game theory
  meets network security and privacy,'' \emph{ACM Comput. Surv.}, vol.~45,
  no.~3, pp. 25:1--25:39, Jul. 2013.

\bibitem{flipit}
M.~van Dijk, A.~Juels, A.~Oprea, and R.~L. Rivest, ``Flipit: The game of
  ``stealthy takeover'','' \emph{Journal of Cryptology}, no.~4, Oct 2013.

\bibitem{dynamic_game_linan_zhu}
L.~Huang and Q.~Zhu, ``A dynamic games approach to proactive defense strategies
  against advanced persistent threats in cyber-physical systems,''
  \emph{Computers \& Security}, vol.~89, p. 101660, 11 2019.

\bibitem{tao_info}
T.~Li, Y.~Zhao, and Q.~Zhu, ``{The role of information structures in
  game-theoretic multi-agent learning},'' \emph{Annual Reviews in Control},
  vol.~53, pp. 296--314, 2022.

\bibitem{kamhoua2021game}
C.~Kamhoua, C.~Kiekintveld, F.~Fang, and Q.~Zhu, \emph{Game Theory and Machine
  Learning for Cyber Security}.\hskip 1em plus 0.5em minus 0.4em\relax Wiley,
  2021.

\bibitem{honeypot_game}
K.~Durkota, V.~Lisy, B.~Bo\v{s}ansky, and C.~Kiekintveld, ``Optimal network
  security hardening using attack graph games,'' in \emph{Proceedings of the
  24th International Conference on Artificial Intelligence}, 2015.

\bibitem{DBLP:journals/compsec/HorakBTKK19}
K.~Hor{\'{a}}k, B.~Bosansk{\'{y}}, P.~Tom{\'{a}}sek, C.~Kiekintveld, and C.~A.
  Kamhoua, ``Optimizing honeypot strategies against dynamic lateral movement
  using partially observable stochastic games,'' \emph{Comput. Secur.},
  vol.~87, 2019.

\bibitem{hammar_stadler_tnsm_23}
K.~Hammar and R.~Stadler, ``Learning near-optimal intrusion responses against
  dynamic attackers,'' \emph{IEEE Transactions on Network and Service
  Management}, vol.~21, no.~1, pp. 1158--1177, 2024.

\bibitem{kim_gamesec23}
------, ``Scalable learning of intrusion response through recursive
  decomposition,'' in \emph{Decision and Game Theory for Security}, J.~Fu,
  T.~Kroupa, and Y.~Hayel, Eds.\hskip 1em plus 0.5em minus 0.4em\relax Cham:
  Springer Nature Switzerland, 2023, pp. 172--192.

\bibitem{ZHAO2020106878}
Y.~Zhao, L.~Huang, C.~Smidts, and Q.~Zhu, ``Finite-horizon semi-markov game for
  time-sensitive attack response and probabilistic risk assessment in nuclear
  power plants,'' \emph{Reliability Engineering \& System Safety}, vol. 201, p.
  106878, 2020.

\bibitem{8691466}
J.~Chen and Q.~Zhu, ``Interdependent strategic security risk management with
  bounded rationality in the internet of things,'' \emph{IEEE Transactions on
  Information Forensics and Security}, vol.~14, no.~11, pp. 2958--2971, 2019.

\bibitem{notpetya_us}
U.~D. of~Justice, ``Six russian gru officers charged in connection with
  worldwide deployment of destructive malware and other disruptive actions in
  cyberspace,'' 2020,
  \url{https://www.justice.gov/opa/pr/six-russian-gru-officers-charged-connection-worldwide-deployment-destructive-malware-and}.

\bibitem{notpetya_mitre_sandworm_2}
T.~M. Corporation, ``Notpetya,'' 2024,
  \url{https://attack.mitre.org/software/S0368/}.

\bibitem{bertsekas2021rollout}
D.~Bertsekas, \emph{Rollout, Policy Iteration, and Distributed Reinforcement
  Learning}, ser. Athena scientific optimization and computation series.\hskip
  1em plus 0.5em minus 0.4em\relax Athena Scientific, 2021.

\bibitem{berk_nash}
I.~Esponda and D.~Pouzo, ``Berk-nash equilibrium: A framework for modeling
  agents with misspecified models,'' \emph{Econometrica}, vol.~84, no.~3, pp.
  1093--1130, 2023/10/13/ 2016.

\bibitem{kagel_mechanism_design}
J.~H. Kagel and D.~Levin, ``The winner's curse and public information in common
  value auctions,'' \emph{The American Economic Review}, vol.~76, no.~5, pp.
  894--920, 1986.

\bibitem{rabin_psychology}
M.~Rabin, ``Inference by believers in the law of small numbers,'' \emph{The
  Quarterly Journal of Economics}, vol. 117, no.~3, pp. 775--816, 2002.

\bibitem{simons_bounded_rationality}
H.~A. Simon, ``Theories of bounded rationality,'' \emph{Decision and
  Organization}, pp. 161--176, 1972.

\bibitem{samuelson_bounded_ratinality}
L.~Samuelson, ``{Bounded rationality and game theory},'' \emph{The Quarterly
  Review of Economics and Finance}, vol.~36, no. Supplemen, pp. 17--35, 1996.

\bibitem{Rosenthal1989}
R.~W. Rosenthal, ``A bounded-rationality approach to the study of
  noncooperative games,'' \emph{International Journal of Game Theory}, vol.~18,
  no.~3, pp. 273--292, Sep 1989.

\bibitem{9144263}
A.~Sanjab, W.~Saad, and T.~Başar, ``A game of drones: Cyber-physical security
  of time-critical uav applications with cumulative prospect theory perceptions
  and valuations,'' \emph{IEEE Transactions on Communications}, vol.~68,
  no.~11, pp. 6990--7006, 2020.

\bibitem{bounded_rational_stackelberg_1}
R.~Gabrys, M.~Bilinski, J.~Mauger, D.~Silva, and S.~Fugate, ``Casino rationale:
  Countering attacker deception in zero-sum stackelberg security games of
  bounded rationality,'' in \emph{Decision and Game Theory for Security},
  F.~Fang, H.~Xu, and Y.~Hayel, Eds.\hskip 1em plus 0.5em minus 0.4em\relax
  Cham: Springer International Publishing, 2023, pp. 23--43.

\bibitem{8362263}
J.~Chen and Q.~Zhu, ``Security investment under cognitive constraints: A
  gestalt nash equilibrium approach,'' in \emph{2018 52nd Annual Conference on
  Information Sciences and Systems (CISS)}, 2018, pp. 1--6.

\bibitem{ijcai2018p775}
A.~Sinha, F.~Fang, B.~An, C.~Kiekintveld, and M.~Tambe, ``Stackelberg security
  games: Looking beyond a decade of success,'' in \emph{Proceedings of the
  Twenty-Seventh International Joint Conference on Artificial Intelligence,
  {IJCAI-18}}.\hskip 1em plus 0.5em minus 0.4em\relax International Joint
  Conferences on Artificial Intelligence Organization, 7 2018, pp. 5494--5501.

\bibitem{9559403}
Z.~Wan, J.-H. Cho, M.~Zhu, A.~H. Anwar, C.~A. Kamhoua, and M.~P. Singh,
  ``Foureye: Defensive deception against advanced persistent threats via
  hypergame theory,'' \emph{IEEE Transactions on Network and Service
  Management}, vol.~19, no.~1, pp. 112--129, 2022.

\bibitem{8750848}
C.~Bakker, A.~Bhattacharya, S.~Chatterjee, and D.~L. Vrabie, ``Learning and
  information manipulation: Repeated hypergames for cyber-physical security,''
  \emph{IEEE Control Systems Letters}, vol.~4, no.~2, pp. 295--300, 2020.

\bibitem{behavioral_gt_1}
M.~Abdallah, P.~Naghizadeh, A.~R. Hota, T.~Cason, S.~Bagchi, and S.~Sundaram,
  ``Behavioral and game-theoretic security investments in interdependent
  systems modeled by attack graphs,'' \emph{IEEE Transactions on Control of
  Network Systems}, vol.~7, no.~4, pp. 1585--1596, 2020.

\bibitem{behavioral_gt_2}
C.~N. Mavridis, A.~Kanellopoulos, K.~G. Vamvoudakis, J.~S. Baras, and K.~H.
  Johansson, ``Attack identification for cyber-physical security in dynamic
  games under cognitive hierarchy,'' \emph{IFAC-PapersOnLine}, vol.~56, no.~2,
  pp. 11\,223--11\,228, 2023, 22nd IFAC World Congress.

\bibitem{posg_cyber_deception_network_epidemic}
O.~Tsemogne, Y.~Hayel, C.~Kamhoua, and G.~Deugoue, ``Partially observable
  stochastic games for cyber deception against network epidemic,'' in
  \emph{Decision and Game Theory for Security}, Q.~Zhu, J.~S. Baras,
  R.~Poovendran, and J.~Chen, Eds.\hskip 1em plus 0.5em minus 0.4em\relax Cham:
  Springer International Publishing, 2020, pp. 312--325.

\bibitem{hammar_stadler_cnsm_20}
K.~Hammar and R.~Stadler, ``Finding effective security strategies through
  reinforcement learning and {Self-Play},'' in \emph{International Conference
  on Network and Service Management (CNSM 2020)}, Izmir, Turkey, 2020.

\bibitem{9328143}
A.~Aydeger, M.~H. Manshaei, M.~A. Rahman, and K.~Akkaya, ``Strategic defense
  against stealthy link flooding attacks: A signaling game approach,''
  \emph{IEEE Transactions on Network Science and Engineering}, vol.~8, no.~1,
  pp. 751--764, 2021.

\bibitem{nework_security_alpcan}
T.~Alpcan and T.~Basar, \emph{Network Security: A Decision and Game-Theoretic
  Approach}, 1st~ed.\hskip 1em plus 0.5em minus 0.4em\relax USA: Cambridge
  University Press, 2010.

\bibitem{learning_in_games_fudenberg}
D.~Fudenberg and D.~K. Levine, \emph{The theory of learning in games}.\hskip
  1em plus 0.5em minus 0.4em\relax MIT Press, Cambridge, MA., 1998.

\bibitem{young_strategic_2004}
H.~P. Young, \emph{Strategic learning and its limits}.\hskip 1em plus 0.5em
  minus 0.4em\relax Oxford University Press, 2004, cited by 0313.

\bibitem{nash_q_learning}
J.~Hu and M.~P. Wellman, ``Nash q-learning for general-sum stochastic games,''
  \emph{J. Mach. Learn. Res.}, vol.~4, no. null, p. 1039–1069, dec 2003.

\bibitem{markov_game_q_littman}
M.~L. Littman, ``Markov games as a framework for multi-agent reinforcement
  learning,'' in \emph{Proceedings of the Eleventh International Conference on
  International Conference on Machine Learning}, ser. ICML'94.\hskip 1em plus
  0.5em minus 0.4em\relax San Francisco, CA, USA: Morgan Kaufmann Publishers
  Inc., 1994, p. 157–163.

\bibitem{ge_li_zhu_infocomm_workshop}
Y.~Ge, T.~Li, and Q.~Zhu, ``Scenario-agnostic zero-trust defense with
  explainable threshold policy: A meta-learning approach,'' \emph{IEEE INFOCOM
  2023 - IEEE Conference on Computer Communications Workshops (INFOCOM
  WKSHPS)}, pp. 1--6, 2023.

\bibitem{hammar_stadler_tnsm}
K.~Hammar and R.~Stadler, ``Intrusion prevention through optimal stopping,''
  \emph{IEEE Transactions on Network and Service Management}, vol.~19, no.~3,
  pp. 2333--2348, 2022.

\bibitem{r1_ref1}
S.~Acharya, Y.~Dvorkin, and R.~Karri, ``Causative cyberattacks on online
  learning-based automated demand response systems,'' \emph{IEEE Transactions
  on Smart Grid}, vol.~12, no.~4, pp. 3548--3559, 2021.

\bibitem{csle_docs}
CSLE, ``Cyber security learning environment,'' 2023, documentation:
  \url{https://limmen.dev/csle/}, traces:
  \url{https://github.com/Limmen/csle/releases/tag/v0.4.0}, source code:
  \url{https://github.com/Limmen/csle}, video demonstration:
  \url{https://www.youtube.com/watch?v=iE2KPmtIs2A&}.

\bibitem{snort}
M.~Roesch, ``Snort - lightweight intrusion detection for networks,'' in
  \emph{Proceedings of the 13th USENIX Conference on System Administration},
  ser. LISA '99.\hskip 1em plus 0.5em minus 0.4em\relax USA: USENIX
  Association, 1999, p. 229–238.

\bibitem{r1_ref4}
S.~Moothedath, D.~Sahabandu, J.~Allen, A.~Clark, L.~Bushnell, W.~Lee, and
  R.~Poovendran, ``A game-theoretic approach for dynamic information flow
  tracking to detect multistage advanced persistent threats,'' \emph{IEEE
  Transactions on Automatic Control}, vol.~65, no.~12, pp. 5248--5263, 2020.

\bibitem{horak_solving_one_sided_posgs}
K.~Horák, B.~Bošanský, V.~Kovařík, and C.~Kiekintveld, ``Solving zero-sum
  one-sided partially observable stochastic games,'' \emph{Artificial
  Intelligence}, vol. 316, p. 103838, 2023.

\bibitem{217638}
Y.~Ji, S.~Lee, M.~Fazzini, J.~Allen, E.~Downing, T.~Kim, A.~Orso, and W.~Lee,
  ``Enabling refinable {Cross-Host} attack investigation with efficient data
  flow tagging and tracking,'' in \emph{27th USENIX Security Symposium (USENIX
  Security 18)}.\hskip 1em plus 0.5em minus 0.4em\relax Baltimore, MD: USENIX
  Association, Aug. 2018, pp. 1705--1722.

\bibitem{NIPS2007_3435c378}
J.~Goldsmith and M.~Mundhenk, ``Competition adds complexity,'' in
  \emph{Advances in Neural Information Processing Systems}, J.~Platt,
  D.~Koller, Y.~Singer, and S.~Roweis, Eds., vol.~20.\hskip 1em plus 0.5em
  minus 0.4em\relax Curran Associates, Inc., 2007.

\bibitem{horak_thesis}
K.~Horák, ``Scalable algorithms for solving stochastic games with limited
  partial observability,'' Ph.D. dissertation, Czech Technical University in
  Prague, 2019.

\bibitem{kuhn1953}
H.~W. Kuhn, \emph{Extensive games and the problem of information}, H.~W. Kuhn
  and A.~W. Tucker, Eds.\hskip 1em plus 0.5em minus 0.4em\relax Princeton, NJ:
  Princeton University Press, 1953.

\bibitem{stochastic_systems_kumar}
P.~R. Kumar and P.~Varaiya, \emph{Stochastic systems: estimation,
  identification and adaptive control}.\hskip 1em plus 0.5em minus 0.4em\relax
  USA: Prentice-Hall, Inc., 1986.

\bibitem{krishnamurthy_2016}
V.~Krishnamurthy, \emph{Partially Observed Markov Decision Processes: From
  Filtering to Controlled Sensing}.\hskip 1em plus 0.5em minus 0.4em\relax
  Cambridge University Press, 2016.

\bibitem{cem_rubinstein}
R.~Rubinstein, ``The cross-entropy method for combinatorial and continuous
  optimization,'' \emph{Methodology And Computing In Applied Probability},
  vol.~1, no.~2, pp. 127--190, Sep 1999.

\bibitem{nash51}
J.~F. Nash, ``Non-cooperative games,'' \emph{Annals of Mathematics}, vol.~54,
  pp. 286--295, 1951.

\bibitem{vonNeumann_1928:TGG}
J.~von Neumann, ``\BIBforeignlanguage{German}{{Zur Theorie der
  Gesellschaftsspiele}. ({German}) [{On} the theory of games of strategy]},''
  \emph{\BIBforeignlanguage{German}{j-MATH-ANN}}, vol. 100, pp. 295--320, 1928.

\bibitem{fudenberg}
D.~Fudenberg and J.~Tirole, \emph{Game Theory}.\hskip 1em plus 0.5em minus
  0.4em\relax MIT Press, 1991.

\bibitem{bayesian_perfect_equilibria}
I.-K. Cho and D.~M. Kreps, ``Signaling games and stable equilibria,'' \emph{The
  Quarterly Journal of Economics}, vol. 102, no.~2, pp. 179--221, 1987.

\bibitem{Banach1922}
S.~Banach, ``\BIBforeignlanguage{fre}{Sur les opérations dans les ensembles
  abstraits et leur application aux équations intégrales},''
  \emph{\BIBforeignlanguage{fre}{Fundamenta Mathematicae}}, 1922.

\bibitem{posg_equilibria_existence_finite_horizon}
J.~Hespanha and M.~Prandini, ``Nash equilibria in partial-information games on
  markov chains,'' in \emph{Proceedings of the 40th IEEE Conference on Decision
  and Control (Cat. No.01CH37228)}, vol.~3, 2001.

\bibitem{horak_bosansky_hsvi}
K.~Horák, B.~Bošanský, and M.~Pěchouček, ``Heuristic search value
  iteration for one-sided partially observable stochastic games,''
  \emph{Proceedings of the AAAI Conference on Artificial Intelligence}, Feb.
  2017.

\bibitem{smallwood_1}
E.~J. Sondik, ``The optimal control of partially observable markov processes
  over the infinite horizon: Discounted costs,'' \emph{Operations Research},
  vol.~26, no.~2, pp. 282--304, 1978.

\bibitem{puterman}
M.~L. Puterman, \emph{Markov Decision Processes: Discrete Stochastic Dynamic
  Programming}, 1st~ed.\hskip 1em plus 0.5em minus 0.4em\relax USA: Wiley,
  1994.

\bibitem{pomdp_rollout}
S.~Bhattacharya, S.~Badyal, T.~Wheeler, S.~Gil, and D.~Bertsekas,
  ``Reinforcement learning for pomdp: Partitioned rollout and policy iteration
  with application to autonomous sequential repair problems,'' \emph{IEEE
  Robotics and Automation Letters}, vol.~5, no.~3, pp. 3967--3974, 2020.

\bibitem{kl_divergence}
S.~Kullback and R.~A. Leibler, ``On information and sufficiency,'' \emph{The
  Annals of Mathematical Statistics}, vol.~22, no.~1, pp. 79--86, 1951.

\bibitem{berk}
R.~H. Berk, ``{Limiting Behavior of Posterior Distributions when the Model is
  Incorrect},'' \emph{The Annals of Mathematical Statistics}, vol.~37, no.~1,
  pp. 51 -- 58, 1966.

\bibitem{esponda21berk_mdp}
I.~Esponda and D.~Pouzo, ``{Equilibrium in misspecified Markov decision
  processes},'' \emph{Theoretical Economics}, vol.~16, no.~2, pp. 717--757,
  2021.

\bibitem{netem}
S.~Hemminger, ``Network emulation with netem,'' \emph{Linux Conf}, 2005.

\bibitem{4365686}
H.~P. Reiser and R.~Kapitza, ``Hypervisor-based efficient proactive recovery,''
  in \emph{2007 26th IEEE International Symposium on Reliable Distributed
  Systems (SRDS 2007)}, 2007, pp. 83--92.

\bibitem{cve}
T.~M. Corporation, ``Cve database,'' 2022, \url{https://cve.mitre.org/}.

\bibitem{cwe}
------, ``Cwe list,'' 2023, \url{https://cwe.mitre.org/index.html}.

\bibitem{strom2018mitre}
B.~E. Strom, A.~Applebaum, D.~P. Miller, K.~C. Nickels, A.~G. Pennington, and
  C.~B. Thomas, ``Mitre att\&ck: Design and philosophy,'' in \emph{Technical
  report}.\hskip 1em plus 0.5em minus 0.4em\relax The MITRE Corporation, 2018.

\bibitem{478761}
M.~Kuhl, J.~Wilson, and M.~Johnson, ``Estimation and simulation of
  nonhomogeneous poisson processes having multiple periodicities,'' in
  \emph{Winter Simulation Conference Proceedings, 1995.}, 1995, pp. 374--383.

\bibitem{google_failure_model}
D.~Ford, F.~Labelle, F.~Popovici, M.~Stokely, V.-A. Truong, L.~Barroso,
  C.~Grimes, and S.~Quinlan, ``Availability in globally distributed storage
  systems,'' in \emph{Proceedings of the 9th USENIX Symposium on Operating
  Systems Design and Implementation}, 2010.

\bibitem{opponent_modeling}
M.~Shen and J.~P. How, ``Robust opponent modeling via adversarial ensemble
  reinforcement learning,'' \emph{Proceedings of the International Conference
  on Automated Planning and Scheduling}, vol.~31, no.~1, pp. 578--587, May
  2021.

\bibitem{NisaRougTardVazi07}
N.~Nisan, T.~Roughgarden, E.~Tardos, and V.~V. Vazirani, \emph{Algorithmic Game
  Theory}.\hskip 1em plus 0.5em minus 0.4em\relax New York, NY, USA: Cambridge
  University Press, 2007.

\bibitem{self_play_cyclic}
D.~{Hernandez}, K.~{Denamganaï}, Y.~{Gao}, P.~{York}, S.~{Devlin},
  S.~{Samothrakis}, and J.~A. {Walker}, ``A generalized framework for self-play
  training,'' in \emph{2019 IEEE Conference on Games (CoG)}, 2019, pp. 1--8.

\bibitem{brown_fictious_play}
G.~W. Brown, ``Iterative solution of games by fictitious play,'' 1951, activity
  analysis of production and allocation.

\bibitem{ppo}
J.~Schulman, F.~Wolski, P.~Dhariwal, A.~Radford, and O.~Klimov, ``Proximal
  policy optimization algorithms,'' \emph{CoRR}, 2017,
  \url{http://arxiv.org/abs/1707.06347}.

\bibitem{heinrich_thesis}
J.~Heinrich, ``Reinforcement learning from self-play in imperfect-information
  games,'' Ph.D. dissertation, University College London, 2017.

\bibitem{tambe}
M.~Tambe, \emph{Security and Game Theory: Algorithms, Deployed Systems, Lessons
  Learned}, 1st~ed.\hskip 1em plus 0.5em minus 0.4em\relax USA: Cambridge
  University Press, 2011.

\bibitem{r1_ref2}
J.~Tan, H.~Jin, H.~Zhang, Y.~Zhang, D.~Chang, X.~Liu, and H.~Zhang, ``A survey:
  When moving target defense meets game theory,'' \emph{Computer Science
  Review}, vol.~48, p. 100544, 2023.

\bibitem{r5_ref3}
T.~T. Nguyen and V.~J. Reddi, ``Deep reinforcement learning for cyber
  security,'' \emph{IEEE Transactions on Neural Networks and Learning Systems},
  vol.~34, no.~8, pp. 3779--3795, 2023.

\bibitem{r3_ref2}
L.-X. Yang, P.~Li, Y.~Zhang, X.~Yang, Y.~Xiang, and W.~Zhou, ``Effective repair
  strategy against advanced persistent threat: A differential game approach,''
  \emph{IEEE Transactions on Information Forensics and Security}, vol.~14,
  no.~7, pp. 1713--1728, 2019.

\bibitem{r3_ref3}
H.~Sun, X.~Yang, L.-X. Yang, K.~Huang, and G.~Li, ``Impulsive artificial
  defense against advanced persistent threat,'' \emph{IEEE Transactions on
  Information Forensics and Security}, vol.~18, pp. 3506--3516, 2023.

\bibitem{9923774}
T.~Zhu, D.~Ye, Z.~Cheng, W.~Zhou, and P.~S. Yu, ``Learning games for defending
  advanced persistent threats in cyber systems,'' \emph{IEEE Transactions on
  Systems, Man, and Cybernetics: Systems}, 2022.

\bibitem{altman_jamming_1}
E.~Altman, K.~Avrachenkov, and A.~Garnaev, ``A jamming game in wireless
  networks with transmission cost,'' in \emph{NET-COOP}, 2007.

\bibitem{r1_ref3}
T.~Halabi, O.~A. Wahab, R.~Al~Mallah, and M.~Zulkernine, ``Protecting the
  internet of vehicles against advanced persistent threats: A bayesian
  stackelberg game,'' \emph{IEEE Transactions on Reliability}, vol.~70, no.~3,
  pp. 970--985, 2021.

\bibitem{r1_ref5}
J.~Tan, H.~Jin, H.~Hu, R.~Hu, H.~Zhang, and H.~Zhang, ``Wf-mtd: Evolutionary
  decision method for moving target defense based on wright-fisher process,''
  \emph{IEEE Transactions on Dependable and Secure Computing}, vol.~20, no.~6,
  pp. 4719--4732, 2023.

\bibitem{stocahstic_games_security_indep_nodes_nguyen_alpcan_basar}
K.~C. Nguyen, T.~Alpcan, and T.~Basar, ``Stochastic games for security in
  networks with interdependent nodes,'' in \emph{2009 International Conference
  on Game Theory for Networks}, 2009, pp. 697--703.

\bibitem{zhu_basar_dynamic_policy_ids_config}
Q.~Zhu and T.~Başar, ``Dynamic policy-based ids configuration,'' in
  \emph{Proceedings of the 48h IEEE Conference on Decision and Control (CDC)
  held jointly with 2009 28th Chinese Control Conference}, 2009.

\bibitem{9096400}
H.~Hu, Y.~Liu, C.~Chen, H.~Zhang, and Y.~Liu, ``Optimal decision making
  approach for cyber security defense using evolutionary game,'' \emph{IEEE
  Transactions on Network and Service Management}, vol.~17, no.~3, pp.
  1683--1700, 2020.

\bibitem{5270307}
S.~A. Zonouz, H.~Khurana, W.~H. Sanders, and T.~M. Yardley, ``Rre: A
  game-theoretic intrusion response and recovery engine,'' in \emph{2009
  IEEE/IFIP International Conference on Dependable Systems \& Networks}, 2009,
  pp. 439--448.

\bibitem{r2_ref3}
Y.~Zhang, J.~Liu, and A.~S. Namin, ``Optimal decision-making approach for cyber
  security defense using game theory and intelligent learning,'' \emph{Sec. and
  Commun. Netw.}, vol. 2019, jan 2019.

\bibitem{r2_ref4}
H.~Zhang, Y.~Mi, X.~Liu, Y.~Zhang, J.~Wang, and J.~Tan, ``A differential game
  approach for real-time security defense decision in scale-free networks,''
  \emph{Computer Networks}, vol. 224, p. 109635, 2023.

\bibitem{r2_ref2}
X.~Liu, H.~Zhang, Y.~Zhang, L.~Shao, and J.~Han, ``Active defense strategy
  selection method based on two-way signaling game,'' \emph{Security and
  Communication Networks}, vol. 2019, pp. 1--14, 11 2019.

\bibitem{10.1145/2764468.2764478}
M.-F. Balcan, A.~Blum, N.~Haghtalab, and A.~D. Procaccia, ``Commitment without
  regrets: Online learning in stackelberg security games,'' in
  \emph{Proceedings of the Sixteenth ACM Conference on Economics and
  Computation}, ser. EC '15.\hskip 1em plus 0.5em minus 0.4em\relax New York,
  NY, USA: Association for Computing Machinery, 2015, p. 61–78.

\bibitem{Lisy_Davis_Bowling_2016}
V.~Lisy, T.~Davis, and M.~Bowling, ``Counterfactual regret minimization in
  sequential security games,'' \emph{Proceedings of the AAAI Conference on
  Artificial Intelligence}, vol.~30, no.~1, Feb. 2016.

\bibitem{HUANG2020101660}
L.~Huang and Q.~Zhu, ``A dynamic games approach to proactive defense strategies
  against advanced persistent threats in cyber-physical systems,''
  \emph{Computers \& Security}, vol.~89, p. 101660, 2020.

\bibitem{harsanyi_2}
J.~Harsanyi, ``Games with incomplete information played by "bayesian" players,
  i-iii part i. the basic model,'' \emph{Management Science}, vol.~14, no.~3,
  pp. 159--182, 1967.

\bibitem{simon_original_br}
H.~A. Simon, ``A behavioral model of rational choice,'' \emph{The Quarterly
  Journal of Economics}, vol.~69, no.~1, pp. 99--118, 1955.

\bibitem{8603817}
Z.~Ni and S.~Paul, ``A multistage game in smart grid security: A reinforcement
  learning solution,'' \emph{IEEE Transactions on Neural Networks and Learning
  Systems}, vol.~30, no.~9, pp. 2684--2695, 2019.

\bibitem{yin2023zeroshot}
M.~Yin, T.~Li, H.~Lei, Y.~Hu, S.~Rangan, and Q.~Zhu, ``Zero-shot wireless
  indoor navigation through physics-informed reinforcement learning,'' 2023,
  \url{https://arxiv.org/abs/2306.06766}.

\bibitem{r5_ref2}
G.~Farina and T.~Sandholm, ``Model-free online learning in unknown sequential
  decision making problems and games,'' \emph{Proceedings of the AAAI
  Conference on Artificial Intelligence}, vol.~35, no.~6, pp. 5381--5390, May
  2021.

\bibitem{bellman1957markovian}
R.~Bellman, ``A markovian decision process,'' \emph{Journal of Mathematics and
  Mechanics}, vol.~6, no.~5, pp. 679--684, 1957.

\bibitem{bertsekas2019reinforcement}
D.~Bertsekas, \emph{Reinforcement learning and optimal control}.\hskip 1em plus
  0.5em minus 0.4em\relax Athena Scientific, 2019.

\bibitem{Ash:2000uj}
R.~B. Ash and C.~Doléans-Dade, \emph{\BIBforeignlanguage{English}{{Probability
  and Measure Theory}}}, ser. Academic Press.\hskip 1em plus 0.5em minus
  0.4em\relax Academic Press, 2000.

\bibitem{Tulcea49}
C.~T. {Ionescu Tulcea}, ``Mesures dans les espaces produits,''
  \emph{Lincei--Rend. Sc. fis. mat. e nat.}, vol.~7, pp. 208--211, 1949.

\bibitem{pollard_2001}
D.~Pollard, \emph{A User's Guide to Measure Theoretic Probability}, ser.
  Cambridge Series in Statistical and Probabilistic Mathematics.\hskip 1em plus
  0.5em minus 0.4em\relax Cambridge University Press, 2001.

\bibitem{billing}
P.~Billingsley, \emph{Convergence of probability measures}, 2nd~ed., ser. Wiley
  Series in Probability and Statistics: Probability and Statistics.\hskip 1em
  plus 0.5em minus 0.4em\relax New York: John Wiley \& Sons Inc., 1999, a
  Wiley-Interscience Publication.

\bibitem{aliprantis06}
C.~D. Aliprantis and K.~C. Border, \emph{Infinite Dimensional Analysis: a
  Hitchhiker's Guide}.\hskip 1em plus 0.5em minus 0.4em\relax Berlin; London:
  Springer, 2006.

\end{thebibliography}
\end{document}